# THE ROLE OF STRANGENESS IN ULTRARELATIVISTIC NUCLEAR COLLISIONS[a]

**Josef Sollfrank**
*Research Institute for Theoretical Physics, P.O. Box 9,*
*FIN–00014 University of Helsinki, Finland*

and

**Ulrich Heinz**
*Institut für Theoretische Physik, Universität Regensburg,*
*D–93040 Regensburg, Germany*

## ABSTRACT

We review the progress in understanding the strange particle yields in nuclear collisions and their role in signalling quark-gluon plasma formation. We report on new insights into the formation mechanisms of strange particles during ultrarelativistic heavy-ion collisions and discuss interesting new details of the strangeness phase diagram. In the main part of the review we show how the measured (multi-)strange particle abundances can be used as a testing ground for chemical equilibration in nuclear collisions, and how the results of such an analysis lead to important constraints on the collision dynamics and space-time evolution of high energy heavy-ion reactions.

---

# Contents

# THE ROLE OF STRANGENESS IN ULTRARELATIVISTIC NUCLEAR COLLISIONS

**Josef Sollfrank**
*Research Institute for Theoretical Physics, P.O. Box 9,
FIN-00014 University of Helsinki, Finland*

and

**Ulrich Heinz**
*Institut für Theoretische Physik, Universität Regensburg,
D-93040 Regensburg, Germany*

## ABSTRACT

We review the progress in understanding the strange particle yields in nuclear collisions and their role in signalling quark-gluon plasma formation. We report on new insights into the formation mechanisms of strange particles during ultrarelativistic heavy-ion collisions and discuss interesting new details of the strangeness phase diagram. In the main part of the review we show how the measured (multi-)strange particle abundances can be used as a testing ground for chemical equilibration in nuclear collisions, and how the results of such an analysis lead to important constraints on the collision dynamics and space-time evolution of high energy heavy-ion reactions.

## 1 Introduction

The high energy heavy-ion program at the Brookhaven Alternating Gradient Synchrotron (AGS) and at the CERN Super Proton Synchrotron (SPS) is devoted to the investigation of nuclear matter under extreme conditions (high temperatures and/or densities). The most exciting prospect is the possible observation of a phase transition from normal nuclear matter to a so-called *Quark-Gluon Plasma*[1] (QGP). A large amount of experimental and theoretical effort has gone into working out clear signatures for the QGP. Although the ultimate hope is to eventually see the QGP "shine" via direct electromagnetic radiation of photons and lepton pairs[2], for practical reasons (larger signals and better understanding of backgrounds) experimental efforts have up to now focussed on the production of hadrons with newly produced quark flavors (strangeness, charm, etc.). At present the largest body of data with the so far most promising indications for unconventional physics has been accumulated in the sector of strange particles, following the early suggestions of Rafelski[3] well over 10 years ago.

There are two major aspects of the so-called "strangeness signal": The more speculative one is connected with the search for exotic strange matter, which requires the intermediate production of a QGP state. The list of these so far undetected phenomena contains strangelets[4-6] (small droplets of metastable strange quark matter - the ultimate proof of the existence the quark-gluon plasma), but also the H-dibaryon[7] and more recently[8] the so-called MEMOs (Metastable Exotic Multistrange Objects). For this aspect of the strangeness signal we refer the reader to Refs. 9, 10 and to the contribution by C. Greiner and J. Schaffner in this volume.

The other, more conventional aspect of strangeness as a QGP signature was introduced by Rafelski who predicted an enhanced production of strange hadrons in heavy-ion collisions when compared to nucleon-nucleon or nucleon-nucleus collisions at the same energy[3,11]. The idea is simple and is based on the different production mechanisms of strangeness in a QGP (individual strange quark pair production from a dense system of gluons and light quarks) and in a hadron gas (production of hadrons with opposite strangeness in inelastic hadron-hadron interactions), which have strongly different intrinsic time scales. But the quantitative interpretation of the observed enhanced nuclear strangeness production as a QGP signal is a subtle issue. Clear and unambiguous statements require a quantitative theoretical understanding of strangeness production in a QGP and in a hadronic environment (including the initial pre-equilibrium stages), of the evolution of strangeness during the fireball life time, in particular during the difficult hadronization stage, and of the variations of the *non*-strange multiplicities as one goes from nucleon-nucleon to nuclear collisions. Some of these issues are still not satisfactorily resolved, but significant progress has been made in the last few years which we will review in this article.

An important, not always appreciated advantage of strangeness is the large variety of strange particles species. Due to the long life time of strange quarks against weak decay, one has by now been able to identify and measure the yields and spectra of 10 different species of strange hadrons in heavy-ion experiments ($K^+$, $K^-$, $K^0_s$, $\phi$, $\Lambda$, $\bar{\Lambda}$, $\Xi^-$, $\bar{\Xi}^+$, $\Omega^-$, $\bar{\Omega}^+$). An up-to-date list of references is given in Tables 8 and 9 in the Appendix; an account of the latest results can be found in Refs. 12, 13. These data triggered a lot of studies within various models for nuclear collisions. Especially thermal models[14-26] which are characterized by a rather small number of adjustable parameters and thus predict various restrictive relations between the various particle species can thus be tested very precisely. We will devote a large fraction of this article to these investigations, not only because large progress has been made in particular in this sector, but also because these studies provide a simple and intuitive picture which gives valuable qualitative and quantitative insight into the physics of hadron production in nuclear collisions. The thermal picture allows for a straightforward investigation of the question of chemical equilibrium in the strange and non-strange sectors and provides a simple characterization of the final stages of the the heavy-ion collision in terms of freeze-out parameters, entropy production, and collective flow dynamics. It provides an intuitive bridge between the data on the one hand and complete kinetic simulations of the phase-space evolution of the collision region on the other hand.

In this review we concentrate on the developments after the publication of the first volume of this book[1]; for the earlier work we refer to the reviews by Koch[9,27]. In addition to the discussion of strange hadron yields and momentum spectra we also report on some recent progress on the the microscopic mechanisms of strange particle production in a hot and dense environment and on the general structure of the strangeness phase diagram.

The article is divided into the following parts. In Section 2 we cover some new developments on strangeness production mechanisms. Section 3 provides the basic equations for the thermodynamics of hadronic matter. These are used in Section 4 to study the phase diagram of strange matter. In Section 5 we discuss additional ingredients like collective flow and deviations from chemical equilibrium which are necessary when testing the thermal model against experimental data. This is then done in Section 6 for various heavy experiments on strange particle production. In Section 7 we shortly discuss the aspect of entropy production in these heavy-ion experiments. A comprehensive discussion of the outcome of the thermal analysis follows in Section 8. In Section 9 we draw some speculative conclusions from this discussion on the nature of the hadronization and freeze-out processes and discuss the internal consistency of our interpretation of the strange particle abundances. We conclude with an outlook in Section 10.

## 2 Strangeness Production Mechanisms

The usefulness of strange particles as probes for the physics of a nuclear collision arises from certain characteristic features in their production and kinetic evolution. In contrast to pions, for example, which are the most efficient carriers of entropy and whose final abundance is thus more or less determined already in the very first, hard collision stage of the reaction where most of the entropy is produced[28] – the transparency of nuclear matter is the important ingredient –, the abundance of strange quarks continues to evolve until the very end of the collision fireball. Thus all stages contribute to the final strangeness yield and must be considered.

The first pre-equilibrium stage is dominated by the physics of hard processes. The amount of strange particles created in these hard collisions is rather uncertain. For the present highest energy range ($\sqrt{s} = 20$ A GeV) this phase is difficult to handle quantitatively because it is still dominated by non-perturbative QCD. At higher energies the parton model can be more reliably used; at $\sqrt{s} = 200$ A GeV, for example, the amount of strangeness produced in the very early collisions has been calculated in the Parton Cascade Model to be about 40 % of the equilibrium value[28]; in the HIJING model, which uses a different soft momentum cutoff procedure, the initial strangeness fraction is somewhat lower[29]. For the lower AGS and SPS energies one must instead make an educated guess, and one generally assumes an initial fraction of strange particles as observed in the final state of $p+p$ collisions at the same energy, after correcting for resonance decays[30], which corresponds to roughly 20% of the equilibrium value[20].

If the collision region is initially in the deconfined phase, thermal equilibration occurs at a very rapid rate[28,31]. Chemical equilibration of gluons and light quarks is expected to follow shortly after[31] although this question is still debated among the various high energy event generators[28,29]. Ideally one would thus after about 2-3 fm/$c$ create a thermally and chemically equilibrated QGP consisting of (mostly) light quarks and gluons. Due to the strange quark mass threshold, strangeness production only follows later. Given the above scenario, the production rate and strangeness equilibration time scale can be calculated with the methods of thermal equilibrium field theory, and if the temperature is sufficiently high, even perturbative QCD can be applied. While the first calculations used bare quark and gluon propagators[32,33], recent investigations have begun to incorporate the new knowledge of how to include certain collective and non-perturbative effects at high temperature via resummation of the perturbative series. These new developments will be described in Section 2.1.

An alternative scenario is a nuclear collision without a deconfining phase transition. Then the relevant degrees of freedom are the hadrons. High temperatures and densities will affect the hadron dynamics, their masses and their resonance widths, and thus the strangeness production. Of particular relevance is a possible decrease of the hadron masses. In Section 2.2 we will report some calculations which take into account this effect for strangeness production.

*2.1 Quark-Gluon Production Mechanisms*

The production rates of strange quarks in a QGP are calculated by perturbative QCD. At finite temperature, however, any massless theory is harmed by severe infrared singularities which are difficult to deal with, especially for gauge theories. Braaten and Pisarski[34] developed a resummation technique for the endangered low momentum modes at finite temperature and density calculations. The main idea is to use for light particles with momenta of order $q \lesssim gT$ ($g \ll 1$) a resummed propagator which contains self-energy corrections due to the medium. In lowest order of the $g$ the low momentum particles in a thermal environment thereby acquire a thermal mass $m_{\text{th}}$, which is given by

$$m_g = (N_c + N_f/2)g^2 T^2/9 \qquad (1)$$

for gluons and by

$$m_q = m_q^0 + g^2 T^2/6 \qquad (2)$$

for quarks. These "massive" quark and gluon modes of finite temperature QCD are analogues of the well-known plasmon in the theory of electron gases and electromagnetic plasmas. At next to leading order in $g$ these modes also acquire a thermal width. In addition to the low-momentum propagators, also some vertices in which all momenta are soft of order $gT$ must be resummed[34].

The new technique was used to recalculate the production rates for strange quarks in a QGP[35,36]. The Feynman diagrams for $s\bar{s}$ production in lowest order QCD are shown in Figure 1. There are three different groups: quark anti-quark collisions (a), gluon fusion (b) and thermal gluon decay (c). The third process, which is not allowed for a bare gluon propagator ($m_g = 0$), is now possible because of the finite mass and width of the thermal gluons.

The expressions for the rates of the processes depicted in Figure 1 are given by[36]:

$$R_{q\bar{q}\to s\bar{s}} = \int \frac{d^3 p_q}{(2\pi)^3 2E_q} \frac{d^3 p_{\bar{q}}}{(2\pi)^3 2E_{\bar{q}}} \frac{d^3 p_s}{(2\pi)^3 2E_s} \frac{d^3 p_{\bar{s}}}{(2\pi)^3 2E_{\bar{s}}} (2\pi)^4 \delta(P_q + P_{\bar{q}} - P_s - P_{\bar{s}})$$
$$\times f_F(E_q) f_F(E_{\bar{q}})(1 - f_F(E_s))(1 - f_F(E_{\bar{s}})) \sum |M(q\bar{q} \to s\bar{s})|^2, \qquad (3)$$

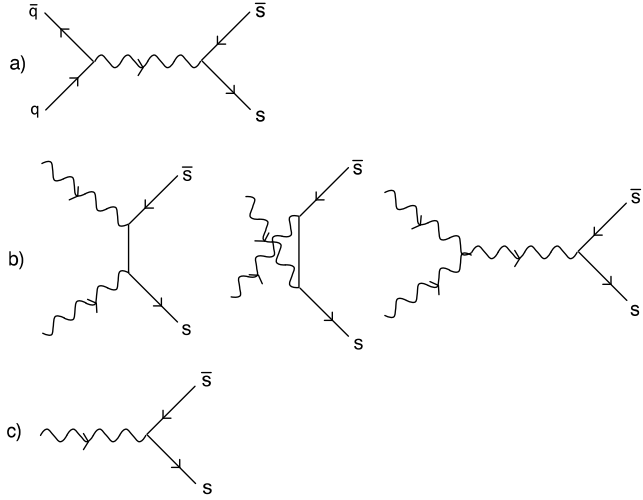

**Figure 1:** Feynman diagrams for $s\bar{s}$ production in lowest order QCD.
(a) $q\bar{q} \longrightarrow s\bar{s}$
(b) $gg \longrightarrow s\bar{s}$
(c) $g \longrightarrow s\bar{s}$

$$R_{gg \to s\bar{s}} = \frac{1}{2} \int \frac{d^3p_1}{(2\pi)^3 2E_1} \frac{d^3p_2}{(2\pi)^3 2E_2} \frac{d^3p_s}{(2\pi)^3 2E_s} \frac{d^3p_{\bar{s}}}{(2\pi)^3 2E_{\bar{s}}} (2\pi)^4 \delta(P_1 + P_2 - P_s - P_{\bar{s}})$$
$$\times f_B(E_1) f_B(E_2)(1 - f_F(E_s))(1 - f_F(E_{\bar{s}})) \sum |M(gg \to s\bar{s})|^2 \quad (4)$$

and

$$R_{g \to s\bar{s}} = \int \frac{d^3q}{(2\pi)^3 2E_g} \frac{d^3p_s}{(2\pi)^3 2E_s} \frac{d^3p_{\bar{s}}}{(2\pi)^3 2E_{\bar{s}}} (2\pi)^4 \delta(Q - P_s - P_{\bar{s}})$$
$$\times f_B(E_g)(1 - f_F(E_s))(1 - f_F(E_{\bar{s}})) \sum |M(g \to s\bar{s})|^2. \quad (5)$$

Capital letters correspond to four-momenta and $f_B$ ($f_F$) denote the Bose-Einstein (Fermi-Dirac) distributions, respectively.

An extensive discussion of the calculation of the first two processes of Figures 1a) and 1b) with bare propagators and vertices can be found in the early literature (see Refs. 9, 27, 33). A complete new calculation of the relevant matrix elements with resummed propagators and vertices is not easy and therefore has so far not been presented; work in this direction is in progress[37]. The new calculations published so far[35-38] simply replace all bare propagators in Figure 1 by thermal propagators with a thermal mass and width. In this approximation the thermal gluon decay, Figure 1c), is then the most interesting new phenomenon and has therefore received most of the attention. In view of the incompleteness of these calculations it is not surprising that the results are controversial. The thermal gluon decay into strange quark pairs was first evaluated in Ref. 38. The motivation for a finite gluon mass was at that time taken from lattice QCD results[39] rather than from resummed thermal perturbation theory. Gluons with masses around 500 MeV fit the data on energy density and pressure of a $SU(3)_c$ lattice calculation[39] above $T_c$ very well. The rate

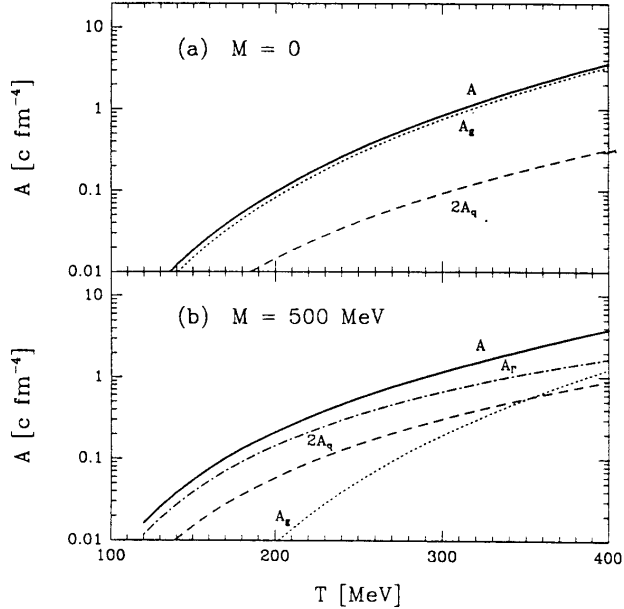

**Figure 2:** Partial and total rates $A = d^4 N_s / dx^4$ for strange quark production, according to the calculation in Ref. 38: $q\bar{q} \to s\bar{s}$ (dashed line), $gg \to s\bar{s}$ (dotted line), $g \to s\bar{s}$ (dashed-dotted line) and $M$ denotes the used constant thermal gluon mass.

(5) was calculated with a constant, temperature independent gluon mass, and the quarks have been assumed to have their constituent masses.

The result is shown in Figure 2. The upper plot shows the old calculation[32,40] with $m_g = 0$ (bare gluon propagator), the lower plot a calculation with $m_g = 500$ MeV. In the first case the gluon fusion process is by far the dominant source of strange quark production, dominating the production from light quarks by an order of magnitude. A finite gluon mass decreases the gluon fusion rate, because of the strongly reduced Bose-factors (see Eq. (4)) for the incoming gluons. On the other hand it raises the strangeness production rate from light $q\bar{q}$ pairs slightly because the intermediate gluon is now less off-shell due to its thermal mass. The dominant process, however, is now the thermal gluon decay, because (in contrast to the gluon fusion) only one thermal distribution function appears in the incoming channel (see Eq. (5)). Note also that this process is of order $g$, while the other processes are of order $g^2$. Surprisingly, in spite of these drastic changes in the individual production channels, the total rate

$$A = R_{tot} = R_{q\bar{q} \to s\bar{s}} + R_{gg \to s\bar{s}} + R_{g \to s\bar{s}} \qquad (6)$$

stays nearly unchanged[38]. The authors conclude that the statements about the equilibration time scales in Ref. 33 are nearly unchanged and the gluons remain the main strangeness producer.

Two recent calculations[35,36] tried to improve this analysis by using the thermal propagators of Ref. 34. We will shortly describe the calculation of Ref. 36. The thermal gluon propagator may be written as

$$iD^{ab}_{\mu\nu}(q_0,q) = -i\delta^{ab}\left[P^T_{\mu\nu}\Delta_T(q_0,q) + P^L_{\mu\nu}\Delta_L(q_0,q)\right], \qquad (7)$$

where $P^T_{\mu\nu}$ and $P^L_{\mu\nu}$ are transverse and longitudinal projectors, respectively, and the corresponding propagators are

$$\Delta_{T,L}(q_0,q) = \frac{1}{Q^2 - \Pi_{T,L}(q_0,q)}. \qquad (8)$$

$\Pi_{T,L}(q_0,q)$ denotes the self-energy correction due to a thermal loop[34]. The poles in the propagator (8) are determined by the dispersion relation

$$q_0^2 = q^2 + \Pi_{T,L}(q_0,q). \qquad (9)$$

The poles are located at

$$q_0 = \omega_{T,L} + i\gamma_{T,L}, \qquad (10)$$

where the imaginary shift of the pole $\gamma_{T,L}$ is related to the imaginary part of the self energy through

$$\gamma_{T,L} = \text{Res}(\Delta_{T,L})\,\text{Im}\,\Pi_{T,L}, \qquad (11)$$

with the residue of the propagator at the pole given by

$$\text{Res}(\Delta_{T,L})^{-1} = \left.\frac{\partial \Delta_{T,L}^{-1}}{\partial q_0}\right|_{\omega_{T,L}}. \qquad (12)$$

The expression for the gluon propagator can then be rewritten as

$$\Delta(Q) = \frac{Q^2 - \text{Re}\,\Pi}{(Q^2 - \text{Re}\,\Pi)^2 + \gamma^2\text{Res}(\Delta)^{-2}} + \frac{i\gamma\text{Res}(\Delta)^{-1}}{(Q^2 - \text{Re}\,\Pi)^2 + \gamma^2\text{Res}(\Delta)^{-2}}, \qquad (13)$$

where we suppressed the subscripts $T, L$. This expression is used to replace the mass-shell $\delta$-function for thermal gluons:

$$\delta(Q^2 - m_g^2) \to \frac{1}{\pi}\frac{\gamma_{T,L}\text{Res}(\Delta_{T,L})^{-1}}{(Q^2 - \text{Re}\,\Pi_{T,L})^2 + \gamma_{T,L}^2\text{Res}(\Delta_{T,L})^{-2}}. \qquad (14)$$

The evaluation of Eq. (5) is then done[36] neglecting the Pauli blocking factors. The result is[35,36]

$$R^T_{g\to s\bar{s}} = \frac{2g^2}{3\pi^4}\int_{2m_s}^{\infty} dq_0\, f_B(q_0)\int_0^{\sqrt{q_0^2 - 4m_s^2}} dq\, q^2\, (Q^2 + 2m_s^2)\sqrt{1 - \frac{4m_s^2}{Q^2}}$$

$$\times \frac{\gamma_T\text{Res}(\Delta_T)^{-1}}{(Q^2 - \text{Re}\,\Pi_T)^2 + \gamma_T^2\text{Res}(\Delta_T)^{-2}}, \qquad (15)$$

plus a similar expression for the decay rate of the longitudinal gluons.

The biggest uncertainty in such a calculation is the determination of the imaginary part of the pole $\gamma_{T,L}$ in the gluon propagator. $\gamma_T$ is the damping rate of transverse plasma oscillations in a QGP at high temperature. For plasmons with nonzero spatial momentum relative to the heat bath $\gamma_T$ is not perturbatively calculable, but can be related[41] to the so-called magnetic mass $m_{\text{mag}}$ (inverse magnetic screening length) in the limit $m_{\text{mag}} \gg \gamma_T$ by the following self-consistency condition:

$$\gamma_T = \frac{g^2 N_c T}{8\pi} \left[ \ln\left(\frac{m_g^2}{m_{\text{mag}}^2 + 2 m_{\text{mag}} \gamma_T}\right) + 1.1 \right]. \tag{16}$$

$m_{\text{mag}}$ itself is given at high temperatures by

$$m_{\text{mag}} = c_{N_c} g^2 T, \tag{17}$$

where $c_{N_c}$ is a perturbatively uncalculable number which so far can only be determined from lattice QCD calculations.

The further evaluation of (15) is done differently in Refs. 35, 36. In both papers the approximation $\gamma_L = \gamma_T \equiv \gamma$ is assumed, but in Ref. 35 Eq. (16) was approximated by the $m_{\text{mag}}$ independent expression

$$\gamma = -\alpha_s N_c T/2 \ln \alpha_s, \tag{18}$$

while in Ref. 36 Eq. (16) was expanded in powers of $\gamma/m_{\text{mag}}$

$$\gamma_T = (1+\eta)^{-1} \frac{g^2 N_c T}{8\pi} \left[ \ln\left(\frac{m_g^2}{m_{\text{mag}}^2}\right) + 1.1 \right], \tag{19}$$

where $\eta = 0$ corresponds to leading order and $\eta = N_c/(4\pi c_{N_c})$ to next to leading order. In the calculation $c_{N_c} = c_3 = 0.405$ was chosen[36]. Further Altherr et al. used for $\text{Re}\,\Pi_T(q) = 3 m_g^2/2$ the asymptotic limit for large $q$ and $q_0$ while Bilić et al. evaluated the dispersion relation (9) numerically in the high temperature (hard thermal loop) approximation.

The results of both calculations are shown in Figure 3, with $g = 2$ and $N_c = 3$. Figure 3a shows a clear dominance of the thermal gluon decay in the mass region relevant for strangeness production ($m_s/T \lesssim 1$). But the other two channels are probably underestimated. They were calculated in the limit $M \gg T$ from (3) and (4), respectively. An exact numerical treatment[36] of Eq. (3) and (4) shows a higher rate for these processes as seen in Figure 3b. The solid line corresponds to the sum of light quark-antiquark scattering and gluon fusion; the gluon decay is shown as dashed lines for various approximations for the damping rate. We see that the $q\bar{q}$ and gluon fusion processes still dominate the strangeness production rate. According to Ref. 36, the gluon fusion process is the more important one of the two channels, exactly as in the original tree-level calculation[32]. Note the large influence of the so far unknown gluon damping rate on the gluon decay into $s\bar{s}$ pairs: a large damping rate, i.e. a large gluon width, enhances the strangeness production rate drastically. In view of this sensitivity the approximation $\gamma_T = \gamma_L$ should be reinvestigated and, if necessary, improved.

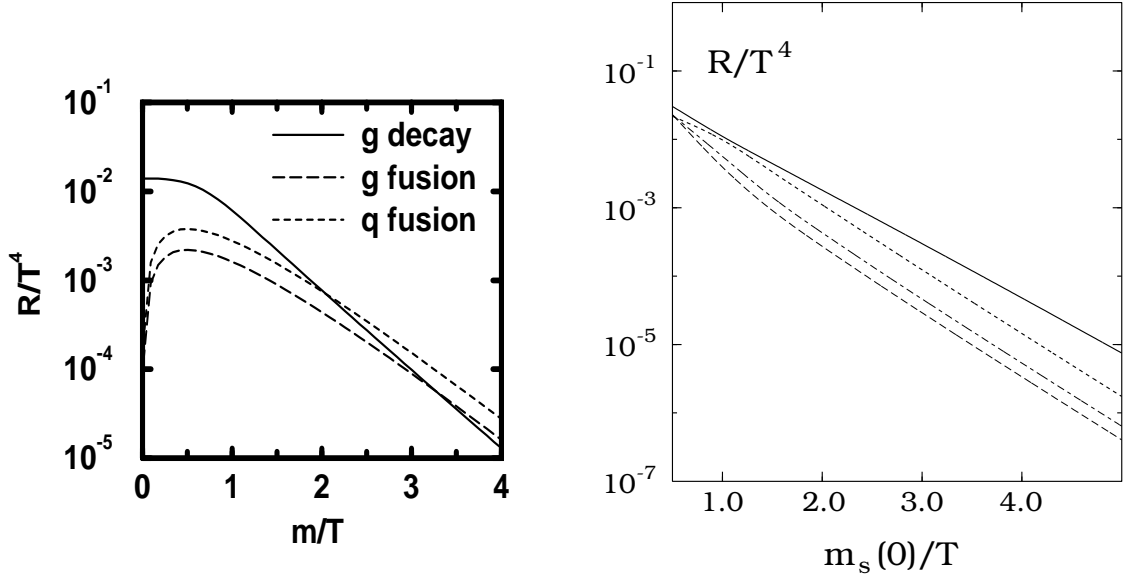

**Figure 3a:** Strange quark production rates for different channels according to the calculations of Ref. 35, from where this figure was taken. For details see text.

**Figure 3b:** Strange quark production rates for thermal gluon decay with different damping rates $\gamma$. The upper short dashed line corresponds to the damping rate of Ref. 35, Eq. (18), the next lower dashed line to $\eta = 0$ and the lowest dashed one to $\eta = N_c/(4\pi c_{N_c})$ in Eq. (19). The solid line is the sum of the gluon fusion and quark anti-quark collision rates. Figure from Ref. 36.

The physically most relevant quantity is the total strangeness equilibration time $\tau_s$. The additional channel of thermal gluon decay can only shorten $\tau_s$, but quantitatively its influence is not really very drastic[35−38]. For temperatures around 200 MeV, $\tau_s$ is of the order of $\gtrsim 10$ fm/c; it strongly depends on $T$, the strange quark mass and the coupling constant $\alpha_s$. The controversy about the importance of the thermal gluon decay does not affect the conclusion that it won't be able to reduce the strangeness equilibration time to values well below 10 fm/c. Thus, even in a heavy-ion induced QGP, full strangeness equilibration remains questionable. Calculating the thermal gluon decay is a first attempt to include non-perturbative features in the strangeness production. At temperatures around 200 - 300 MeV such non-perturbative aspects play an important role, and therefore many more detailed investigations of this question (in particular the consistent implementation of the real and imaginary parts of the thermal propagators also in the diagrams 1a,b[37]) are necessary to obtain a reliable picture.

*2.2 Hadronic Production Mechanisms*

It is generally assumed that with the beam energies available at the AGS and SPS we may reach energy densities close to, but certainly not very far above the deconfinement phase transition. It is therefore a possibility that in the present heavy-ion experiments the collision zone spends all or a large fraction of its lifetime in the confined hadronic phase. With the critical temperature and energy density from lattice QCD now apparently having settled firmly[42] at $T_c \lesssim 150$ MeV and $\varepsilon_c \lesssim 1$ GeV/fm$^3$, we must, however, expect some deviations from the often used picture of a non-interacting mixture of hadronic resonances[43]. Being close to the point of deconfinement and chiral symmetry restoration, the masses and widths of the hadrons as well as their interaction cross sections may already be strongly modified by the medium compared to their vacuum values.

Most authors working on this problem agree that the masses of most of the hadrons decrease with density and temperature[44-47]. The Goldstone bosons of chiral symmetry breaking (i.e. the pions and kaons) are an exception: their character protects their masses from large changes with temperature[48,49] and density[50]. The reason for that is an approximate compensation of two medium effects. One can view these particles as bound states of constituent quarks with a very large negative binding energy. This results in a very low total energy, i.e. mass of these particles. The constituent quark mass is generated by chiral symmetry breaking. By heating the system the constituent quark mass decreases, because chiral symmetry gets restored, but so does the absolute value of the binding energy. Altogether the Goldstone masses are nearly stable, and if anything their masses tend to slightly increase[48-50] as the chiral phase transition is approached.

Another controversial particle is the $\rho$ meson whose mass some authors[45,51] expect to drop in the medium while others[52] expect it to rise. Decreasing masses of heavier particles and resonances open the phase space for their production. This leads to the expectation that the mostly heavy strange particles are produced with larger rates in hot and dense hadronic matter[53] compared to calculations using free cross sections and mass thresholds.

The effect of the medium on the hadronic production of strange particles has been studied mainly by Ko and his group; their work includes the production of kaons[54], $\phi$ mesons[55], and $\Lambda$'s and $\bar{\Lambda}$'s[56].

*Kaon production*

Hadronic kaon production in heavy ion collisions proceeds mainly through the following three production channels ($M$: non-strange meson; $B$: non-strange baryon)

$$\begin{aligned}
&\text{(a)} \quad BB \longrightarrow KYB \\
&\text{(b)} \quad MB \longrightarrow KY \\
&\text{(c)} \quad MM \longrightarrow K\bar{K} \, .
\end{aligned} \qquad (20)$$

The channel (a) is the dominant production mechanism in collisions of protons with hydrogen or light nuclear (e.g. Be) targets. There the $K^+/\pi^+$-ratio at AGS energies (10 - 14 A GeV incident momentum) is about[57] 0.04 . For heavier targets (e.g. Au) the ratio rises at the same incident momentum to[129] $K^+/\pi^+ \approx 0.1$ , indicating that probably also channel (20b) begins to contribute. Channel (20c) will become important once a substantial pion density has emerged. This is supposed to be the case in heavy ion collisions, such as Si+Au, where indeed $K^+/\pi^+ \approx 0.2$ at the AGS[128]. Semi-hydrodynamical[b] calculations[58] and also some Monte Carlo calculations[59], using free strangeness production cross sections from experiment, fail to reproduce the $K^+/\pi^+$-ratio of the AGS Si+Au experiment. The event generators for nuclear collisions either reproduce the pions correctly, underestimating somewhat the $K^+$ yield, as e.g ARC[60], or correctly reproduce the kaon yields at the expense of slightly overpredicting the pions, as RQMD[61]. So far these calculations have ignored medium effects on the involved interactions; it is conceivable that somewhat larger elementary cross sections might give better agreement with the data.

Ko et al. have parametrized the dependence of the hadronic masses on $T$ and $\rho_B$ (baryon density) by[54]

$$\frac{m^*}{m} = \left[1 - \left(\frac{T}{T_c}\right)^2\right]^n \left[1 - \frac{\lambda}{2}\left(\frac{\rho_B}{\rho_0}\right)\right] , \qquad (21)$$

where the detailed behavior is model dependent. $T_c$ is the deconfinement temperature (taken as $T_c = 195$ MeV in the studies described below), $\rho_0$ is the nuclear ground state density ($\rho_0 = 0.17$ fm$^{-3}$), and $\lambda$ is an empirical parameter, determined as $\lambda \approx 0.2 - 0.4$ from $K$, $p$, and $e^-$ scattering on nuclei[62]. The power $n$ depends on the model, in which the temperature dependence is studied. $n = 1/6$ is obtained from QCD sum rule calculations[46], while $n = 1/2$ is favored by chiral models[47].

The density dependence of the kaon mass[63,53] differs from the mass scaling (21). Within linear chiral perturbation theory the kaon mass behaves like[53]

$$\frac{m_K^*}{m_K} = \left(1 - \frac{\rho_B}{\rho_c}\right)^{1/2} , \qquad (22)$$

where $\rho_c = f_K^2 m_K^2/\Sigma^{KN}$ is the critical density for kaon condensation ($f_K = 93$ MeV is the kaon decay constant, $\Sigma^{KN} = 45$ MeV is the kaon-nucleon sigma term). Eq. (22) reflects an exceptional behavior of the kaons: although the kaon belongs to the Goldstone octet in a SU(3)$_f$ representation for the flavor group, its mass *decreases* with density in contrast to the density scaling of Goldstone bosons discussed before. According to the authors[53] the reason is the large explicit chiral symmetry breaking by the strange quark mass. This issue is, however, still under debate. Within the Nambu-Jona-Lasinio model the tendency for kaon condensation (22) is not supported: Lutz et al. get an increase in mass for the $K^+$, while the $K^-$ stays nearly unchanged as a function of density[63]. More recent calculations in the framework of chiral perturbation theory lead to a rising $K^+$ and a decreasing $K^-$ mass[64]; including higher orders of the approximation within a coupled channels approach reduces somewhat the decrease of the $K^-$ mass[65], and additionally coupling in the $\Lambda^*(1405)$ resonance essentially destroys any indications for $\bar{K}$ condensation at high baryon densities[66].

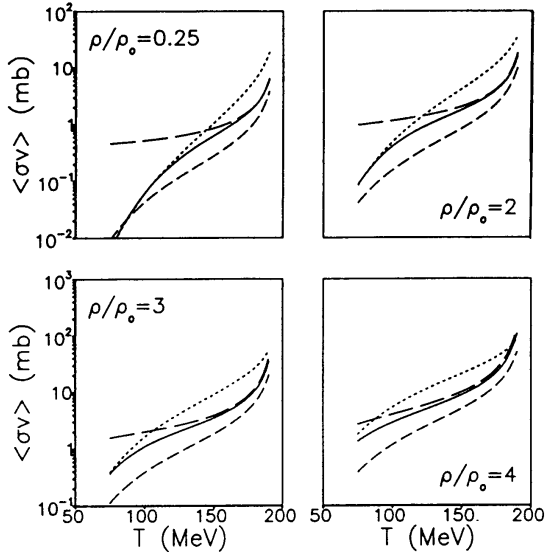

**Figure 4:** Temperature and density dependence of $\langle \sigma v \rangle$ for $\pi\pi \to K\bar{K}$ (dotted curve), $\pi\rho \to K\bar{K}$ (short-dashed curve), and $\rho\rho \to K\bar{K}$ (long-dashed curve). The solid curve is for $MM \to K\bar{K}$, averaging over initial meson distributions. Note the different vertical scales in the upper and lower row. The figure was taken from Ref. 54.

Thermally averaged cross sections $\langle \sigma v \rangle$ of the strangeness producing reactions are needed in the calculation described below[54]. For a process $1 + 2 \to 3 \ldots n$ they are defined by

$$\langle \sigma_{12 \to 3 \ldots n} v_{12} \rangle = \int d^3 k_1 \int d^3 k_2 \; f_1(\vec{k_1}) f_2(\vec{k_2}) \; \sigma_{12 \to 3 \ldots n}(\vec{k_1}, \vec{k_2}) \, v_{12} \;, \qquad (23)$$

where $f_i$ denote the (normalized) thermal distribution functions for the incoming particles, and $v_{12}$ is their relative velocity. The elementary medium cross section $\sigma_{12 \to 3 \ldots n}(\vec{k_1}, \vec{k_2})$ is calculated using established form factors and coupling constants, but masses which satisfy the scaling relations (21) and (22). An example is shown in Figure 4 from Ref. 54, where the temperature and density dependence of mesonic production cross sections are plotted. One sees a clear rise of the cross sections with temperature and baryon density.

The medium cross sections may now be used to calculate the final strange particle abundances in a heavy-ion collision. This requires a model for the time evolution of the nuclear collision. In Ref. 54 the hydrodynamical model of Biro et al.[67] was used. It employs an ideal hadron gas equation of state without phase transition. In addition to the hydrodynamical equations, which determine the overall time evolution of the initial fireball (carried mainly by pions and nucleons), rate equations for the evolution of the strange particle densities were solved. The rate equation for a particle species $s$ has the general form

$$\frac{1}{V}\frac{d(V\rho_s)}{dt} = \sum_{i+j\to s+X} \langle \sigma_{i,j\to s+X} v \rangle \rho_i \rho_j \left( 1 - \frac{\rho_i^0 \rho_j^0 \rho_s \prod_f^{N_{\text{final}}-1} \rho_f}{\rho_i \rho_j \rho_s^0 \prod_f^{N_{\text{final}}-1} \rho_f^0} \right) . \quad (24)$$

---

[b]The expression semi-hydrodynamical will be defined below.

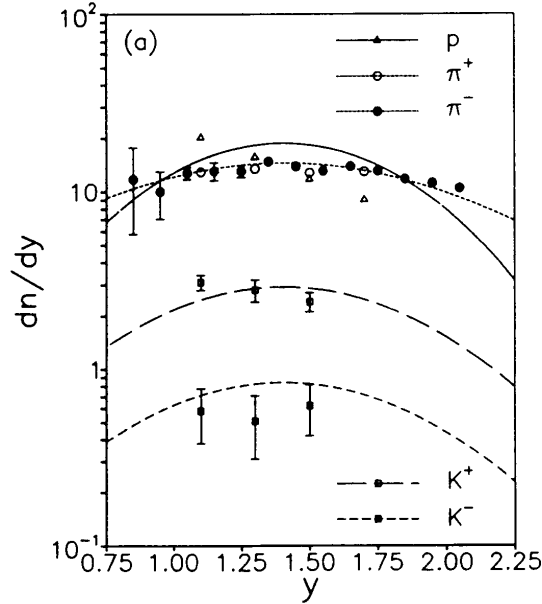

Figure 5: The rapidity distributions of different particles at AGS. The figure is from Ref. 54 where the calculation (various curves) was made, while the data are from Ref. 128.

The summation is meant to extend over all considered processes for producing (eliminating) particles of kind $s$. The superscript 0 indicates the equilibrium densities. Eq. (24) contains gain and loss terms and takes care of the expansion of the fireball volume $V$. Since hydrodynamical models for nuclear collisions usually assume complete chemical equilibrium (even for strange particles) throughout the dynamical evolution, we use the expression *semi-hydrodynamical* for these calculations which decouple some particle species from the bulk evolution and treat them separately.

As initial conditions for Si+Au collisions at 14.6 A GeV a baryon density of $4\rho_0$ and an energy density of 1.4 GeV/fm$^3$ were chosen, corresponding to an initial temperature of about 180 MeV. Figure 5 shows the rapidity distribution of pions, kaons and protons after freeze-out. Freeze-out is assumed to occur spontaneously at a common temperature of $T_f = 125$ MeV.

The agreement with the data is reasonable; the kaon abundances are well reproduced. The authors conclude (in contrast to the earlier work in Ref. 58) that no QGP phase transition is needed in order to understand the kaon yields at the AGS.

It should be stressed that the main mechanism for kaon production in the described analysis is the process $\rho\rho \to K\bar{K}$[54]. The assumed reduction of the $\rho$-mass gives the necessary high initial $\rho$-density; however, as pointed out above, this behavior of the $\rho$ must still be considered as controversial.

*$\phi$ production*

The strangeness enhancement from p+A to A+A collisions is also clearly seen in the $\phi$-meson yield. The NA38 collaboration reported a rise of the $\phi/(\rho+\omega)$-ratio from p+U to S+U by a factor of 3 in the dimyon spectrum (most central collisions)[68]. As originally suggested by Shor[70] it is possible to explain this enhancement as a consequence of QGP formation[71]. However, hadronic rescattering mechanisms with different absorption cross sections for the $\rho$, $\omega$ and $\phi$ mesons by the nuclear medium[71,72] can also account for the observed systematics of NA38, although the required rather large fireball densities and lifetimes present some problems, on the conceptional level[71] as well as in practice, since the expected significant rise of the $\rho$ dilepton yield with the fireball lifetime[73] is in contradiction with the observed independence of the $\rho/\omega$ dilepton peak with collision centrality[74].

There are also theoretical studies predicting a drastic mass reduction of the $\phi$-meson[69] in the medium. In combination with reduced kaon masses it is speculated that this could drastically enhance $\phi$ production[55]. The main hadronic production mechanisms are here

$$\begin{aligned}&(a) \quad K\Lambda \quad \longrightarrow \quad \phi N\,; \\ &(b) \quad K\bar{K} \quad \longrightarrow \quad \phi\rho\,;\end{aligned} \qquad (25)$$

these are the only channels taken into account in Ref. 55 in the rate equation (24) for the $\phi$. The channel $MM \to K\bar{K}$ violates the OZI (Okubo-Zweig-Izuka) rule and is therefore suppressed.

The calculation of Ref. 55 takes similar initial conditions as in Sec. 2.2, but assumes a higher freeze-out temperature of $T_f \approx 140$ MeV at CERN SPS energies. The result is a ratio of $\phi/\omega \approx 20\%$ which was claimed[55] to be in agreement with the measured one of NA38[68] in 200 A GeV S+U collisions. But a direct comparison with data obtained from the leptonic decay channel is problematic: First, the $\rho$ and $\omega$ cannot be resolved experimentally; therefore the $\rho$ should be included in the theoretical calculation to predict the $\phi/(\rho+\omega)$ ratio. Secondly, dilepton spectra are integrated over the whole history of the collision (which is important in particular for the dilepton yield under the $\rho$ peak[73]) while the calculation in Ref. 55 considers only ratios at freeze-out. Thus hadronic mechanisms for the observed $\phi$ enhancement remain a subtle issue.

It is expected that in the medium also the widths of the particles change. The calculation of Ref. 69 within the vector dominance model shows a large broadening of the $\phi$-width with increasing density, and in addition $\phi$ mesons and kaons have reduced effective masses. Contrary, the authors of Ref. 75 claim that within the Nambu-Jona-Lasinio model the width of the $\phi$ decreases with increasing temperature, because the $\phi$ mass goes down, while the kaon mass stays nearly unchanged, such that the decay phase space for the dominant channel $\phi \to K\bar{K}$ closes. On the other hand, an increase of the particle widths in the medium by collision broadening[76] is a generic feature which should be taken into account in any realistic calculation.

In the dilepton spectra no change in mass or width of the $\phi$ has been seen so far within the (rather poor) experimental resolution. In the invariant mass spectra of $K^+K^-$ pairs from Si+Au collisions at the AGS an indication for a small mass shift of the $\phi$ peak has recently been reported[153]; the observed shift amounts to a mass decrease by 3–5 MeV for low-$p_T$ $\phi$'s from very central nuclear collisions. A surprising feature of this reported mass shift is that in the selected kinematic region *all* $\phi$'s appear to be shifted, i.e. the width of the shifted $\phi$ peak is smaller than the mass shift and shows no traces of an unshifted component. Clearly this situation requires further clarification.

*$\Lambda$-$\bar{\Lambda}$ production*

In the context of strangeness enhancement as a QGP signal, the special role of strange antibaryons was pointed out very early[11]. For producing antibaryons an energy threshold of at least 2 GeV must be overcome (taking free hadronic masses). The observed high $\bar{\Lambda}$ yield in 200 A GeV collisions is thus hard to explain by ordinary hadronic mechanisms. Hydrodynamical calculations[77] based on a hadron gas equation of state with a freeze-out temperature[c] around 150 MeV fail to reproduce the $\bar{\Lambda}$'s. The same is true for event generators based on string models, unless string fusion (the formation of color ropes[78] or of quark-gluon clusters[79]) is taken into account. Thus an explanation of the $\bar{\Lambda}$ yield on a microscopic level would give much insight to the nuclear reaction dynamics.

Ko[56] et al. have investigated the $\bar{\Lambda}$ multiplicities from the NA35 experiment[157,165]. The analysis was done in the same manner as in Sections 2.2 and 2.2. An enhancement of $\bar{\Lambda}$'s is expected mainly due to a mass decrease with rising density and temperature.

The mass scaling of the $\bar{\Lambda}$ ($\Lambda$) was deduced from the $SU(3)_f$ Walecka model[80] as

$$\frac{m_\Lambda^*}{m_\Lambda} = 1 - \frac{g_{\sigma\Lambda\Lambda}g_{\sigma NN}}{m_\sigma^2}\rho_S , \qquad (26)$$

with the couplings $(3/2)g_{\sigma\Lambda\Lambda} = g_{\sigma NN} = \sqrt{48}$ and a $\sigma$-meson mass of $m_\sigma = 550$ MeV. $\rho_S$ is the scalar nuclear density.

The main contribution to the $\bar{\Lambda}$ production rate comes from meson collisions which are not suppressed by the OZI rule:

$$\begin{array}{lrcl} (a) & K\pi & \longrightarrow & \bar{\Lambda}N ; \\ (b) & K\rho & \longrightarrow & \bar{\Lambda}N . \end{array} \qquad (27)$$

As initial conditions[56] a temperature of $T_0 = 195$ MeV and a density of twice the nuclear ground state density were chosen. The model results in a multiplicity of 1.5 $\bar{\Lambda}$'s per S+S collision (NA35, Ref. 157: 1.5±0.4 ; Ref. 165: 2.2±0.4), but underestimates the pion, kaon and $\Lambda$ multiplicities by 15%, 30%, and 50% respectively.

---

[c]Higher temperatures would solve the problem, but they are hard to justify.

We see that with certain initial condition and a mass reduction for the $\bar{\Lambda}$ ($\Lambda$) a reasonable value for the $\bar{\Lambda}$ multiplicity results. But the calculation does not give a consistent overall picture. Medium effects for the $\bar{\Lambda}$'s are certainly an important ingredient, but whether the mass reduction alone can account for the observed final yields remains to be proven.

*Further remarks on hadronic strangeness production*

The introduced model raises the question how the particles get on the mass-shell at freeze-out. At the chosen freeze-out temperature the particles still experience a considerable reduction of their free masses. Strictly speaking this is inconsistent with the naive freeze-out concept which assumes a more or less sudden transition from interacting to free-streaming particles. The refined picture behind these calculations assumes that after the "freeze-out" point all two-body collisions have ceased, but the particles still feel an attractive mean field which arises from virtual processes. It was argued[53,81] that on their way out the particles recover their vacuum masses by giving up kinetic energy as they move up the mean field potential well. In this process their momentum changes according to the equation

$$m^{*2} + k^{*2} = m_0^2 + k^2 ,  \qquad (28)$$

where the quantities with a star correspond to the in-medium values at freeze-out. This leads to an accumulation of particles with low asymptotic momenta. Shuryak has shown that such an assumption can explain the low-$p_\perp$ rise of pions[81], and for the recently observed "cold" low momentum kaons at the AGS[147] the same mechanism was suggested[82]. But there are also alternative explanations for the low-$p_\perp$ enhancement discussed in the literature[83], and in particular for the "cold" kaons the experimental situation is not yet completely settled[13]. Thus for the time being the direct experimental evidence for medium effects on the particles widths and masses remains shaky.

The calculations of Ko et al. show that such effects are able to account for a large part of measured strange particle yields, without assuming QGP formation. In these calculations the chemical equilibration is driven by the strongly decreased masses, which was not taken into account in the earlier calculations of Ref. 33. However, up to now these investigations suffer from the fact that each calculation focuses on one particular aspect of the observations, without reaching a coherent and internally consistent description of all observables and their characteristic mutual relationships simultaneously. For example, an additional test for the model would be the multi-strange (anti-)baryons, measured by the NA36 and WA85 collaborations at CERN (for references see Table 9). In spite of the extremely high mass thresholds, their multiplicities are high and present a particular challenge to any hadronic kinetic model.

Other points need improvement, too: the study of Ko et al. suffers from a thermodynamical inconsistency[54]. The reduction of the masses has only been taken into account in the particle distribution functions and cross sections for the kinetic rate equations, but not in the equation of state for the hydrodynamic evolution. One expects that the mass reduction should also lower the energy density and the pressure of the system.

Last but not least the subject of medium effects on hadron widths and masses is altogether still very controversial. Different approaches lead to qualitatively different and mutually contradictory results. A more fundamental and comprehensive approach to the problem of in-medium masses (based, for example, on QCD-inspired low-energy effective hadron Lagrangians) should eventually replace the rather phenomenological scaling law Eq. (21). Since these effective masses enter the chemical equilibration rates in a crucial way and the whole issue of strangeness equilibration as a signature for interesting new physics is mostly a question of time scales, this should be a high priority project.

## 3 Thermal Models

In this chapter we explain in more detail the physics of thermal models and their application to relativistic nuclear collisions, and discuss the conclusions that can be drawn by looking at strange particle production yields.

### 3.1 Thermal Parameters

The thermal model starts with the assumption of a locally equilibrated, not necessarily stationary fireball. The fireball could consist of a QGP, a hadron gas, or an equilibrated mixture of both. The following set of thermal parameters can be used for all these phases.

The fireball is described by a volume $V$, a temperature $T$ and chemical potentials $\mu$ for the conserved quantum numbers. In general the intensive thermal parameters are local quantities, i.e. depend on space and time. This description corresponds to a grand canonical ansatz, i.e. the quantum number densities parametrized by the chemical potentials are not conserved exactly, but only in the average. The corrections compared to a canonical description with exact quantum number conservation are of order $1/\sqrt{N}$, with $N$ being the number of particles with corresponding quantum numbers. Thus the corrections are expected to be negligible for heavy nuclei.

On the time scale of a nuclear reaction the important forces are the strong and electro-magnetic interactions, both conserving the quark flavors. Flavor changing weak interactions can be completely neglected in nuclear collisions. (This is a first example of the entrance of time scales: in the cosmological context, where the dynamics happens on the time scale of microseconds, weak interactions can be considered as chemically equilibrated, and the corresponding parametrization in terms of chemical potentials and conserved quantum numbers is correspondingly different.) Therefore we introduce chemical potentials $\mu_\mathrm{u}$, $\mu_\mathrm{d}$ and $\mu_\mathrm{s}$ for the three flavors considered here. Note that this ansatz automatically takes care of baryon number and electric charge conservation. In view of the (approximate) isospin symmetry of the colliding nuclei, which carries over to the fireball, it is common to combine the two light flavors

$$\mu_\mathrm{q} = \frac{1}{2}\left(\mu_\mathrm{u} + \mu_\mathrm{d}\right). \tag{29}$$

The small breaking of the isospin symmetry can be parametrized by

$$\delta\mu = \mu_d - \mu_u \,. \tag{30}$$

The nuclei considered here will be either exactly isospin symmetric (e.g. $^{32}$S), or the breaking is small, leading to $\delta\mu \approx 0$.[25] Therefore isospin effects can usually be neglected, and we use always the averaged light quark chemical potential (29). Flavor conservation is expressed on the level of hadrons as baryon number and strangeness (as well as isospin) conservation. It is also common to introduce the chemical potentials on this level, i.e. a baryon chemical potential $\mu_B$ and a chemical potential for the quantum number *strangeness* $\mu_S$. These two equivalent sets of chemical potentials are related by

$$\begin{aligned} \mu_q &= \mu_B/3 \,, \\ \mu_s &= \mu_B/3 - \mu_S \,, \\ \mu_S &= \mu_q - \mu_s \,, \end{aligned} \tag{31}$$

where the minus signs are due to the conventional assignment of strangeness $-1$ to the strange quark.

If $B_h, S_h$ are the baryon number and strangeness of hadron $h$, its chemical potential can thus be written either as

$$\mu_h = B_h \, \mu_B + S_h \, \mu_S \tag{32}$$

in terms of $\mu_B$ and $\mu_S$, or as

$$\mu_h = \nu_h^q \mu_q + \nu_h^s \mu_s \,, \tag{33}$$

where $\nu_h^q, \nu_h^s$ count the number of light and strange valence quarks inside the hadron, respectively, with anti-quarks counted with a minus sign.

The particle numbers are more directly given in terms of the fugacities, related to the chemical potentials by

$$\lambda_i = e^{\mu_i/T} \,. \tag{34}$$

The fugacity of each hadronic species is simply the product of the valence quark fugacities, *viz.* $\lambda_p = \lambda_u^2 \lambda_d$, $\lambda_{K^+} = \lambda_u \lambda_{\bar{s}}$, etc.

*3.2 Relative and Absolute Chemical Equilibrium*

On the nuclear time scale, the state of *absolute chemical equilibrium* is defined as the state in which all inelastic processes which transform different hadron species into each other, subject only to the conservation laws for baryon number and strangeness, have come to equilibrium. On much larger time scales, this is actually only a state of relative chemical equilibrium, because weak flavor changing processes have not been equilibrated. Since the strong and weak interaction time scales are so widely separated, the concept chemical equilibrium with respect to strong interaction processes only is very useful. In such a state the chemical potentials for particles and anti-particles are opposite to each other, implying for the respective fugacities

$$\lambda_{\bar{i}} = \lambda_i^{-1} \,. \tag{35}$$

However, the lifetime of a nuclear collision is actually so short that not all allowed chemical processes among the different species of hadrons may have time to fully equilibrate. The assumption of absolute chemical equilibrium is thus, even at freeze-out, the most questionable one in a thermal fireball approach. Thermal equilibration is much easier to justify since *every* type of collision (elastic and inelastic ones) involves exchange of momentum and thus leads to thermalization of the momentum spectrum. No thresholds or small cross sections inhibit thermal equilibration. Chemical processes, on the other hand, involve only certain inelastic channels with sometimes small branching ratios (cross sections) and may be suppressed by large mass thresholds (in particular in the case of baryon-antibaryon or strange-antistrange pair production).

Looking at the systematics of hadronic chemical reactions[33] it is, however, possible to subdivide them into two classes with generically well-separated time constants: The first class of processes involves the creation or destruction of only light quarks (or hadrons containing only light quarks); these processes have low mass thresholds and happen relatively fast (a few fm/$c$). Such processes include the exchange of existing strange quarks or antiquarks among different hadron species, e.g. in the process $\bar{K}N \to \pi Y$, which is characterized by a strong absorption cross section of antikaons by nucleons. Much slower are processes involving the creation or destruction of one or more strange quark-antiquark pairs (several 10 fm/$c$). A quite analogous situation is found in a QGP environment: $q\bar{q}$ pairs are created from gluons with a much faster rate than $s\bar{s}$ pairs from gluons or light quark pairs.

Given these different time scales it is thus sensible to assume that the state of absolute chemical equilibrium with respect to strong interactions is reached in stages: first a state of *relative chemical equilibrium* is reached in which gluons and light quarks and antiquarks are thermally and chemically equilibrated, but strange quarks and antiquarks are only thermally equilibrated, but haven't completely reached their equilibrium density. On the hadronic level this state corresponds to chemical equilibration of all non-strange hadrons as well as a partial chemical equilibrium with respect to all strangeness *exchanging* processes, while at the same time the absolute amount of strange valence quarks and antiquarks remains below its saturation limit. In other words, all fast chemical processes are equilibrated, while the slow strange pair production processes are out of equilibrium. Only in a second, much later stage also these latter processes equilibrate and a state of full chemical equilibration occurs.

The crucial question is how these various chemical time scales compare to the lifetime of a nuclear collision fireball. As far as we know from the discussions in Section 2, this hierarchy between light and strange quark producing processes is generic and true in both hadronic and QGP environments; the absolute time scales for both types of processes are, however, much faster in the QGP than in the hadron gas, both because of the larger densities of scatterers and lower mass thresholds. This is the deeper reason why strangeness is such a useful observable: with an initial QGP phase, relative chemical equilibrium in the above sense should be a valid concept in any case, and the nuclear collision lifetimes may even be at the threshold for complete chemical equilibration. Without the QGP as a "catalyst" absolute chemical equilibrium seems out of reach in nuclear collisions, and even the concept of relative chemical equilibrium may not be very well realized.

The idea of relative chemical equilibrium was first introduced heuristically by Rafelski[84] and later exploited in more detail by several groups (see Refs. 16, 25, 21, 22, 24, and 17). Only recently it was put on a solid theoretical foundation and generalized to other situations (e.g. the possible breaking of chemical equilibrium between mesons and baryons after sudden hadronization of a QGP followed by rapid freeze-out[25]) by C. Slotta[85]. He used the principle of maximum entropy to derive the distribution function for relative chemical equilibrium. In the case of relative strangeness equilibration it turns out to be sufficient to introduce one common strangeness saturation factor $\gamma_s$ with $0 < \gamma_s \leq 1$, where $\gamma_s = 1$ means full chemical equilibrium resp. complete saturation of the strangeness phase space. The strange quark and antiquark densities are then regulated by the effective fugacities

$$\lambda_s^{\text{eff}} = \gamma_s \lambda_s \,,$$
$$\lambda_{\bar{s}}^{\text{eff}} = \gamma_s \lambda_s^{-1} \,. \tag{36}$$

In the hadron language these enter the hadron fugacities according to Eqs. (33,34). The individual hadron fugacities $\lambda_h$ are thus given by a product of effective valence quark fugacities,

$$\lambda_h = \prod_j \lambda_j^{\text{eff}} \,, \tag{37}$$

and the approach to equilibrium is controlled by the factor:

$$\gamma_h = \gamma_s^{n(h,s)} \,, \tag{38}$$

where the power $n(h,s)$ counts the total number of valence strange quarks *plus* antiquarks in the hadron species $h$. From the treatment in Ref. 85 it follows that $\gamma_s$ must be treated as an additional fugacity, i.e. it can be written in the form

$$\gamma_s = e^{\mu_{|s|}/T} \,, \tag{39}$$

where $\mu_{|s|}$ is an additional chemical potential for strangeness carrying quarks and equal for strange and anti-strange quarks. Its deviation from zero parametrizes the deviation of the thermodynamic state from absolute chemical equilibrium. The hidden temperature dependence on $\gamma_s$ in Eq. (39) is important for the calculation of the entropy of the system[85].

Since $\gamma_s$ parametrizes the relative equilibrium between the two subgroups of non-strange and strange particles, it can be used as a quantitative measure for the strangeness enhancement from p+p over p+A to A+A collisions[20]. Due to its thermal definition, however, this requires an application of the thermal model even to low-multiplicity p+p collisions. The viability of such an approach has been discussed for more than 30 years (see Ref. 43 for a historical overview and Ref. 86 for a recent careful investigation of the $p\bar{p}$-annihilation system at rest) and can up to now only be justified by its phenomenological success.

*3.3 The Partition Function*

*Partition function for the QGP*

The calculation of all thermodynamical variables starts from the grand canonical partition function $\mathcal{Z}$. It is related to the thermodynamical potential $\Omega$ through

$$\Omega = -T \ln \mathcal{Z} . \tag{40}$$

This can then further be expressed by the pressure $P = -\Omega/V$. For the QGP with two massless flavors ($u$ and $d$) and a massive ($m_s$) strange quark we have

$$\begin{aligned}
P^Q(T, \mu_q, \mu_s, \gamma_s) &= \frac{T}{V} \ln \mathcal{Z}^Q(V, T, \mu_q, \mu_s, \gamma_s) \\
&= -B + \frac{37}{90}\pi T^4 + \frac{1}{2\pi^2}\mu_q^4 + \mu_q^2 T^2 \\
&\quad + \frac{1}{\pi^2} \int_{m_s}^{\infty} dE \, (E^2 - m_s^2)^{3/2} \left( \frac{1}{\gamma_s \, e^{\beta(E-\mu_s)} + 1} + \frac{1}{\gamma_s \, e^{\beta(E+\mu_s)} + 1} \right) .
\end{aligned} \tag{41}$$

Here $B$ is the MIT-bag constant which is needed to simulate the background confinement pressure when we investigate the strangeness phase space diagram. For other thermal quantities it has no relevance. From the partition function one derives the net particle densities for light quarks $\rho_q$, strange quarks $\rho_s$, and the baryon density $\rho_B$ as

$$\rho_q = \frac{T}{V}\left(\frac{\partial \ln \mathcal{Z}^Q}{\partial \mu_q}\right) = \frac{\lambda_q}{V}\left(\frac{\partial \ln \mathcal{Z}^Q}{\partial \lambda_q}\right) = \left(\frac{\partial P^Q}{\partial \mu_q}\right) , \tag{42}$$

$$\rho_s = \frac{T}{V}\left(\frac{\partial \ln \mathcal{Z}^Q}{\partial \mu_s}\right) = \frac{\lambda_s}{V}\left(\frac{\partial \ln \mathcal{Z}^Q}{\partial \lambda_s}\right) = \left(\frac{\partial P^Q}{\partial \mu_s}\right) , \tag{43}$$

$$\rho_B = \frac{1}{3}(\rho_q + \rho_s) . \tag{44}$$

*Partition function for the hadron gas*

The partition function for the hadron gas is given as a product of the one-particle partition functions of the different hadrons $h$:

$$\mathcal{Z}^H(T, V, \mu_q, \mu_s, \gamma_s) = \prod_h \exp\left[Z_h(T, V, \mu_h, \gamma_h)\right] . \tag{45}$$

The latter is given by

$$\ln Z_h(T,V,\mu_h,\gamma_h) = \beta V P_h^{\text{pt}} = \frac{g_h \beta V}{6\pi^2} \int_{m_h}^{\infty} dE \, \frac{(E^2 - m_h^2)^{3/2}}{\gamma_h e^{\beta(E-\mu_h)} \pm 1} \, . \qquad (46)$$

$g_h$ is the degeneracy factor, $m_h$ is the mass, $\mu_h$ is the chemical potential according to Eq. (32,33), and $\gamma_h$ is the strangeness suppression factor of hadron species $h$ given by Eq. (38). The upper index "pt" stands for *point-like*, because so far we treated the hadrons as point-like particles. The real pressure $P^{\text{H}}$ involves an additional proper volume correction[87] and is given by

$$P^{\text{H}} = \frac{1}{1 + \varepsilon^{\text{pt}}/4B} \sum_h P_h^{\text{pt}} \, , \qquad (47)$$

where $\varepsilon^{\text{pt}}$ is the energy density calculated for point-like particles.

In the analysis of measured strange particle yields the Boltzmann approximation is sufficient. Quantum statistical corrections are important only for the very light pions or for fermions at very large baryon densities (which will not occur in the actual applications discussed here). The emphasis of a chemical analysis of the data lies on the fugacities and the strangeness suppression factor $\gamma_s$. We spell these factors out explicitly and obtain in Boltzmann approximation

$$\ln \mathcal{Z}^{\text{H}} = \frac{VT^3}{2\pi^2} \Big[ F_{\text{M}} + (\lambda_q^3 + \lambda_q^{-3})F_{\text{N},\Delta} + (\lambda_s \lambda_q^{-1} + \lambda_s^{-1}\lambda_q)\gamma_s F_{\text{K}} \qquad (48)$$
$$+ (\lambda_s \lambda_q^2 + \lambda_s^{-1}\lambda_q^{-2})\gamma_s F_{\text{Y}} + (\lambda_s^2 \lambda_q + \lambda_s^{-2}\lambda_q^{-1})\gamma_s^2 F_{\Xi} + (\lambda_s^3 + \lambda_s^{-3})\gamma_s^3 F_{\Omega} \Big] \, .$$

This expression takes into account all non-strange mesons (M) (for mesons with hidden strangeness see below), kaons (K), non-strange baryons, i.e. nucleons (N) and deltas ($\Delta$), hyperons (Y), cascades ($\Xi$) and the omegas ($\Omega$), together with their antiparticles and all their resonance excitations. The factors

$$F_f = \sum_h g_h \left(\frac{m_h}{T}\right)^2 K_2(m_h/T) \qquad (49)$$

(where $K_2$ is the modified Bessel function of second order) result from a summation over all hadrons (ground state and resonances) $h$ in the hadron family $f$ where $f =$ M, N, $\Delta$, K, Y, $\Xi$, or $\Omega$.

In order to describe consistently a strongly interacting hadron gas, the full (divergent) hadron spectrum should be taken into account, as described by Hagedorn[43]. For a comparison with data we need, however, the final decays into stable (measurable) particles. These branching ratios are only known for the lowest states. Therefore one must cut the hadronic mass spectrum somewhere. The bias from this artificial cut will be investigated when we analyze the NA35 data from S+S collisions.

The effect of the strangeness suppression factor $\gamma_s$ on mesons with hidden strangeness via Eq. (38) requires a careful discussion. According to the SU(3)$_{\text{f}}$ quark model classification[88] we take $\phi$, $f_0(980)$, $f_1(1510)$, $f'_2(1525)$, $\phi(1680)$ and $\phi_3(1850)$ as pure $s\bar{s}$-states, with a suppression factor $\gamma_s^2$. The physical $\eta$- and $\eta'$-states are mixed states consisting of $u\bar{u}$, $d\bar{d}$ and $s\bar{s}$. We take a mixing angle between $\eta$ and $\eta'$ of $-20°$[88]. This results in an $s\bar{s}$-content of 32% in the $\eta$ and 68% in the $\eta'$. It follows that the multiplicities of these two mesons incorporate the following strangeness saturation factors:

$$\gamma_\eta = 0.68 + 0.32\gamma_s^2, \qquad \gamma_{\eta'} = 0.32 + 0.68\gamma_s^2. \qquad (50)$$

Therefore the term $F_M$ contains some $\gamma_s$ factors not explicitly written down in Eq. (48).

## 4  The Phase Diagram of Strange Matter

In this Section we describe the phase diagram for strange hadronic matter. For systems with vanishing net strangeness it was first studied in Refs. 89, 90. Due to the conservation of strangeness by the strong interaction this is the most relevant case for heavy-ion collisions. Due to surface radiation of strange particles from the expanding collision fireball it is, however, possible that the net strangeness of the system fluctuates around zero or even acquires an appreciable non-zero value[91]. Furthermore, hadronic systems with finite net strangeness are relevant for early cosmology and the structure of neutron stars (with or without a quark core) where weak interactions are equilibrated. Therefore the authors of Ref. 92 generalized the phase diagram to systems with non-zero net strangeness.

Such systems can be characterized by their strangeness fraction

$$f_s = \frac{\rho_s(T, \mu_q, \mu_s)}{\rho_B(T, \mu_q, \mu_s)}, \qquad (51)$$

the net number of strange quarks (or minus the strangeness) per baryon.

Strange hadronic matter, like matter with zero net strangeness, undergoes a phase transition to a quark gluon plasma if its density is sufficiently high. High particle density can be achieved either via quark pair creation by heating the system or via compression (high net baryon density). In the latter case it doesn't matter whether the baryon density is generated by strange or non-strange baryons, nor is the sign of the net baryon density important; only its modulus must be large. In the following subsection we discuss this phase transition.

### 4.1  The Strange Matter Igloo

Good model equations of state for hadronic matter near this phase transition are hard to obtain. The usual phenomenological procedure is to estimate the location of the phase transition by matching two different model equations of state, one for the hadronic phase which is valid at low temperatures and normal baryon densities where we have information from conventional nuclear physics, and one for the quark-gluon plasma above the phase transition which is derived from QCD at very high temperatures where it can be calculated perturbatively. Such a matching procedure produces almost by necessity a first order phase transition.

Two commonly used equations of state are in this context the ones defined in Eqs. (41) and (45)–(47), respectively. The phase transition line is then obtained from the Gibbs stability conditions for phase coexistence:

$$P_{\rm H} = P_{\rm Q}\,, \qquad T_{\rm H} = T_{\rm Q}\,, \qquad \mu_{\rm q,H} = \mu_{\rm q,Q}\,, \quad \text{and} \quad \mu_{\rm s,H} = \mu_{\rm s,Q}\,. \tag{52}$$

By setting the temperatures and chemical potentials in the two phases equal, one obtains the pressure balance relation

$$P_{\rm H}(T,\mu_{\rm q},\mu_{\rm s}) = P_{\rm Q}(T,\mu_{\rm q},\mu_{\rm s})\,. \tag{53}$$

Its solution defines a 2-dimensional phase coexistence surface in the 3-dimensional half-space defined by $T$ ($>0$), $\mu_{\rm q}$, and $\mu_{\rm s}$, the so-called "strange matter igloo"[90]. Along the igloo surface hadronic and quark matter can coexist with variable volume fractions, but in general with different strangeness fractions $f_{\rm s}^{\rm H}$ and $f_{\rm s}^{\rm Q}$ in the two subphases. If the hadronic subphase occupies the volume fraction $\alpha = V_{\rm H}/(V_{\rm H} + V_{\rm Q})$, the strangeness fraction of the total system depends on $\alpha$ and is given by

$$f_{\rm s}(T,\mu_{\rm q},\mu_{\rm s}) = \alpha f_{\rm s}^{\rm H}(T,\mu_{\rm q},\mu_{\rm s}) + (1-\alpha) f_{\rm s}^{\rm Q}(T,\mu_{\rm q},\mu_{\rm s})\,. \tag{54}$$

Systems with a fixed strangeness fraction $f_{\rm s}$ will thus move along the igloo surface as they hadronize: as $\alpha$ increases from 0 (QGP) to 1 (hadron gas), the condition of constant strangeness fraction (54) cuts out a strip from the igloo surface, as shown in Figure 6. If the system is further constrained to possess a certain fixed baryon number $B$ and total entropy $S$ and is allowed to hadronize slowly and adiabatically (i.e. at constant specific entropy $S/B$), it will follow a specific line, the so-called isentropic expansion trajectory, through that strip.

*4.2 Isentropic Expansion Trajectories*

A discussion of these trajectories is very interesting. Suppose that our system reaches a state of thermodynamic equilibrium somewhere in the quark phase (characterized by a point $P_0$ in Figure 6 outside the igloo) and begins to expand hydrodynamically. In the absence of shock discontinuities the hydrodynamic equations conserve entropy. Since baryon number and strangeness are also conserved by the strong interactions, the system will expand at constant strangeness fraction $f_{\rm s}$ and specific entropy $S/B$. These two conditions together define a line through the three-dimensional phase diagram. This is the isentropic expansion trajectory just mentioned. It begins at $P_0$ in the quark phase, reaches the igloo surface at a point $P_1$ at the edge of the strip on that surface which corresponds to the chosen $f_{\rm s}$, and then begins to cross that strip. Along the trajectory both the temperature and baryon density are continuously decreasing functions. If allowed to stay in local thermal equilibrium forever (in reality the system will decouple once the density becomes too low), the hydrodynamic expansion will only stop at zero temperature and zero baryon density, after the system has expanded over an infinite volume. There is no other possibility to reach a state of zero temperature at finite total entropy.

The interesting question is which point on the $T=0$ plane the trajectory will finally approach. There are two possibilities:

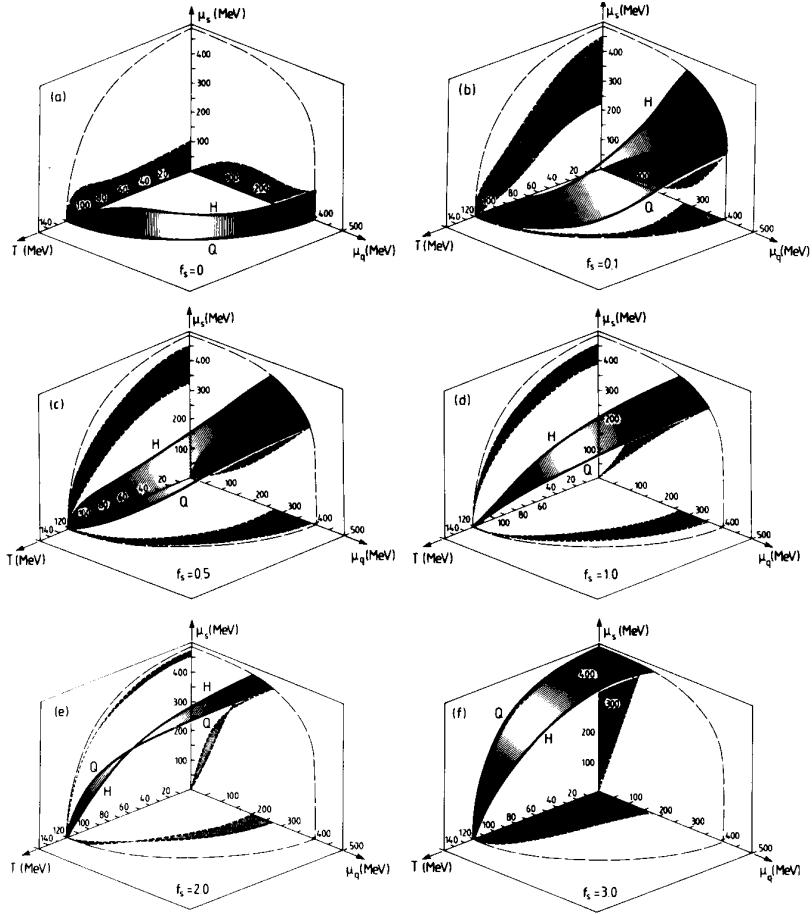

**Figure 6:** The phase diagram of strongly interacting matter in $(T, \mu_q, \mu_s)$ space for $B^{1/4} = 180$ MeV. The igloo-type surface describes the phase-coexistence region (mixed phase) between quark matter (outside) and hadronic matter (inside). Figures (a) through (f) show various sections through this surface corresponding to systems with fixed strangeness fraction $f_s$. Also shown are the projections of the mixed phase region onto the three coordinate planes. From Ref. 92.

1. The trajectory reaches the other side of the phase-coexistence strip before hitting the $T = 0$ plane. In this case it will leave the igloo surface at a point $P_2$ on the other side of the strip and proceed into the interior of the igloo until it finally reaches a point $P_3$ on the $T = 0$ plane.

2. The trajectory reaches the $T = 0$ plane before the crossing is completed, at a point $P_3'$ somewhere on the line in which the strip and the zero temperature plane intersect.

These two cases correspond to two completely different physical situations. In order to discuss them we must understand the $T = 0$ limit of the phase diagram in more detail.

### 4.3 The $T \to 0$ Limit of the Phase Diagram

Non-strange hadronic matter has at zero temperature a quite simple structure: mesons are absent (because they don't carry any conserved charges and thus their chemical potentials vanish), and only those baryons exist whose mass is smaller than the baryon chemical potential $\mu_B$ (i.e. the Fermi energy). If $\mu_B$ is smaller than the mass of the lightest baryon, the nucleon, then no particles exist at all. The hadronic state corresponding to $T = 0$ and $\mu_B < m_N$ is thus the vacuum; it has zero pressure. In other words: as $T \to 0$, any state with finite net baryon number has to approach a point with $\mu_B > m_N$.

If strangeness is added, things get more complicated. The $T = 0$ phase diagram is now a 2-dimensional plane spanned by $\mu_q = \mu_B/3$ and $\mu_s$. Its structure (phase coexistence line, lines of constant strangeness fraction, etc.) is very rich and highly nontrivial. A detailed discussion is given in Ref. 92. Here we can only summarize the most important features.

Generally speaking, mesons are still absent at $T = 0$. But there are exceptions: since kaons are bosons which carry a conserved charge (strangeness), they can form a Bose condensate if their chemical potential is equal to their rest mass:

$$\mu_K = \mu_q - \mu_s = m_K \quad \text{or} \quad \mu_{\bar{K}} = \mu_s - \mu_q = m_K . \tag{55}$$

These conditions define two straight lines in the $\mu_q$-$\mu_s$ plane, where the first corresponds to kaon, the second to antikaon condensation (Figure 7). These equations delineate the physical limits of the hadron phase at $T = 0$ which can only exist between these two lines.

Baryons exist at $T = 0$ only if their chemical potentials exceed their rest masses. The baryon chemical potentials have the form $\mu_i = s_i \mu_s + (3 - s_i)\mu_q$ where $s_i$ is the number of strange quarks in the baryon. For $s_i=0$, 1, 2, 3 (nucleons, hyperons, cascades and $\Omega$-baryons) the condition $\mu_i > m_i$ defines four straight lines. Their envelope separates the region with zero baryon density (and zero pressure!) from the region with finite baryon density (see Figure 7).

One can now solve the pressure balance equation (53) in this plane. For the three values $B^{1/4} = 145, 195, 235$ MeV for the QCD bag constant one obtains the three solid curves shown in Figure 7. One sees that for the two smaller bag constants part or all of the coexistence line lies inside the region of zero baryon density in the hadronic phase; these parts of the line thus correspond to a mixed phase, where the hadronic subphase is just the vacuum. Another way of describing such a state is that it consists of quark matter bubbles separated by vacuum! This state is stable, because it solves the pressure balance equation: the pressure of the hadronic phase (=vacuum) is zero, and the quark bubbles have also zero pressure, because the Pauli pressure of the quarks is balanced by the bag pressure on the surface. If the quark matter in this state has $f_s \neq 0$, these bubbles are called strangelets.

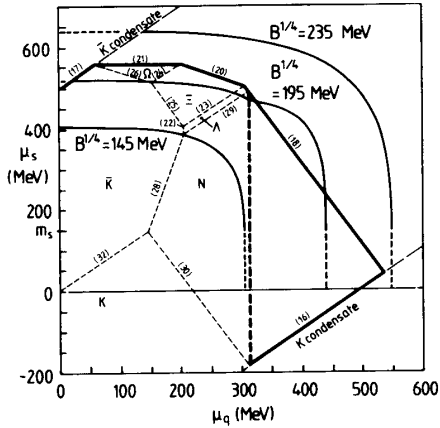

**Figure 7:** Phase structure at $T = 0$. The three solid curves are the phase coexistence lines for $B^{1/4} = 145, 195, 235$ MeV. The thick solid and dashed lines separate the regions of zero baryon density (left) from regions of finite baryon density (right). The left region, which at $T = 0$ is empty, is subdivided into 6 smaller regions, indicating which particle species dominate at small, but non-zero temperature. The small numbers along the lines refer to equation numbers in Ref. 92 from where this figure was taken.

One can show[92] that for $B^{1/4} \lesssim 200$ MeV some part of the phase coexistence line lies in the region of zero baryon density in the hadronic phase. Actually, practically all points corresponding to a non-zero strangeness fraction lie on this section of the phase coexistence line. For these values of the bag constant cold strange quark matter is thus mechanically stable and stops expanding; it cannot hadronize.

*4.4 Hadronization of Strange Quark Matter at $T \neq 0$*

Another way of plotting these results is shown in Figure 8. Here we show the phase diagram in the $(T, \rho_B)$ plane, eliminating $\mu_q$ and $\mu_s$. The previous phase coexistence line is here resolved into a whole (shaded) domain, the mixed phase (M). Different points in this mixed phase at the same temperature $T$ correspond to different volume fractions $\alpha$ of the hadron and quark subphases.

One sees that, for the particular value of the bag constant chosen in this Figure, at low temperatures the mixed phase extends all the way to zero baryon density even for systems with very small amounts of net strangeness. This is just the situation discussed at the end of the previous subsection. If we start at finite temperature and let the system expand adiabatically, we see that some of the isentropic expansion trajectories never exit the mixed phase. This corresponds to the situation 2 mentioned at the end of Section 4.2. Strangelets form. On the other hand, if the specific entropy $S/B$ was sufficiently large to begin with, this does not happen and the system hadronizes completely; this corresponds to situation 1 mentioned at the end of Section 4.2.

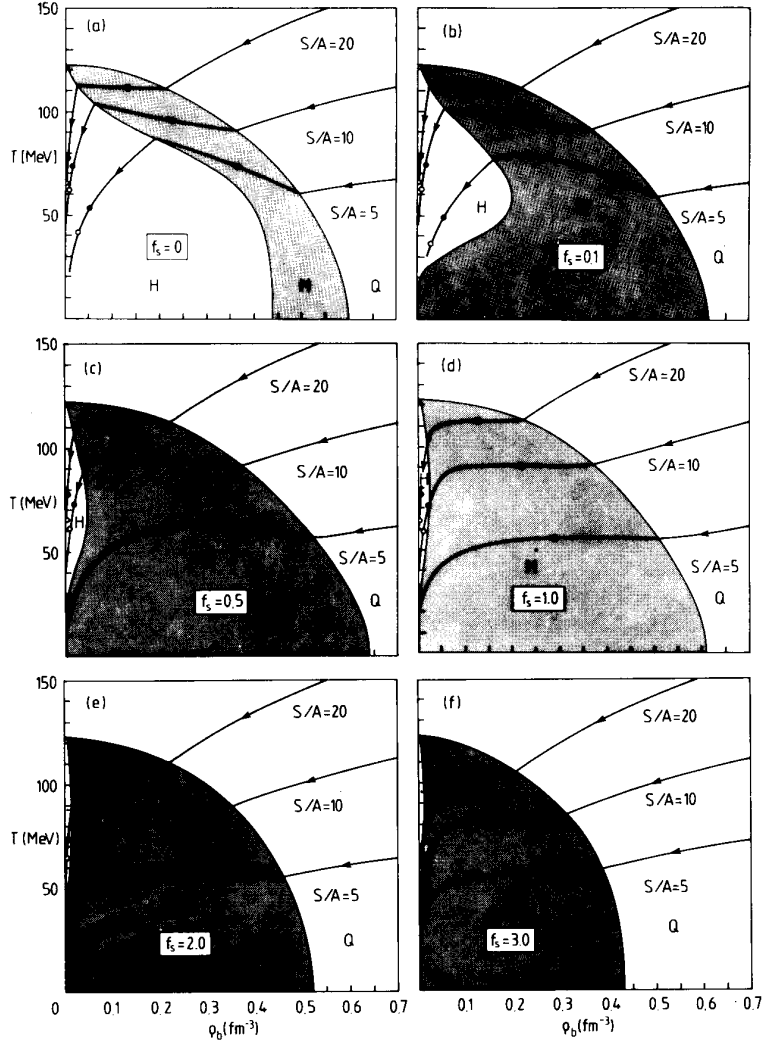

**Figure 8:** Isentropic expansion trajectories in the $(T, \rho_B)$ plane, for $B^{1/4} = 180$ MeV and several $f_s$ values. For $f_s \neq 0$, at $T = 0$ the mixed phase extends to zero baryon density. Systems with sufficiently large $f_s$ and/or small specific entropy $S/B$ reach $T = 0$ without leaving the mixed phase, forming strangelets. The full and open circles indicate pion freeze-out for spherical fireballs with 4 and 8 fm radius, respectively. For $B^{1/4} > 200$ MeV, the diagrams look like Figure (a) for *all* values of $f_s$, and strangelet formation is impossible. From Ref. 92.

It is, of course, interesting to ask which are the largest allowed values for $S/B$ which still lead to strangelet formation in a system with given $f_s$. We already know that we need bag constants below $B^{1/4} = 200$ MeV ($B < 200$ MeV/fm$^3$) for strangelets to form at any value of $S/B$ and $f_s$. For $B^{1/4} \lesssim 145$ MeV ($B < 60$ MeV/fm$^3$), on the other hand, even non-strange quark matter would be stable, which contradicts nuclear phenomenology. Within this narrow window of the bag constant, the values of $f_s$ and $S/B$ play a crucial role[92]. E.g. for $B^{1/4} = 180$ MeV, strange matter with $f_s$=1 (hyperon matter) completely hadronizes as soon as $S/B > 6$. Typical values for $S/B$ from heavy ion collisions at Brookhaven or CERN range from 13 to 45 (see Section 7).

We would like to close this section with one important remark: if (strange) quark matter undergoes adiabatic hadronization at fixed specific entropy $S/B$ and strangeness fraction $f_s$ (which may be zero), the chemical potentials usually change during the phase transition. Thus a strangeness neutral QGP will start to hadronize with $\mu_s = 0$, but at the end of the hadronization process $\mu_s$ will take on a different value which depends on $T$ and $\mu_B$ and corresponds to the equilibrium value in a strangeness neutral hadron gas. To draw conclusions from measured values of the chemical potentials in the final hadronic state about their values in the preceding QGP state is thus highly nontrivial. In particular, a vanishing strange quark chemical potential, $\mu_s = 0$, in the final state can have survived from a primordial QGP phase only if hadronization did *not* happen adiabatically, but suddenly.

## 5  Ingredients and Extensions of the Thermal Model

### 5.1  *The Strangeness Neutrality Condition*

An important constraint on the thermal model parameters arises from the fact that in strong interactions strange quarks are only created in pairs, and that therefore the nuclear collision fireball (up to small fluctuations due to possibly asymmetric surface radiation of strange and antistrange hadrons[6] or due to "strangeness distillation" during hadronization of a QGP[89]) always remains approximately strangeness neutral.

The condition of vanishing total strangeness according to Eq. (43) takes the form

$$0 = \rho_s = \langle s \rangle - \langle \bar{s} \rangle = \lambda_s \frac{\partial}{\partial \lambda_s} \ln \mathcal{Z} \,. \tag{56}$$

For the hadron gas Eq. (56) is an implicit equation relating $\lambda_s$ to $\lambda_q$ in a way which strongly depends on the temperature[89,93].

In Figure 9 we show the relation between $\mu_q$ and $\mu_s$ for different temperatures which ensures vanishing net strangeness in a hadron gas. One recognizes a slight dependence on the number of hadronic resonances taken into account. The curves have a characteristic behavior and, for not to high temperatures, intersect the $\mu_s = 0$ line twice.

At high temperature the heavy resonance states become important. Unfortunately, the high mass resonance states are poorly known, especially in the mesonic sector where the states above about 1.5 GeV become very broad and are difficult to identify experimentally. This triggers an artificial influence of the known resonance mass spectrum on the strangeness neutrality relation between mesons and baryons at high temperature. We believe that the strangeness neutrality curves are unreliable above temperatures of about 250 MeV.

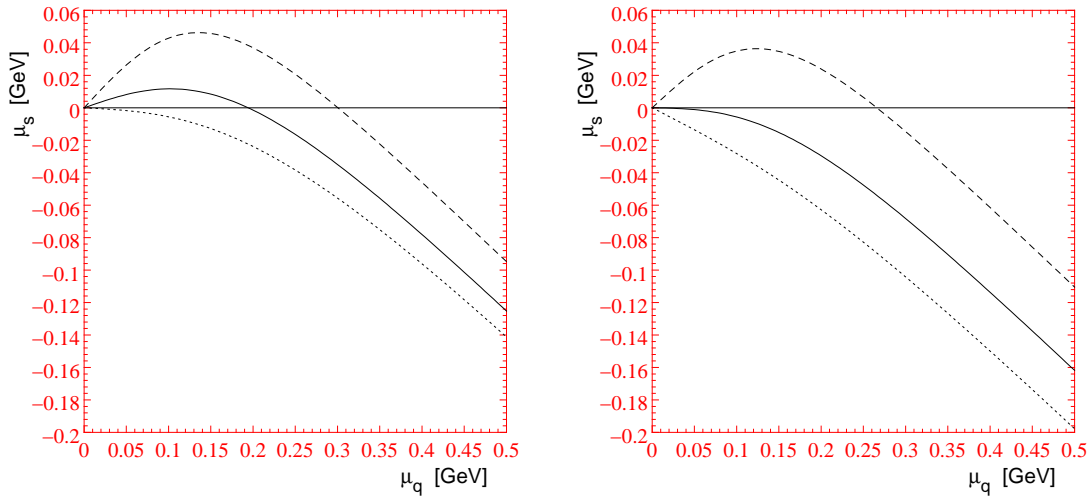

**Figure 9:** Strangeness neutrality curves for a hadron gas for different temperatures. The dashed line corresponds to $T = 150$ MeV, the solid line to $T = 200$ MeV, and the dotted line to $T = 250$ MeV. In the left figure the hadron mass spectrum was cut at $m_{\text{cut}} = 1.6$ GeV, in right figure at $m_{\text{cut}} = 2.0$ GeV.

Generally Eq. (56) relates the three parameters $\lambda_s$, $\lambda_q$ and $T$ (resp. $\mu_s$, $\mu_q$, and $T$). If two of them are given, the third can be determined from this equation which thus reduces the number of independent parameters. The strangeness neutrality condition depends also somewhat on $\gamma_s$, but this influence is weak because it really enters only through the multistrange hadrons. They carry a factor $\gamma_s^n$ ($n > 1$) which survives in Eq. (56) after the leading linear $\gamma_s$-dependence has been factored out.

For a QGP strangeness neutrality is a matter of the strange quarks alone, independent of the amount of light quarks, because the strange quarks are no longer bound together with other non-strange quarks into hadrons. Strangeness neutrality in the QGP is thus given by $\mu_s = 0$ independent of $\mu_q$. This can be easily derived by inserting Eq. (41) into Eq. (56).

As we will see later, the experiments at 200 A GeV suggest a value for $\mu_s$ very close to zero. While for a strangeness neutral hadron gas this solution is only possible for $T \lesssim 200$ MeV and a particular value of $\mu_q$, a strangeness neutral QGP leads to this solution naturally and independent of $\mu_q$ and $T$. This is the characteristic difference in the thermodynamics of a hadron gas and a QGP.

A more detailed discussion of this issue will follow later. In order to be able to discuss violations of the strangeness neutrality condition we define

$$\varepsilon = \frac{\langle \bar{s} \rangle - \langle s \rangle}{\langle s \rangle} \qquad (57)$$

as a measure for the net strangeness.

*5.2 Transverse Flow and Transverse Momentum Spectra*

If the system is in local thermal equilibrium, then the observed momentum spectra should reflect the (average) temperature. However, since a thermalized source which is surrounded only by vacuum must necessarily begin to expand, the thermal motion is superimposed in the spectra by a dynamical component arising from the collective expansion. This is most clearly seen in the measured longitudinal momentum or rapidity spectra of hadrons from heavy ion collisions which for both Brookhaven and CERN energies are much broader than a thermal distribution from a stationary fireball[26,94,96]; the strong collective flow component here is partially due to incomplete stopping of the colliding nuclei and partially due to additional hydrodynamical expansion generated in the later stages of the collision. In the transverse momentum spectra, which look approximately exponential like a thermal distribution, the identification of a collective flow component is more involved; the observed convex shape of the spectra at low momenta, which does not agree with the concave behavior of a thermal distribution near $p_\perp = 0$, can be associated both with transverse collective flow[95] and with non-thermal contributions from resonance decays after freeze-out[83]. This ambiguity was systematically studied and resolved in Ref. 96; theoretical arguments based on the consistency of the observations with the freeze-out criterium[95,97] and direct experimental reconstruction of the contribution of $\Delta$ decays to the pion spectrum[26,147] confirm the existence of a strong collective flow component ($\langle \beta_f \rangle \simeq 0.3 - 0.4$) also in the transverse momentum spectra.

For a purely thermal distribution the $m_\perp$ spectra and the double differential distribution $dN/(dy\,dm_\perp^2)$ are given in the Boltzmann approximation by

$$\frac{dN_h}{dy\,dm_\perp^2} = \frac{g_h V}{8\pi^2} m_\perp \cosh y \, e^{-\beta(m_\perp \cosh y - \mu_h)} ; \tag{58}$$

$$\frac{dN_h}{dm_\perp^2} = \frac{g_h V}{4\pi^2} e^{\beta \mu_h} m_\perp \, \mathrm{K}_1\left(\frac{m_\perp}{T}\right) \stackrel{T \ll m_\perp}{\approx} \frac{g_h V}{4\pi\sqrt{2\pi}} e^{\beta \mu_h} \sqrt{m_\perp T} \, e^{-m_\perp/T} , \tag{59}$$

where $\mathrm{K}_1(x)$ is a modified Bessel function. If one fits the measured $m_\perp$ spectra with the expression (59), an "apparent temperature" can directly be read off from the exponential fall off. This is a very common procedure which is easily applied by both theorists and experimentalists. But due to the problem of resonance decay and flow contamination just mentioned, care should be taken not to denote these slope parameters indiscriminately by "temperatures". At AGS and SPS energies the resulting apparent temperatures are always of the order of 200 MeV or higher which is well above Hagedorn's limiting temperature[43] for a hadron gas and above the deconfinement temperature from lattice QCD[42]. This is true even though the contributions from resonance decays tend to steepen the spectra and thus reduce the apparent temperatures, as will be discussed in Section 5.3. It appears to be theoretically inconsistent to associate such a large temperature with the observed final hadronic state from heavy ion collisions.

Transverse collective flow provides a simple resolution for this dilemma. It is a necessary dynamical consequence of early thermalization of the collision zone[97]. The amount of hydrodynamically generated transverse flow depends, among other things, on the initial energy (entropy) density and the equation of state. These are poorly known so far, and the resulting theoretical freedom must be removed by the experiment.

Transverse momentum spectra from a collectively expanding thermalized source can be parametrized by[96]

$$\frac{dN_h^{\text{flow}}}{dm_\perp^2} = \frac{L}{2\pi} g_h \gamma_h \lambda_h \int_0^{R_f} r\, dr\, m_\perp\, \mathrm{K}_1\left(\frac{m_\perp \cosh \rho(r)}{T}\right) \mathrm{I}_0\left(\frac{p_\perp \sinh \rho(r)}{T}\right), \quad (60)$$

where $\mathrm{I}_0(x)$ is also a modified Bessel function, $R_f$ the freeze-out radius and $L$ is some normalization length. $\rho(r)$ stands for the transverse flow rapidity

$$\rho(r) = \tanh^{-1} \beta_\perp(r), \quad (61)$$

and the transverse flow velocity profile $\beta_\perp(r)$ is usually assumed to have the form

$$\beta_\perp(r) = \beta_f \left(\frac{r}{R_f}\right)^\alpha. \quad (62)$$

If the system expands longitudinally in a boost-invariant way[99], the double differential spectrum is simply given by

$$\frac{dN^{\text{flow}}}{dy\, dm_\perp^2} \sim \frac{dN^{\text{flow}}}{dm_\perp^2}. \quad (63)$$

Eq. (60) represents an approximation which assumes a particular shape of the freeze-out hypersurface[96,97]. A detailed study within a global hydrodynamical evolution model[96,97] shows that (60) represents well the transverse momentum spectra even if they are calculated with a more sophisticated kinematical freeze-out criterium. For not too heavy particles and not too low values of $p_\perp$ the fit is also rather insensitive to the power $\alpha$ in Eq. (62)[95,96,97]. Thus we take here for simplicity a constant transverse flow velocity, $\alpha = 0$.

For large $m_\perp$ the spectrum (60) falls off exponentially, but with a slope which does not reflect the true temperature $T$ of the locally equilibrated system at freeze-out, but rather an apparent "blue shifted" temperature $T_{\text{app}}$ which in the above approximation is simply given by[95,96]

$$T_{\text{app}} = T \sqrt{\frac{1+\beta_f}{1-\beta_f}}. \quad (64)$$

In the case $\rho \to 0$ ($\beta_f \to 0$) Eq. (60) reduces the purely thermal spectrum Eq. (59). But since the experimental spectra fix only the asymptotic inverse slope (64) accurately, while at lower values of $p_\perp$ the details of the spectra (in particular the observed strongly concave shape of the pion spectrum) are governed by an intricate interplay of transverse flow and resonance decays, the true temperature and transverse flow velocity cannot be easily separated by the fit procedure and must be determined by additional theoretical input. Figure 10 shows this ambiguity of the fitting procedure for the transverse momentum spectra of five different hadron species measured in 200 A GeV S+S collisions by the NA35 collaboration[98,157]. The experimental $m_\perp$ spectra can be reproduced within a large range of temperatures (100 – 200 MeV) by adjusting $\beta_f$ accordingly. Hydrodynamical studies[97] show that dynamical consistency between the longitudinal and transverse momentum spectra and with the freeze-out kinetics restrict the allowed freeze-out parameters to a rather narrow domain around $T = 150$ MeV and $\langle \beta_f \rangle = 0.3c$. In detail there remains, however, certainly some model dependence to these statements, mostly related to the equation of state used for the hydrodynamical expansion. In spite of our theoretical prejudice towards freeze-out temperatures in the 150 MeV region[97,42] we will therefore treat for the chemical analysis $T$ as a free parameter.

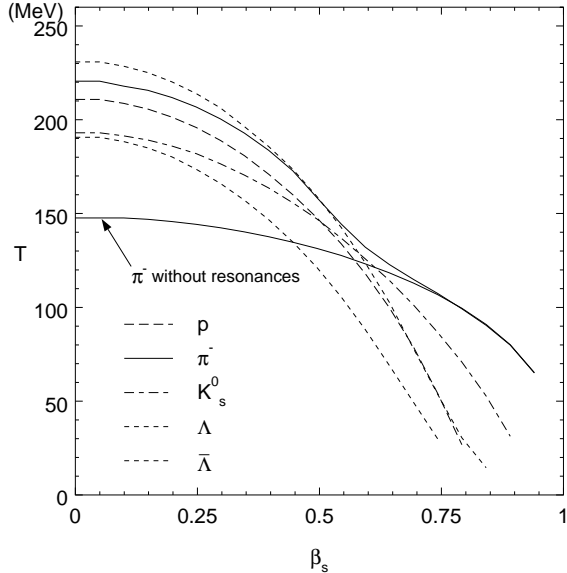

**Figure 10:** All fit pairs $(T, \beta_f)$ compatible with the measured $m_\perp$-spectra. Every point on these curves results in a good fit of the computed $m_\perp$ spectrum (including resonance decays and transverse flow) to the respective measured[98,157] particle spectra. From Ref. 96.

*5.3 Resonance Decays*

At temperatures above 150 MeV hadronic resonance production becomes very important. After freeze-out the resonances decay into stable particles which are indistinguishable from their directly emitted brothers and sisters. The measured distribution of a particle species $i$ is thus given by

$$\frac{dN_i}{dy\,dm_\perp^2} = \frac{dN_i^{\text{thermal/flow}}(T)}{dy\,dm_\perp^2} + \sum_R b_{R \to i} \frac{dN_i^R(T)}{dy\,dm_\perp^2}\,. \qquad (65)$$

with the direct thermal contribution given by Eq. (58) or (60). The contribution from decays of resonances $R \to i + 2 + \ldots + n$ (with branching ratio $b_{R \to i}$ into the observed channel $i$) is calculated according to[83]

$$\frac{dN_i^R}{dy\, dm_\perp^2} = \int_{s_-}^{s_+} ds\, g_n(s) \int_{Y_-}^{Y_+} dY \int_{M_\perp^-}^{M_\perp^+} dM_\perp^2 \qquad (66)$$

$$\frac{M}{\sqrt{P_\perp^2 p_\perp^2 - [ME^* - M_\perp m_\perp \cosh(Y-y)]^2}} \left(\frac{dN_R}{dY\, dM_\perp^2}\right).$$

Capital letters indicate variables associated with the resonance $R$, $\sqrt{s}$ is the invariant mass of the unobserved decay products $2, \ldots, n$, and the kinematic limits are given by

$$s_- = \left(\sum_{k=2}^n m_k\right)^2, \qquad s_+ = (M - m_i)^2, \qquad (67)$$

$$Y_\pm = y \pm \sinh^{-1}\frac{p^*}{m_\perp}, \qquad (68)$$

$$M_\perp^\pm = M\frac{E^* m_\perp \cosh(Y-y) \pm p_\perp \sqrt{p^{*2} - m_\perp^2 \sinh^2(Y-y)}}{m_\perp^2 \sinh^2(Y-y) + m_i^2}, \qquad (69)$$

$$E^* = \frac{1}{2M}(M^2 + m_i^2 - s), \qquad p^* = \sqrt{E^{*2} - m_i^2}. \qquad (70)$$

In Eq. (66) $g_n(s)$ is the decay phase space for the $n$-body decay $R \to i + 2 + \ldots + n$; for the dominant 2-body decay (assuming isotropic decay in the resonance rest frame) it is given by

$$g_2(s) = \frac{1}{4\pi p^*} \delta(s - m_2^2). \qquad (71)$$

For $n > 2$ see Ref. 83.

The resonance spectrum $dN_R/dy\, dM_\perp^2$ entering the r. h. s. of Eq. (66) is in general itself a sum of a thermal or flow spectrum like Eq. (58) or Eq. (60) and of decay spectra from still higher lying resonances; for example, most decays of high-lying nucleon resonances into protons or anti-protons proceed through the $\Delta(1232)$ or its anti-particle (e.g. $N(1440) \to \Delta(1232) + \pi \to p + 2\pi$). This has been taken into account in the calculations. Extracting the degeneracy factors and fugacities of the decaying resonances, we write shortly

$$N_i^R \equiv \gamma_R \lambda_R \tilde{N}_i^R \equiv \gamma_R \lambda_R \int_{y-\text{window}} dy \int_{m_\perp^{\text{cut}}}^\infty dm_\perp^2 \sum_R g_R\, b_{R \to i} \frac{d\tilde{N}_i^R}{dy\, dm_\perp^2}, \qquad (72)$$

with the sum now including only resonances with identical quantum numbers; if these quantum numbers agree with those of species $i$, the sum is meant to also include the thermal contribution. We will use this notation below.

Between particles and anti-particles we have the relation

$$N_i^{\bar{R}} = \gamma_R \lambda_R^{-1} \tilde{N}_i^R = \lambda_R^{-2} N_i^R. \qquad (73)$$

## 5.4 Particle Ratios

A quantitative comparison of the experimental (strange) particle yields or ratios with the thermal model prediction must take into account the kinematic acceptance windows of the experiments. The measured ratio of two different particles $a$ and $b$ is in general given by

$$\frac{N_a}{N_b} = \frac{\iint\limits_{\mathcal{R}_a} dy\, dm_\perp \frac{dN_a}{dy\, dm_\perp^2}}{\iint\limits_{\mathcal{R}_b} dy\, dm_\perp \frac{dN_b}{dy\, dm_\perp^2}} \; . \tag{74}$$

Here $\mathcal{R}_i$ denotes the experimental kinematic acceptance window for particle species $i$. The advantage of considering particle ratios is the cancellation of the (poorly known) fireball volume $V$ at freeze-out (assuming that both particle species freeze out simultaneously). Of course, in general one must consider the (different) resonance decay contributions to the numerator and denominator in (74) according to Eq. (65). A first, naive estimate can, however, be obtained by neglecting the resonance contributions and further taking for $dN_i/(dy\, dm_\perp^2)$ the purely thermal distribution (58). Then the integration kernels of (74) are independent of the particle rest masses $m_{a,b}$. If we then choose $\mathcal{R}_a = \mathcal{R}_b$, i.e. cut the data to the same rapidity and $m_\perp$ window, the integrals over the Boltzmann factors cancel in the ratio[84], which then simply reduces to a quotient of fugacities and $\gamma_s$ factors:

$$\left.\frac{N_a}{N_b}\right|_{\text{est.}} = \frac{\gamma_a \lambda_a}{\gamma_b \lambda_b} \; . \tag{75}$$

With proper flow distributions (60) the integral kernels in (74) are no longer independent of the rest masses of the particles, and the simple estimate (75) must be corrected by a proper numerical calculation. Similar comments generally apply to the contributions from resonance decays.

## 5.5 The Specific Entropy

A useful quantity for studying the evolution of a nuclear reaction is the specific entropy $S/B$, as already discussed in Section 4. If we knew $S/B$ and the temperature, we would be able to decide whether we are in the QGP phase or in the hadron phase, because of the characteristic high specific entropy of the plasma compared to the hadron gas at the same temperature. Thus it would be good to have a measure for $S/B$. A rough estimate is provided by the following considerations.

In the ultrarelativistic limit ($T \gg m$) the entropy per particle is independent of temperature and approximately $S/N \approx 4$ (a little less for bosons, a little more for fermions). If the three charge states $(+, 0, -)$, in which the stable final state hadrons arise, are equally distributed, we have $N^{\text{tot}} = 3/2(N^+ + N^-)$. The net charged multiplicity $N^+ - N^-$ is given by the number of incoming protons. Therefore the total baryon number is approximately $B \approx 2(N^+ - N^-)$. Putting all this together one can write

$$\frac{S}{B} = \frac{S}{N^{\text{tot}}} \frac{N^{\text{tot}}}{B} \approx 4 \cdot \frac{\frac{3}{2}(N^+ + N^-)}{\frac{1}{2}(N^+ - N^-)} = 3\frac{N^+ + N^-}{N^+ - N^-} \; . \tag{76}$$

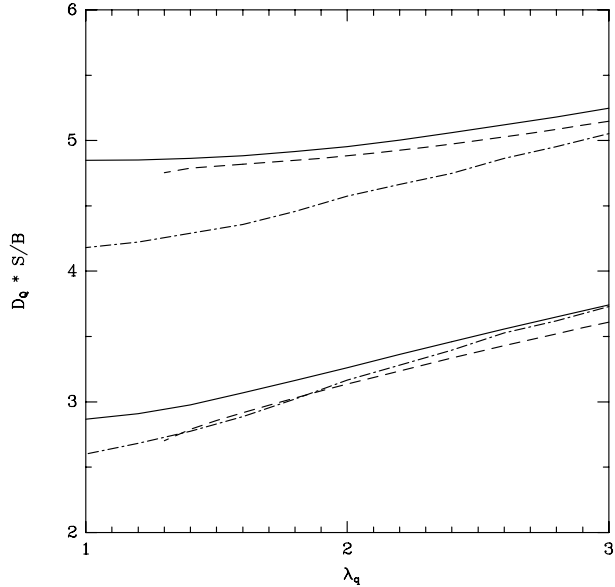

**Figure 11:** The product $D_Q \cdot S/B$ as a function of $\lambda_q$ for different values of $\lambda_s$. Solid line: $\lambda_s = 1.00$; dashed line: $\lambda_s = 0.95$; dash-dotted line: $\lambda_s = 1.05$. The additional parameters are $\gamma_s = 1$, and $T$ is fixed by the strangeness neutrality relation (56). The upper curves are for $D_Q^0$ (without resonance decays), the lower curves include the decay of resonances after freeze-out. Figure from Ref. 25.

Defining the specific net charge $D_Q$ as

$$D_Q = \frac{N^+ - N^-}{N^+ + N^-}, \qquad (77)$$

we can rewrite this as

$$(S/B) \cdot D_Q \approx K \approx 3. \qquad (78)$$

$D_Q$ is an easily measurable quantity and most precisely accessible in emulsion experiments.

Of course, in a realistic hadron resonance gas at $T \sim 150 - 200$ MeV the ultra-relativistic limit $S/N \approx 4$ is not realized. We can test the accuracy of the estimate Eq. (78) numerically. We introduce the notation $D_Q^0$ for the specific net charge in the hadron gas, before post-freeze-out resonance decays. In the upper curves of Figure 11 we plot $S/B \cdot D_Q^0$ for a hadron resonance gas at different values of $\lambda_s$ near $\lambda_s = 1$, as a function of $\lambda_q$. The temperature varies with $\lambda_q$ according to the strangeness neutrality relation (56) and actually covers a wide range over the $\lambda_q$ interval shown. Obviously the product $S/B \cdot D_Q$ is indeed roughly constant as suggested by the estimate (78), but the constant is approximately 5 rather than 3. The reason is the larger specific entropy of the heavier particles compared to the assumed value of 4. However, after adding the resonance decay contributions to the total charged particle multiplicity in the denominator of (77), the lower set of curves in Figure 11 is obtained. It is surprising to see how the decays bring the approximate constant $K$ again down to a value around 3 as in the simple estimate Eq. (78)! We thus have with $D_Q^{-1}$ a reasonably accurate measure for $S/B$ from a thermalized and chemically equilibrated hadron gas. For hadronic systems far away from chemical equilibrium, however, the estimate (78) should not be applied without further scrutiny.

# 6 Chemical Parameters from Strange Particle Ratios

In this part we want to confront the thermal description with experimental data. This will be done in several steps: first, the input parameters for the thermal model, $T$, $\mu_s$, $\mu_q$, $\gamma_s$, $(V)$ will be determined from a sufficiently large subset of the measured quantities. Due to the non-linearity of the equations to be solved (see below), it is not a priori obvious how large such a minimal set of observables must be. Second, the validity of the thermal ansatz will be subjected to a test by making predictions for additional observables for which data exist or can be measured in future experiments. Third, systematic deviations of the model predictions with the experimental data will be analyzed and interpreted.

Before proceeding further we should comment on the determination of the freeze-out temperature $T_f$. It can be extracted either from the hadronic $m_\perp$-spectra[96,97] or from the particle ratios[17,20,19]. However, it should be stressed that these two procedures lead conceptually to different freeze-out temperatures: The momentum spectra determine (modulo possible ambiguities from collective transverse flow) the *thermal freeze-out* temperature, at which *all collisions* stop and the momentum distributions are frozen in. The temperature extraction from particle ratios relies on the assumption of chemical equilibrium and thus determines the point of *chemical freeze-out* where the number-changing *inelastic collisions* cease and the particle abundances are frozen in. Generically chemical freeze-out occurs prior to thermal freeze-out, but in exceptional situations like the sudden hadronization of a QGP into a dilute, free-streaming hadronic system both freeze-out processes can occur simultaneously.

We would like to stress that in the context of the following analysis temperature values far above $T_c \simeq 150$ MeV generally pose a threat to its theoretical consistency: taking such temperatures seriously one is necessarily led to doubt the lattice QCD evidence[42] for a deconfining phase transition around $T_c$=150 MeV, and one must face serious questions about the applicability of the non-interacting hadron resonance gas picture underlying the analysis (at $T$=200 MeV a hadron resonance gas is already beyond its dense packing limit).

*6.1 Thermal and Chemical Parameters from the WA85 Experiment*

The WA85 experiment provides data on multistrange baryon and antibaryon production on sulphur-tungsten collisions at 200 A GeV (see Table 9). The WA85 collaboration has measured $\Lambda$, $\bar{\Lambda}$, $\Xi^-$, $\bar{\Xi}^+$, $\Omega^-$, and $\bar{\Omega}^+$; very recently (see WA85 contributions to Refs. 12, 13 and Ref. 188) they also extracted the spectra and yields of neutral and charged kaons. The data on strange baryons have been analyzed independently by two different groups, Refs. 25, 100 and Refs. 18, 19, with mutually consistent results. We will follow here the major steps of the work of Ref. 25 and comment on the other work later.

We concentrate here on the dataset published in Refs. 185, 186. There we have the four ratios

$$R_\Xi = \frac{\bar{\Xi}^+}{\Xi^-} = 0.41 \pm 0.05 \quad \text{for } y \in (2.3, 2.8) \text{ and } m_\perp > 1.9 \text{ GeV}; \quad (79)$$

$$R_\Lambda = \frac{\bar{\Lambda}}{\Lambda} = 0.20 \pm 0.01 \quad \text{for } y \in (2.3, 2.8) \text{ and } m_\perp > 1.9 \text{ GeV}; \quad (80)$$

$$R_{\Xi,\Lambda} = \frac{\Xi^-}{\Lambda} = 0.19 \pm 0.01 \quad \text{for } y \in (2.3, 2.8) \text{ and } m_\perp > 1.9 \text{ GeV}; \quad (81)$$

$$R_{\bar{\Xi},\bar{\Lambda}} = \frac{\bar{\Xi}^+}{\bar{\Lambda}} = 0.40 \pm 0.04 \quad \text{for } y \in (2.3, 2.8) \text{ and } m_\perp > 1.9 \text{ GeV}. \quad (82)$$

Note that these ratios are all measured in the same kinematic region, and that only three of the four ratios are independent because of the relation

$$R_\Lambda \, R_{\bar{\Xi},\bar{\Lambda}} = R_\Xi \, R_{\Xi,\Lambda} \,. \quad (83)$$

*First estimates for WA85*

The common kinematical region of the ratios in Eqs. (79)–(82) allows a first estimate[84] via Eq. (75). Specifically we obtain

$$R_\Xi = \frac{\lambda_q^{-1}\lambda_s^{-2}}{\lambda_q\lambda_s^2} \,; \quad R_\Lambda = \frac{\lambda_q^{-2}\lambda_s^{-1}}{\lambda_q^2\lambda_s} \,; \quad R_{\Xi,\Lambda} = \frac{\gamma_s^2\lambda_q\lambda_s^2}{\gamma_s\lambda_q^2\lambda_s} \,; \quad R_{\bar{\Xi},\bar{\Lambda}} = \frac{\gamma_s^2\lambda_q^{-1}\lambda_s^{-2}}{\gamma_s\lambda_q^{-2}\lambda_s^{-1}}. \quad (84)$$

This system of equations is easily solved for $\lambda_q$, $\lambda_s$ and $\gamma_s$:

$$\lambda_q^6 = R_\Xi R_\Lambda^{-2} \,; \quad (85)$$

$$\lambda_s^6 = R_\Lambda R_\Xi^{-2} \,; \quad (86)$$

$$\gamma_s^2 = R_{\Xi,\Lambda} R_{\bar{\Xi},\bar{\Lambda}} \,, \quad (87)$$

from which one obtains

$$\lambda_q = 1.47 \pm 0.05 \,; \quad \lambda_s = 1.03 \pm 0.05 \,; \quad \gamma_s = 0.28 \pm 0.03 \,. \quad (88)$$

As we will see in the following subsection, the first two values are roughly correct, but the estimate for $\gamma_s$ is wrong by about a factor 2 due to uncorrected resonance decays.

*Resonance gas analysis for WA85*

Incorporating resonance decays and allowing for collective flow needs a numerical treatment. Because for the particle ratios we need the *chemical freeze-out temperature*, which is not necessarily the same as the *thermal freeze-out temperature* reflected in the momentum spectra, we take $T_f$ as a free parameter and investigate three cases:

**A:** a thermal model without flow, $\beta_f = 0$, where the freeze-out temperature $T_f$ is assumed to correspond directly to the value $T_{app} = 232$ MeV deduced from the slope of the transverse mass spectra of high-$m_\perp$ strange (anti-)baryons[185,186];

**B:** a model with a freeze-out temperature of $T_f = 150$ MeV, i.e. a value consistent with the kinetic freeze-out criterium developed in Ref. 97. This temperature also agrees with the QCD phase transition temperature from lattice Monte Carlo simulations. It entails a flow velocity at freeze-out of $\beta_f = 0.41$;

**C:** in order to maintain zero net strangeness in a hadron gas fireball at relative chemical equilibrium between strange meson and baryon abundances, we consider also freeze-out at $T_f = 190$ MeV with $\beta_f = 0.20$.

The system of equations to be solved may be written as

$$R_\Xi = \left.\frac{\Xi^+}{\Xi^-}\right|_{m_\perp \geq m_\perp^{\rm cut}} = \frac{\gamma_s^2 \lambda_q^{-1} \lambda_s^{-2} \tilde{N}_\Xi^{\Xi^*} + \gamma_s^3 \lambda_s^{-3} \tilde{N}_\Xi^{\Omega^*}}{\gamma_s^2 \lambda_q \lambda_s^2 \tilde{N}_\Xi^{\Xi^*} + \gamma_s^3 \lambda_s^3 \tilde{N}_\Xi^{\Omega^*}}, \tag{89}$$

$$R_\Lambda = \left.\frac{\overline{\Lambda}}{\Lambda}\right|_{m_\perp \geq m_\perp^{\rm cut}} = \frac{\lambda_q^{-3} \tilde{N}_\Lambda^{N^*} + \gamma_s \lambda_q^{-2} \lambda_s^{-1} \tilde{N}_\Lambda^{Y^*} + \gamma_s^2 \lambda_q^{-1} \lambda_s^{-2} \tilde{N}_\Lambda^{\Xi^*}}{\lambda_q^3 \tilde{N}_\Lambda^{N^*} + \gamma_s \lambda_q^2 \lambda_s \tilde{N}_\Lambda^{Y^*} + \gamma_s^2 \lambda_q \lambda_s^2 \tilde{N}_\Lambda^{\Xi^*}}, \tag{90}$$

$$R_s = \left.\frac{\Xi^-}{\Lambda}\right|_{m_\perp \geq m_\perp^{\rm cut}} = \frac{\gamma_s^2 \lambda_q \lambda_s^2 \tilde{N}_\Xi^{\Xi^*} + \gamma_s^3 \lambda_s^3 \tilde{N}_\Xi^{\Omega^*}}{\lambda_q^3 \tilde{N}_\Lambda^{N^*} + \gamma_s \lambda_q^2 \lambda_s \tilde{N}_\Lambda^{Y^*} + \gamma_s^2 \lambda_q \lambda_s^2 \tilde{N}_\Lambda^{\Xi^*}}, \tag{91}$$

where the $\tilde{N}_i^R$ are only functions of the rest masses and the given temperature $T$, as can be seen from (72). The result of the numerical solution of (89)–(91) is given in Table 1.

Since the probability to excite the various resonances contributing to the observed final particle abundances depends on $T_f$, it is not surprising that slightly different values of the chemical parameters result from the same data in the three different scenarios. Comparing the numerical results of Table 1 with the simple analytical approximation of Eq. (88), which contained no contributions from higher resonances, we can draw the following conclusions:

1. The extraction of $\lambda_q$ and $\lambda_s$ according to Eq. (84) is only very weakly affected by resonance decays and by the origin of the slope of the $m_\perp$-spectrum (thermal or flow). (The absolute value of the associated chemical potentials does, of course, depend on the choice of the freeze-out temperature.) The reason is that in Eq. (89) (and respectively (90)) the first (respectively second term) in the sums occurring in the numerator and denominator completely dominates. Neglecting then the other terms leads to the same equation as in (84). Therefore the determination of $\lambda_q$ and $\lambda_s$ via Eqs. (85) and (86) leads already to a precise result.

2. Since the $\Xi^-/\Lambda$ ratio is the crucial ingredient for the determination of the strangeness saturation factor $\gamma_s$, the effects from resonance decays and flow can be clearly seen in this number. From Eq. (88) we had extracted $\gamma_s = 0.28 \pm 0.03$. This could be drastically improved by the simple observation that the $\Sigma^0$, which is approximately as abundant as the $\Lambda$, decays into $\Lambda + \gamma$, and this fast electromagnetic decay cannot be resolved experimentally. Therefore about 50% of the observed $\Lambda$'s result from $\Sigma^0$ decays which improves the naive estimate (88) by a factor of 2 to a value of 0.56. Taking into account the full resonance spectrum raises the value of $\gamma_s$ even further to values around 0.7 – 0.8, surprisingly close to full equilibrium, i.e. $\gamma_s = 1$.

Table 1: Thermal fireball parameters extracted from the WA85 data[185,186] on strange baryon and antibaryon production, for three different interpretations of the measured $m_\perp$-slope. Resonance decays were included. For details see text.

|  | **A** "thermal" $T = 232$ MeV $\beta_f = 0$ | **B** "thermal & flow" $T = 150$ MeV $\beta_f = 0.41$ | **C** strangeness balance $T = 190$ MeV $\beta_f = 0.20$ |
|---|---|---|---|
| $\lambda_s$ $\mu_s/T$ $\mu_s$ (MeV) | 1.03 ± 0.05 0.03 ± 0.05 7 ± 11 | 1.03 ± 0.05 0.03 ± 0.05 4 ± 7 | 1.03 ± 0.05 0.03 ± 0.05 6 ± 9 |
| $\lambda_q$ $\mu_q/T$ $\mu_B$ (MeV) | 1.49 ± 0.05 0.40 ± 0.04 278 ± 23 | 1.48 ± 0.05 0.39 ± 0.04 176 ± 15 | 1.48 ± 0.05 0.39 ± 0.04 223 ± 19 |
| $\gamma_s$ | 0.69 ± 0.06 | 0.79 ± 0.06 | 0.68 ± 0.06 |
| $\varepsilon$ | −0.22 | 0.37 | 0 |
| $S/B$ $D_Q$ | 18.5 ± 1.5 0.135 ± 0.01 | 48 ± 5 0.08 ± 0.01 | 26 ± 2.5 0.12 ± 0.01 |

3. We see from Eq. (88) and Table 1 that $\mu_s$ is small and compatible with zero. This confirms the results of Ref. 84 based on the simple estimates (85), (86) and implies that the WA85 data on strange baryon and antibaryon production from 200 GeV A S+W collisions establish a vanishing strange quark chemical potential. This result is stable against large variations in the temperature and transverse flow velocities and thus insensitive to the production and decay of high mass resonances. The reason is that these parameters are extracted from particle/antiparticle ratios in which the resonance decays largely cancel. A vanishing strange quark chemical potential is natural in a strangeness neutral QGP. Note also that also the light quark fugacity is practically unaffected by the complex pattern of resonance formation and decay as well as by the amount of transverse flow in the spectra.

4. Assuming that at freeze-out the hadronic system is in a state of thermal and relative chemical equilibrium and thus fully characterized by the parameters given in Table 1, all other hadron abundances can now be predicted and the strangeness neutrality can be checked. The second to last row in Table 1 shows that under these assumptions strangeness neutrality is only established in scenario C. Although it is not necessary that strangeness neutrality is satisfied *locally* – the conservation laws ensure strangeness neutrality only for the total fireball – it is hard to see how local variations from strangeness neutrality above $|\varepsilon| \lesssim 0.1$ could be generated dynamically or by fluctuations. Thus we get into serious problems with scenarios A and B.

   One should keep in mind that the strange mesons play a crucial role in the strangeness neutrality relation (56); in order to resurrect, for example, scenario B (because it features a reasonably low freeze-out temperature), it would be necessary to break chemical equilibrium at least between the mesons and baryons. This would require the introduction of at least one further non-equilibrium parameter (in addition to the global strangeness undersaturation factor $\gamma_{\rm s}$). As we will see below and in Section 7, entropy considerations point in a similar direction. Under these conditions the applicability of the concept of relative chemical equilibrium in a thermalized hadron gas state to the observed final state in heavy ion collisions at CERN energies appears highly questionable. We will, however, continue a little further down this line by asking whether the concept works at least within the subspace of baryons and antibaryons. The following subsection will show that, strikingly, the predictions arising from this picture are in qualitative agreement with experiment.

5. A similar extraction of thermal and chemical parameters from the WA85 experiment was done in Refs. 17, 18, 19. Because Refs. 17, 18 deal with older data of the WA85 experiment we only want to comment on Ref. 19 which uses the same input as the calculation described above. The strategy in Ref. 19 is somewhat different. They extrapolate the ratios of Ref. 186 to $p_\perp = 0$ and compare these values to fully integrated thermal particle ratios. They also force the system to be strangeness neutral, using the strangeness balance equation (56) to uniquely determine the chemical freeze-out temperature. Their results are $T = 190 \pm 10$ MeV, $\mu_{\rm B} = 240 \pm 40$ MeV and $\gamma_{\rm s} = 0.7$, in agreement with the scenario C presented in Table 1 above.

**Table 2:** Input (*) and predicted high-$m_\perp$ particle ratios near central rapidity ($2.5 < y < 3.0$) for 200 A GeV S+W collisions. The parameters for the three "thermal", "thermal & flow" and "balanced strangeness" scenarios are as specified in Table 1.

| $N_i/N_j$ $m_\perp^{\text{cut}} = 1.9 \text{ GeV}$ | A "thermal" | B "thermal & flow" | C "strangeness balance" |
|---|---|---|---|
| $\bar{\Xi}^+/\Xi^-$ | 0.41 ± 0.05* | 0.41 ± 0.05* | 0.41 ± 0.05* |
| $\bar{\Lambda}/\Lambda$ | 0.20 ± 0.01* | 0.20 ± 0.01* | 0.20 ± 0.01* |
| $\Xi^-/\Lambda$ | 0.19 ± 0.01* | 0.19 ± 0.01* | 0.19 ± 0.01* |
| $\bar{\Xi}^+/\bar{\Lambda}$ | 0.40 ± 0.04* | 0.40 ± 0.04* | 0.40 ± 0.04* |
| $\Lambda/p$ | 0.60 ± 0.06 | 0.62 ± 0.06 | 0.59 ± 0.06 |
| $\bar{\Lambda}/\bar{p}$ | 1.2 ± 0.1 | 1.2 ± 0.1 | 1.2 ± 0.1 |
| $\Omega^-/\Xi^-$ | 0.53 ± 0.05 | 0.29 ± 0.03 | 0.45 ± 0.04 |
| $\bar{\Omega}^+/\bar{\Xi}^+$ | 1.1 ± 0.1 | 0.60 ± 0.06 | 0.94 ± 0.09 |
| $\Omega^-/\Lambda$ | 0.10 ± 0.02 | 0.05 ± 0.01 | 0.08 ± 0.02 |
| $\bar{\Omega}^+/\bar{\Lambda}$ | 0.42 ± 0.08 | 0.23 ± 0.05 | 0.36 ± 0.07 |
| $\bar{\Omega}^+/\Omega^-$ | 0.85 ± 0.25 | 0.85 ± 0.25 | 0.85 ± 0.25 |
| $\bar{p}/p$ | 0.10 ± 0.02 | 0.10 ± 0.02 | 0.10 ± 0.02 |
| $K^-/K^+$ | 0.53 ± 0.08 | 0.51 ± 0.08 | 0.53 ± 0.08 |
| $(\Omega^- + \bar{\Omega}^+)/(\Xi^- + \bar{\Xi}^+)$ | 0.7 ± 0.2 | 0.4 ± 0.1 | 0.6 ± 0.15 |
| $1.0 < p_\perp < 2.5$ GeV | | | |
| $K_s^0/\Lambda$ | 0.41 ± 0.08 | 1.0 ± 0.2 | 0.56 ± 0.11 |
| $K_s^0/\bar{\Lambda}$ | 2.0 ± 0.4 | 5.1 ± 1.0 | 2.8 ± 0.6 |

*Predictions resulting from the WA85 analysis*

It is evident that the extraction of three thermodynamic parameters ($\lambda_q$, $\lambda_s$, and $\gamma_s$) from three independent particle ratios cannot be considered a test of the generalized thermal model considered here, in particular since the extracted values depend on how the temperature is fixed in the analysis. However, with the thermal parameters now fixed according to scenarios A, B, or C, further particle ratios can be predicted and are given in Table 2.

Some of these ratios can be compared with new data from 200 A GeV S+W collisions. WA85 has published the following (partially preliminary) values[185,188,189,190]:

$$R_\Omega = \frac{\bar{\Omega}^+}{\Omega^-} = 0.57 \pm 0.41 \quad \text{for } y \in (2.5, 3.0) \text{ and } m_\perp > 2.3 \text{ GeV}$$

$$R_{\Xi\Omega} = \frac{\Omega^- + \bar{\Omega}^+}{\Xi^- + \bar{\Xi}^+} = 1.7 \pm 0.9 \quad \text{for } y \in (2.5, 3.0) \text{ and } m_\perp > 2.3 \text{ GeV}$$

$$R_{K\Lambda} = \frac{K_s^0}{\Lambda} = 1.4 \pm 0.1 \quad \text{for } y \in (2.5, 3.0) \text{ and } 1.0 < p_\perp < 2.5 \text{ GeV}$$

$$R_{K\bar{\Lambda}} = \frac{K_s^0}{\bar{\Lambda}} = 6.4 \pm 0.4 \quad \text{for } y \in (2.5, 3.0) \text{ and } 1.0 < p_\perp < 2.5 \text{ GeV}$$

$$R_K = \frac{K^-}{K^+} = 0.60 \pm 0.05 \quad \text{for } y \in (2.3, 3.0) \text{ and } p_\perp > 0.9 \text{ GeV}$$

The $\bar{\Omega}/\Omega$-ratio agrees within error bars with the prediction, but the large experimental uncertainty limits its usefulness. In the model $\bar{\Omega}/\Omega$ is given by $\lambda_s^{-6}$ and thus should be 1 if the strange quark chemical potential vanished exactly. – $R_{\Xi\Omega}$ disagrees with the prediction: only for scenario A with the highest freeze-out temperature the two values overlap within their error bars. For larger $\gamma_s$ the situation would improve. However, again the experimental error bar is very large, and a definite conclusion requires better data statistics. – The $K^-/K^+$ ratio is much more accurately known; its central value is slightly higher than the prediction, but considering the error bars and the fact that the experimental ratio includes all kaons down to $m_\perp \simeq 1$ GeV, where the resonance contribution in particular to the $K^-$ begins to become important, the agreement is good.

The kaon to (anti-)lambda ratios, which are the only two ratios combining mesons with baryons, are quite badly off the predicted values; only for the lowest assumed freeze-out temperature the prediction is anywhere close to the data. This raises the question how strangeness neutrality is realized in the experimental data. Clearly more work is required to clarify these issues.

There is also a recent experimental value for $\bar{\Lambda}/\bar{p}$ for the similar S+Au collision system from the NA35 collaboration[166] at CERN. Unfortunately, it was measured in a different kinematic domain, namely forward of central rapidity ($3 \leq y \leq 5$) and integrated over all $p_\perp \leq 2$ GeV. Therefore the comparison with WA85 is somewhat problematic. The experimental value from NA35 is $\bar{\Lambda}/\bar{p} = 0.8 \pm 0.3$. The ratios predicted from the thermal parameters in Table 1 for the same $p_\perp$ range are $0.59 \pm 0.06$ (A), $0.60 \pm 0.06$ (B) and $0.57 \pm 0.06$ (C), respectively, for the three scenarios. Again there is agreement, but due to the rather large error bars and the different $y$-window of the data the value of this statement is limited.

From this comparison, taking into account the partially still very large experimental uncertainties, we conclude that the concept of relative chemical equilibrium works reasonably well within the baryonic and mesonic sectors separately, but fails badly when mesonic and baryonic yields are compared with each other. We will encounter a similar problem with pions relative to the baryons when we analyse entropy production in Section 7. This indicates a breaking of chemical equilibrium between mesons and baryons.

*6.2 Thermal and Chemical Parameters from the NA35 Experiment*

For the strange particle production data of the NA35 collaboration a similar analysis was done in Ref. 20, but now for the lighter, symmetric system S+S. Although this experiment has not measured any multistrange baryons or antibaryons, it has accumulated a complete set of $m_\perp$ and $y$-spectra for all the hadrons which they could identify. Due to the symmetry of the system, data taken at rapidities $y \leq 3$ can be simply reflected around the nucleon-nucleon center of mass at $y_{cm} = 3$ to obtain a complete rapidity spectrum. Therefore this experiment can be used to perform a global chemical analysis using the phase space integrated particle yields ($4\pi$ data), or to study the rapidity dependence of the chemical parameters.

The advantage of using full phase space data lies in the insensitivity to the flow dynamics at freeze-out. In the WA85 analysis longitudinal and transverse flow components had to be taken into account by cutting the theoretical flow spectra to the experimental kinematic acceptance range. For the $4\pi$ data this is not necessary. However, to achieve this simplification, one must assume that the system is characterized by a single set of thermal freeze-out parameters. While the observed rapidity independence of the slope of the $m_\perp$ spectra suggests a roughly rapidity independent freeze-out temperature and transverse flow velocity, the assumption of rapidity independent chemical potentials cannot be justified. A $4\pi$ analysis can thus only give information on the rapidity averaged chemical potentials.

In Ref. 20 a chemical analysis of the $4\pi$ data was performed and compared with an analysis of the particle ratios from the midrapidity bin only. Except for the extracted values for the baryon chemical potential (whose $4\pi$ average came out larger than the central rapidity value, in agreement with expectations based on the central rapidity dip of the proton distribution seen in this collision system), the results and conclusion were the same. More recently, Slotta et al.[85] performed a fit to the complete rapidity dependence of all particle yields and ratios. We will comment on that work at the end of this subsection and show that it mostly confirms the conclusions of Ref. 20.

Within the global ansatz for the $4\pi$-integrated data one can also easily compare the total hadron production from S+S and from elementary nucleon-nucleon (N+N) collisions at the same collision energy per nucleon. Gaździcki et al.[101] have extracted total multiplicities for an isospin symmetric nucleon-nucleon collision at 200 A GeV by isospin averaging available p+p and p+n (p+d) results, using a reasonable extrapolation to the n+n case. A comparison with the isospin symmetric S+S collision system illustrates very clearly the genuine nuclear (medium) effects on particle production. We will perform such a comparison within the thermal model and parametrize these effects via the thermal and chemical parameters. In spite of all the caveats about applying a thermal model to single N+N collisions, the result provides a very nice intuitive understanding of the basic difference between nucleon-nucleon and nucleus-nucleus collisions.

The relevant data[101,102,157-165] are summarized in Table 3. From the quoted multiplicities we build three ratios, avoiding meson to baryon ratios because of the indications for a meson/baryon non-equilibrium from the WA85 analysis above.

**Table 3:** List of measured particle multiplicities and corresponding ratios. The S+S data are from Refs. 102, 157–165, the N+N data from Ref. 101. The values listed under "S+S midrapidity" are the hadron multiplicities in the rapidity interval $2 < y < 3$.

| Multiplicity & Ratio | S+S $4\pi$ | S+S midrap. | N+N $4\pi$ |
|---|---|---|---|
| $K^+$ | $12.5 \pm 0.4$ | $3.2 \pm 0.5$ | $0.24 \pm 0.02$ |
| $K^-$ | $6.9 \pm 0.4$ | $2.2 \pm 0.5$ | $0.17 \pm 0.02$ |
| $\Lambda$ | $8.2 \pm 0.9$ | $2.05 \pm 0.2$ | $0.096 \pm 0.015$ |
| $\bar{\Lambda}$ | $1.5 \pm 0.4$ | $0.57 \pm 0.2$ | $0.013 \pm 0.005$ |
| $p - \bar{p}$ | $20 \pm 3$ | $3.2 \pm 1$ | $0.90 \pm 0.09$ |
| $N^-$ | $98 \pm 5$ | $26 \pm 1$ | $3.22 \pm 0.06$ |
| $R_\Lambda = \dfrac{\bar{\Lambda}}{\Lambda}$ | $0.18 \pm 0.05$ | $0.28 \pm 0.1$ | $0.135 \pm 0.055$ |
| $R_K = \dfrac{K^+}{K^-}$ | $1.8 \pm 0.1$ | $1.45 \pm 0.4$ | $1.4 \pm 0.2$ |
| $R_{\Lambda,p} = \dfrac{\Lambda}{p - \bar{p}}$ | $0.41 \pm 0.08$ | $0.64 \pm 0.2$ | $0.11 \pm 0.02$ |
| $D_Q$ | $0.09 \pm 0.02$ | $0.06 \pm 0.02$ | $0.12 \pm 0.01$ |

*First estimates for NA35*

We begin again by using the approximation (75) to obtain a first estimate of the thermal parameters. From the two relations

$$R_\Lambda = \lambda_q^{-4} \lambda_s^{-2}, \quad R_K = \lambda_q^2 \lambda_s^{-2} \tag{92}$$

one easily gets

$$\lambda_q = R_\Lambda^{-1/6} R_K^{1/6} = 1.47 \pm 0.02, \quad \lambda_s = R_\Lambda^{-1/6} R_K^{-1/3} = 1.09 \pm 0.02. \tag{93}$$

With the available data $\gamma_s$ cannot be estimated by simple algebra. For this we would need the ratio $R_{\bar{\Lambda},\bar{p}} = \bar{\Lambda}/\bar{p}$ in order to combine it with $R_{\Lambda,p}$ such that the rest masses and the temperature cancel. This ratio was recently published[166], but only in a restricted kinematic region. Therefore we must determine $\gamma_s$ numerically from a complete resonance gas analysis.

*Resonance gas analysis for NA35*

The treatment of particle production in an equilibrated hadron gas is always sensitive to its mass spectrum $\{i : g_i, m_i\}$. It should contain all independent hadronic degrees of freedom at freeze-out. However, resonance data become sparse above 1.8 GeV, and thus we don't have a complete set of hadronic states. In contrast to the WA85 analysis, where we always used all well known[88] resonance states up to a mass of 2 GeV, we will here also explore the sensitivity of the chemical analysis to this limit of our knowledge by restricting the experimental mass spectrum in alternative ways, labeled by A, B and C($m_{\rm cut}$):

A: Only the measured particles $\pi$, $K$, $p$, $(n)$, $\Lambda$, plus the $\Sigma$ are taken into account; the $\Sigma$ is added because a large fraction of the measured $\Lambda$ stems from $\Sigma^0 \to \Lambda + \gamma$.

B: We include the pseudoscalar and pseudovector meson nonet and the baryon spin-1/2 octet and spin-3/2 decuplet.

C: We include the full particle spectrum of all known hadronic resonances with $m_i \leq m_{\rm cut}$.

We also treat the temperature differently than before: Whereas in the WA85 analysis it was essentially taken as a free parameter, we fix it here from the condition of strangeness neutrality, since certainly within the full phase space strangeness must be conserved.

To obtain the parameters $T$, $\lambda_{\rm q}$, $\lambda_{\rm s}$ and $\gamma_{\rm s}$, we now must solve a similar set of equations as in (89)–(91) plus the strangeness neutrality condition (56). The contributions from strong and electromagnetic decays after freeze-out are taken into account. Decay products from weak decays can be neglected because they are strongly discriminated against in the experiment by the requirement that the reconstructed tracks point back to the target. The results are given in Table 4. Although for the measured particle species the resonance decay contributions are appreciable, the simple first estimates (93) are again not strongly affected, because of a large amount of cancellation of these contributions from particle/antiparticle ratios. $\gamma_{\rm s}$ is mostly sensitive to the experimental $\Lambda/(p - \bar{p})$ ratio for which no such cancellation occurs. Accordingly we see a strong sensitivity of $\gamma_{\rm s}$ to the chosen resonance spectrum. This is also graphically presented in Figure 12. The errors on the freeze-out parameters were calculated by error propagation from the measured multiplicities.

We now shortly discuss these results with respect to their physical and systematic reliability. The physical validity of our procedure can be tested by comparing these full phase space results with a similar analysis of midrapidity ratios. Of course, the midrapidity results are correlated to the full phase space results since the same data set is used; an estimate of the systematic uncertainties of our procedure can, however, still be obtained from this comparison. Since the midrapidity ratios have much larger statistical errors, error propagation becomes a problem, and we thus refrain from giving error estimates for the midrapidity results in Table 4.

**Table 4:** Generalized thermal model parameters resulting from our analysis of the particle ratios given in Table 3, for the various tested scenarios as described in the text.

|  |  | $\lambda_q$ | $\lambda_s$ | $\gamma_s$ | $T$ [MeV] |
|---|---|---|---|---|---|
| S+S 4π | A | 1.47 ± 0.07 | 1.09 ± 0.05 | 0.58 ± 0.13 | 191 ± 13 |
|  | B | 1.48 ± 0.07 | 1.08 ± 0.06 | 0.67 ± 0.15 | 187 ± 16 |
|  | C(1.4) | 1.51 ± 0.07 | 1.04 ± 0.07 | 0.64 ± 0.14 | 217 ± 30 |
|  | C(1.6) | 1.52 ± 0.08 | 1.02 ± 0.08 | 0.84 ± 0.18 | 219 ± 35 |
|  | C(1.8) | 1.54 ± 0.06 | 1.00 ± 0.09 | 0.95 ± 0.20 | 205 ± 39 |
|  | C(2.0) | 1.57 ± 0.06 | 0.99 ± 0.09 | 1.00 ± 0.21 | 197 ± 29 |
| S+S midrap. | A | 1.32 | 1.09 | 0.76 | 185 |
|  | B | 1.32 | 1.08 | 0.88 | 176 |
|  | C(1.4) | 1.33 | 1.06 | 0.83 | 198 |
|  | C(1.6) | 1.34 | 1.05 | 1.05 | 193 |
|  | C(1.8) | 1.35 | 1.04 | 1.16 | 183 |
|  | C(2.0) | 1.36 | 1.04 | 1.19 | 178 |
| N+N 4π | A | 1.48 ± 0.12 | 1.25 ± 0.12 | 0.17 ± 0.04 | 154 ± 22 |
|  | B | 1.48 ± 0.12 | 1.25 ± 0.11 | 0.19 ± 0.04 | 152 ± 21 |
|  | C(1.4) | 1.51 ± 0.12 | 1.18 ± 0.16 | 0.18 ± 0.04 | 174 ± 43 |
|  | C(1.6) | 1.52 ± 0.12 | 1.18 ± 0.15 | 0.21 ± 0.05 | 172 ± 44 |
|  | C(1.8) | 1.51 ± 0.12 | 1.18 ± 0.16 | 0.21 ± 0.05 | 163 ± 35 |
|  | C(2.0) | 1.51 ± 0.12 | 1.18 ± 0.16 | 0.22 ± 0.05 | 161 ± 31 |

The observed differences to the full phase space results are consistent with simple expectations: from the shape of the proton rapidity spectra[102] it is known that the rather small sulphur nuclei cannot fully stop each other, resulting in a baryon number deficiency at midrapidity. On the other hand, excess strangeness production was found to be maximal near midrapidity[157-165]. Correspondingly we find a reduced value of $\lambda_q$ and an increased value of $\gamma_s$ at midrapidity. Values of $\gamma_s$ above unity should not be taken too seriously: the use of the strangeness neutrality condition in a restricted rapidity window is not absolutely safe, and the large experimental error (30%) on the experimental $\Lambda/(p - \bar{p})$ ratio also directly propagates into the determination of $\gamma_s$.

Altogether the differences between the midrapidity and the 4π values are small, and for $\lambda_s$ and $T$ the differences are even below 10%. This was recently confirmed in an analysis by Slotta[85] of the full rapidity spectra from this experiment. Assuming a boost-invariant longitudinal expansion velocity profile with a maximum flow rapidity $\bar{\eta}$ of ±1.75 relative to the center of mass (this value yields an excellent fit to the shape of the measured pion rapidity spectrum), a rapidity independent freeze-out temperature $T$ and strange quark chemical potential $\mu_s = 0$, he was able to reproduce the $p_T$ and rapidity spectra of the measured proton and all strange hadron spectra from the NA35 S+S experiment simultaneously, in shape as well as normalization. The calculation contains three free parameters: the value of the baryon chemical potential at midrapidity $\mu_B^0$, one parameter $A$ describing its dependence on the longitudinal coordinate (resp. the space-time rapidity $\eta$), and a strangeness suppression factor $\gamma_s$. $\mu_B^0$ and $A$ were fixed from the relative normalization and shape of the $\Lambda$ and $\bar{\Lambda}$ rapidity spectrum using the parametrization $\mu_B(\eta) = \mu_B^0 + A\,\eta^4$ where $\eta$ is measured from midrapidity. $\gamma_s$ was determined by adjusting the relative normalization of the proton and $\Lambda$ spectra. Assuming a freeze-out temperature of $T = 200$ MeV, Slotta found $\mu_B^0 = 177 \pm 33$ MeV, $A = 34.5 \pm 1.5$ MeV, and $\gamma_s = 0.75$ (see Figure 13). Averaged over rapidity, the baryon chemical potential comes out as $240 \pm 40$ MeV, in good agreement with the $4\pi$ analysis which gives $264 \pm 72$ MeV (see Table 4). Overall strangeness neutrality is satisfied at the level of a few percent; perfect neutrality could be reached by a slight adjustment of the temperature. Problems arise with the normalization of the pion spectrum which is underpredicted by about a factor 2 – this problem is generic and will be discussed extensively later in this review.

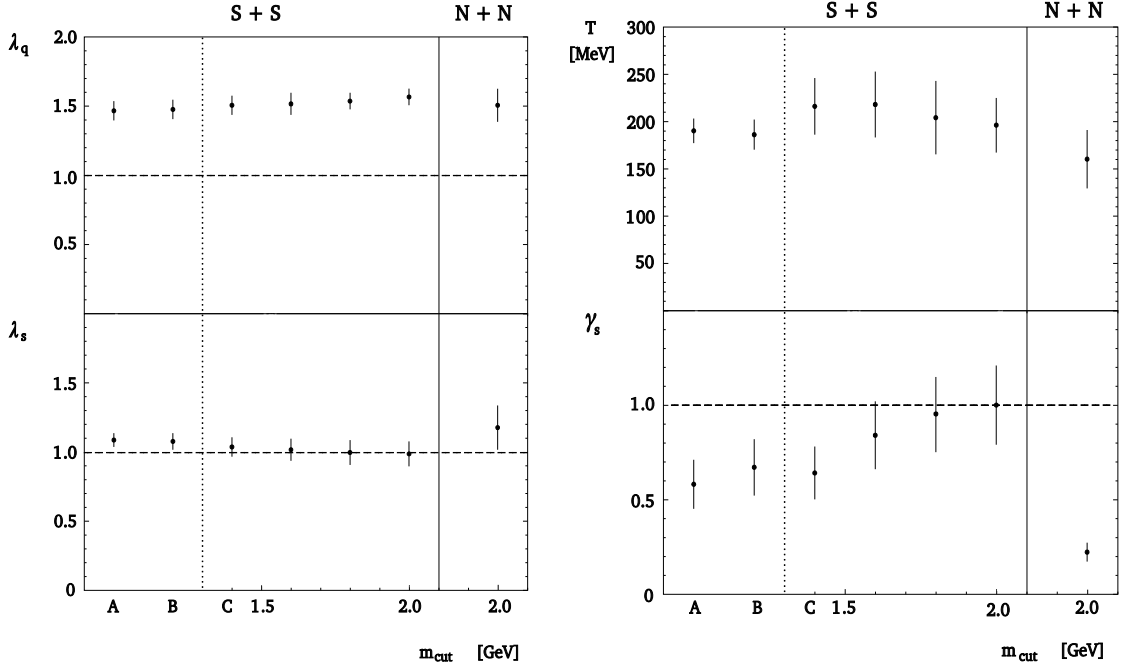

**Figure 12:** The dependence of the extracted thermal parameters $\lambda_q$, $\lambda_s$, $T$ and $\gamma_s$ on the mass spectrum of the hadron system at freeze-out for central S–S collisions at 200 A GeV ($4\pi$ data). (For details see text.) Also one set of values for N+N collisions is plotted.

In Figure 14 an attempt is shown to fit the same rapidity spectra with a lower freeze-out temperature of $T = 150$ MeV, compensating the for the slope of the experimental $m_\perp$ spectra (which for the S+S data corresponds to an apparent temperature of about 200 MeV) by a transverse flow velocity $\beta_f = 0.31$. This is in the spirit of scenario B of the WA85 analysis. The rapidity dependent baryon chemical potential is now given by $\mu_B^0 \simeq 130$ MeV and $A \simeq 24$ MeV. In this case the proton spectra are automatically correctly normalized with $\gamma_s=1$, but the strangeness neutrality and kaon normalization are badly off – very similar to the observations made in the WA85 S+W case. It is striking to see that one additional parameter, namely a meson/baryon suppression factor $\gamma_m = 0.37$, repairs both[85] deficiencies very economically (Figs. 14d,e). The problem of underpredicting the pions[85] by a factor 2-3 remains, however.

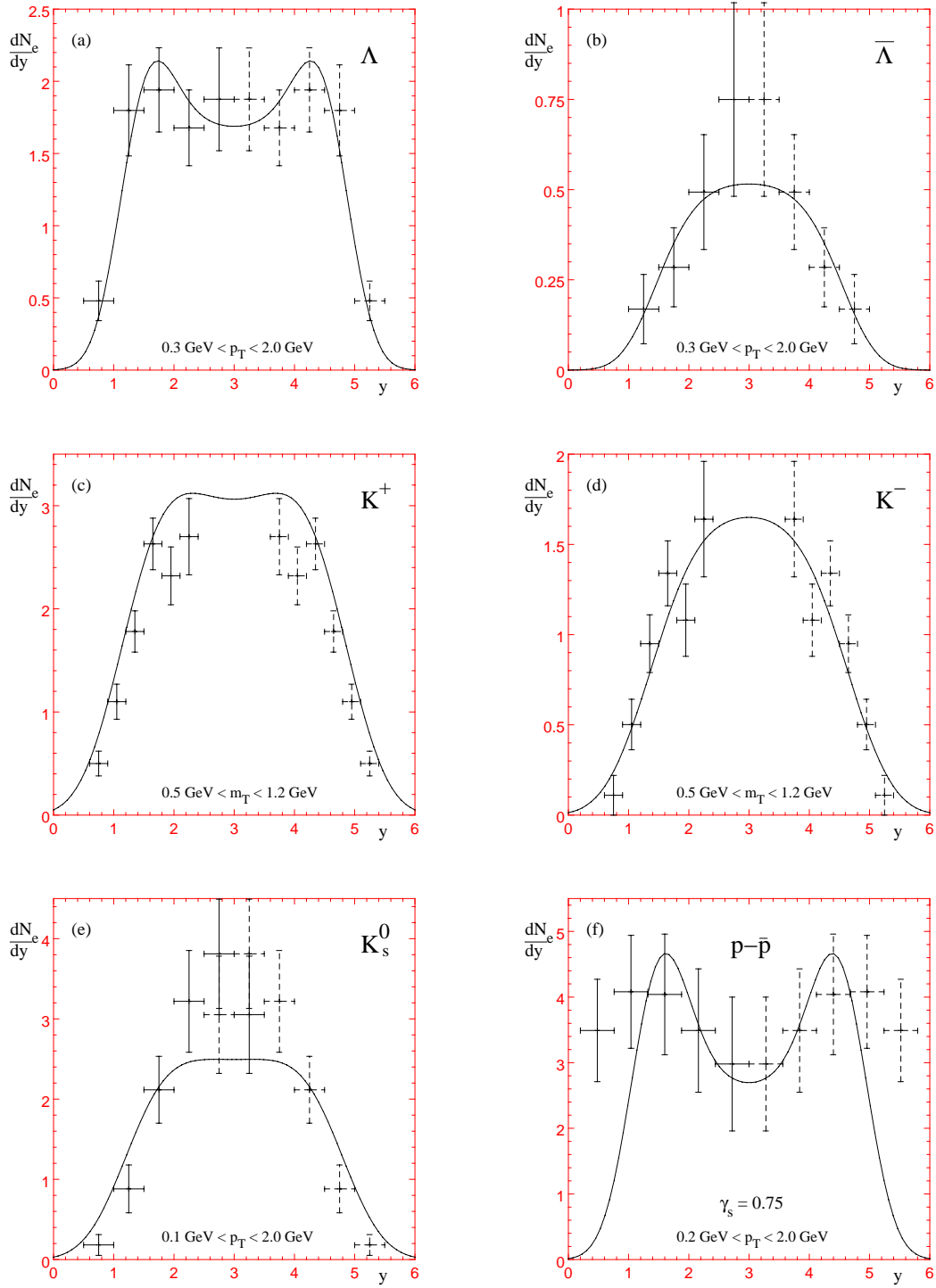

**Figure 13:** Rapidity distributions of $\Lambda$, $\bar{\Lambda}$, $K^+$, $K^-$, $K_s^0$ and $p - \bar{p}$ from the NA35 S+S experiment[102,162,165]. The theoretical curves assume $T = 200\,\text{MeV}$, $\tilde{\eta} = 1.75$, and $\mu_s = 0$ everywhere. The other parameters are described in the text. The theoretical spectra are cut to the experimental acceptance windows, indicated in the figures by the index "e" on the vertical axis.

Let us now return to the sensitivity of the chemical analysis to the hadron mass spectrum. This is illustrated in Figure 12 for the analysis of the $4\pi$ data of NA35. From left to right the assumed spectrum of hadron resonances gets more and more rich. While $T$, $\lambda_q$ and $\lambda_s$ appear to be quite insensitive to this, $\gamma_s$ shows appreciable variations with the hadron mass spectrum. However, as more and more known resonances are included in the spectrum, all values seem to converge to asymptotic limits given by $\lambda_s \simeq 1$, $\lambda_q \simeq 1.6$, $\gamma_s \simeq 1$ and $T \simeq 200$ MeV.

We conclude that chemical freeze-out in central S+S collisions at 200 A GeV is characterized by:

- a vanishing strange quark chemical potential ($\lambda_s \approx 1$ or $\mu_s \approx 0$);

- a largely saturated strange quark phase space ($\gamma_s \simeq 0.75 - 1$).

The chemical freeze-out temperature is around $T = 200$ MeV if relative chemical equilibrium and strangeness neutrality of the hadron gas are assumed. However, we argued before that this temperature is uncomfortably high. Furthermore, the hadron gas picture cannot correctly reproduce the pion multiplicity. Reducing the freeze-out temperature to more "reasonable" values by postulating transverse flow leads to a violation of the strangeness neutrality. This can be repaired by assuming a global meson/baryon suppression (i.e. by giving up relative chemical equilibrium between mesons and baryons), but this does not resolve the pion deficiency problem. All of these features are in qualitative and quantitative agreement between the S+W and S+S collision systems.

An interesting additional result is the difference between the S+S and the N+N data, shown in the right sector of Figure 4 for the C(2.0) scenario. Although one must be careful when applying the thermal description to N+N collisions, we think that the extracted results still show clearly the important qualitative difference between nucleon-nucleon and nucleus-nucleus collisions: While $\lambda_q$ takes similar values, intermediate between the S+S $4\pi$ and midrapidity results, the strange quark fugacity tends to deviate from unity. The temperature is somewhat lower than in S+S case (it is close to the canonical Hagedorn value of 160 MeV); this is also seen in the slope of the $m_\perp$ spectra. A clear qualitative difference is observed, however, for the strangeness saturation factor: it is 4 times larger in central S+S collisions than in N+N interactions. The strangeness enhancement is therefore clearly seen in the thermal description by an increased $\gamma_s$ factor.

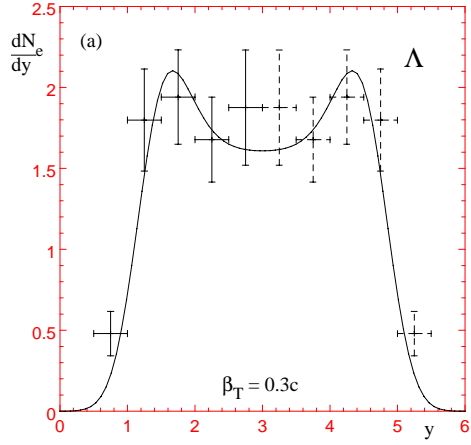
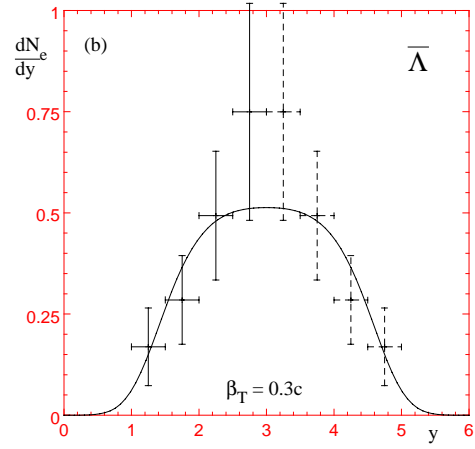
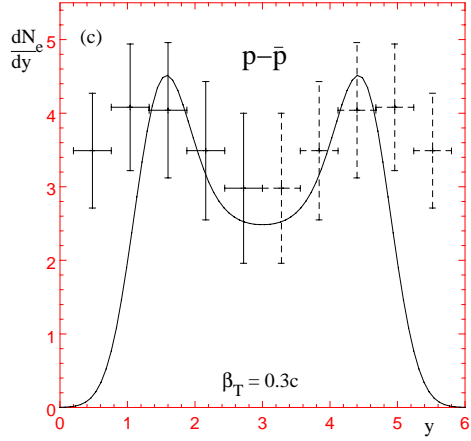
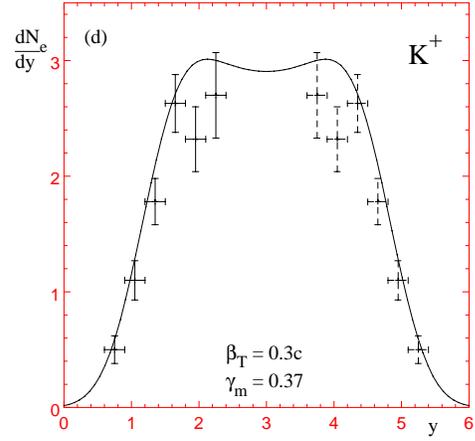
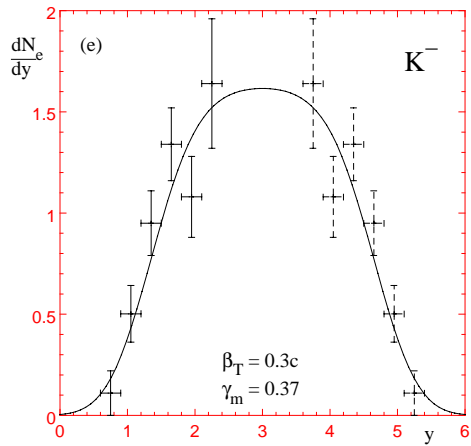

**Figure 14:** Similar to Figure 13, but now assuming $T = 150$ MeV together with a transverse flow velocity $\beta_{\rm f} = 0.3$. In this case the protons are correctly reproduced with $\gamma_s = 1$, but the correct kaon normalization and overall strangeness neutrality require a meson/baryon suppression factor $\gamma_{\rm m} = 0.37$.

**Table 5**: Predicted total particle multiplicities in 200 A GeV S+S collisions for scenario C(2.0), with the freeze-out volume fixed by the total $K^+$ multiplicity. In the lower part of the table we give the abundances of hadronic resonances before their strong decays, but including the feed-down contributions from higher resonances.

| Total multiplicities | S+S $4\pi$ | S+S midrap. | N+N $4\pi$ |
|---|---|---|---|
| $K^+$ | 12.5 | 3.2 | 0.24 |
| $N^-$ | 65 | 15 | 3.9 |
| $\pi^-$ | 54 | 12 | 3.6 |
| $K^-$ | 6.9 | 2.2 | 0.17 |
| $\Lambda$ | 7.5 | 1.4 | 0.097 |
| $\bar{\Lambda}$ | 1.35 | 0.39 | 0.013 |
| $p - \bar{p}$ | 18 | 2.2 | 0.89 |
| $\bar{p}$ | 1.4 | 0.43 | 0.081 |
| $\Xi^-$ | 0.73 | 0.19 | $2.4 \times 10^{-3}$ |
| $\bar{\Xi}^+$ | 0.31 | 0.089 | $5.5 \times 10^{-4}$ |
| $\Omega$ | 0.11 | 0.037 | $7.5 \times 10^{-5}$ |
| $\bar{\Omega}$ | 0.12 | 0.029 | $2.8 \times 10^{-5}$ |
| $\eta$ | 4.9 | 1.29 | 0.34 |
| $\rho^0$ | 6.7 | 1.48 | 0.48 |
| $\omega$ | 5.5 | 1.32 | 0.45 |
| $\phi$ | 1.3 | 0.43 | $5 \times 10^{-3}$ |
| $\Delta$ | 17 | 2.2 | 0.84 |
| $\bar{\Delta}$ | 1.17 | 0.35 | 0.071 |

*Predictions resulting from NA35*

Similar to the WA85 case we can use the extracted thermal parameters to predict further particle ratios[20]. Table 5 lists the predicted total multiplicities from which any ratio can be constructed. The (point particle) volume parameter $V^{pt}$ (46) was fixed by the total $K^+$ yield and then used to normalize all other multiplicities.

Comparing the predictions of Table 5 with the measured values of Table 3 the following points emerge:

- The $K^-$ yield is exact, because the $K^+$ yield and the $K^+/K^-$-ratio were fitted.

- For the baryons $\Lambda$, $\bar{\Lambda}$ and $p - \bar{p}$ data and prediction agree within error bars; this indicates that the $K/\Lambda$ ratio is correctly reproduced – all other ratios were already used in the thermal analysis.

- Table 5 gives a $\bar{\Lambda}/\bar{p}$ ratio of 1.0 ($4\pi$) and 0.9 (midrapidity). The recently published[166] central rapidity value from the NA35 collaboration of $\bar{\Lambda}/\bar{p} = 1.5 \pm 0.4$ is somewhat on the high side, but also affected by large statistical errors.

**Table 6:** Particle ratios and acceptance windows covered by the NA36 S+Pb experiment at CERN. The ratios in the lower part of the table were obtained by extrapolating the data to a common $m_\perp$-window using a thermal ansatz (59) with the measured $m_\perp$-slopes, but without correction for the different $y$ acceptances. All values are taken from Ref. 173.

| Ratio | value | $y$ range | $p_\perp$-range |
|---|---|---|---|
| $R_\Lambda = \bar\Lambda/\Lambda$ | $0.117 \pm 0.011$ | $2.0 < y < 2.5$ | $0.6 < p_\perp < 1.6$ GeV |
| $R_\Xi = \bar\Xi^+/\Xi^-$ | $0.276 \pm 0.108$ | $2.0 < y < 2.5$ | $0.8 < p_\perp < 1.8$ GeV |
| $R_{\Xi,\Lambda} = \Xi^-/\Lambda$ | $0.066 \pm 0.013$ | $1.5 < y < 2.5$ | $0.8 < p_\perp < 1.8$ GeV |
| $R_{\bar\Xi,\bar\Lambda} = \bar\Xi^+/\bar\Lambda$ | $0.127 \pm 0.022$ | $2.0 < y < 3.0$ | $0.6 < p_\perp < 1.6$ GeV |
| Ratio | value | $y$ range | $m_\perp$-range |
| $R_\Lambda = \bar\Lambda/\Lambda$ | $0.093 \pm 0.017$ | $2.0 < y < 2.5$ | $1.5 < m_\perp < 2.2$ GeV |
| $R_\Xi = \bar\Xi^+/\Xi^-$ | $0.276 \pm 0.108$ | $2.0 < y < 2.5$ | $1.5 < m_\perp < 2.2$ GeV |
| $R_{\Xi,\Lambda} = \Xi^-/\Lambda$ | $0.122 \pm 0.024$ | $1.5 < y < 2.5$ | $1.5 < m_\perp < 2.2$ GeV |
| $R_{\bar\Xi,\bar\Lambda} = \bar\Xi^+/\bar\Lambda$ | $0.281 \pm 0.050$ | $2.0 < y < 3.0$ | $1.5 < m_\perp < 2.0$ GeV |

- The negative particle multiplicity $N^-$ is strongly underestimated. This will be discussed in Section 7 in the context of entropy production.

- The resonance yields listed in Table 5 show the importance of resonance contributions. For example the four $\Delta$ isobars contribute about 50% to the final proton yield, and similarly for the antiprotons.

*6.3 Thermal and Chemical Parameters from the NA36 Experiment*

The NA36 collaboration has published[172,173] strange baryon ratios from S+Pb collisions at 200 A GeV. They study the same set of ratios as the WA85 collaboration (Eqs. (79)–(82)), and since the target nuclei are not too different the results should in principle be comparable. Unfortunately, the NA36 ratios (listed in Table 6) are not all given in the same kinematic window, which makes the analysis more difficult. The rapidity spectra from S+S collisions in Figure 13 show that in general the particle ratios have a strong rapidity dependence, and since for the asymmetric S+W and S+Pb collisions the shape of the rapidity spectra is not known, a straightforward comparison between the two experiments is not possible.

Although the ratios may be safe, we should still mention that the absolute normalization of the NA36 strange particle yields has been controversial (see NA35 and NA36 contributions in Ref. 12): for so far unknown reasons the $\Lambda$ rapidity distributions from the NA35 and NA36 experiments disagree by about a factor 2 in normalization and also show a different shape. In view of these unsettled questions all theoretical interpretations of these data sets should be taken with a grain of salt.

*First estimates for NA36*

Similar estimates as in Section 6.1 can be extracted from these data only if the rapidity dependence of the ratios is neglected. As already pointed out, this is very dangerous, and indeed a comparison of the results given in Refs. 172 and 173 shows that at least $R_\Lambda$ is very sensitive to the selected rapidity interval:

$$1.5 < y < 3.0: \quad R_\Lambda = 0.207 \pm 0.012 \quad \text{(Ref. 172)};$$
$$2.0 < y < 2.5: \quad R_\Lambda = 0.117 \pm 0.011 \quad \text{(Ref. 173)}. \tag{94}$$

The $p_\perp$ range for both ratios is the same ($0.6 < p_\perp < 1.6$ GeV).

Since the particle-antiparticle ratios in Table 6 were obtained in the same rapidity interval, one can at least apply Eqs. (85) and (86) to extract the fugacities[173]:

$$\lambda_q = 1.78 \pm 0.09; \quad \lambda_s = 1.03 \pm 0.06. \tag{95}$$

The estimate for $\gamma_s$ from Eq. (87) is less reliable, due to the different rapidity intervals of the required ratios. Indeed, the identity $(R_\Lambda/R_\Xi)(R_{\Xi,\bar\Lambda}/R_{\Xi,\Lambda}) = 1$ is violated by the ratios given in Table 6 by 25%. Neglecting this problem one obtains $\gamma_s = 0.19 \pm 0.02$, and after multiplication by a factor 2 for the $\Sigma^0$ contribution one gets[173] $\gamma_s = 0.38 \pm 0.04$. This is below the naive estimate (88) from the WA85 data, but we already know that it will be corrected upward once resonance decays are included.

*Resonance gas analysis for NA36*

The NA36 collaboration[173] have analyzed their data following the work of Cleymans[17] et al. This incorporates the resonance contributions. In this description strangeness neutrality is taken as a constraint, and full chemical equilibration, $\gamma_s = 1$, is assumed. The result is shown in Figure 15. All ratios are compatible with a fully equilibrated hadron resonance gas at $T = 172 \pm 16$ MeV and $\mu_B = 290 \pm 50$ MeV. This result can be summarized by the parameters

$$\lambda_q = 1.76 \pm 0.15; \quad \lambda_s = 1.07 \pm 0.07; \tag{96}$$
$$T = 172 \pm 16 \text{ MeV}; \quad \gamma_s = 1.0; \tag{97}$$

After the first naive estimate for $\gamma_s$ in Section 6.3 came out below the corresponding one for the WA85 results, it is surprising to see that this resonance gas analysis yields compatibility of the NA36 data with $\gamma_s = 1$, while a similar analysis of the WA85 data[19] finds strong incompatibility with $\gamma_s = 1$ and rather confirms the value $\gamma_s \simeq 0.7$ extracted in Section 6.1. On the other hand, the resulting values for $\lambda_q$ and $T$ are also different. These differences are obviously a consequence of the different experimental particle ratios, in particular of $R_\Lambda$ which differs from the WA85 value by more than a factor 2. As long as quantitative experimental information on the crucial rapidity dependence of these ratios is not available, more detailed statements about the consistency of the two experiments and their respective theoretical interpretations cannot be made. It is, however, interesting to observe that the strange quark chemical potential seems to vanish in all these analyses of heavy ion data at CERN energies, independent of the selected rapidity window, and $\gamma_s$ is always quite large.

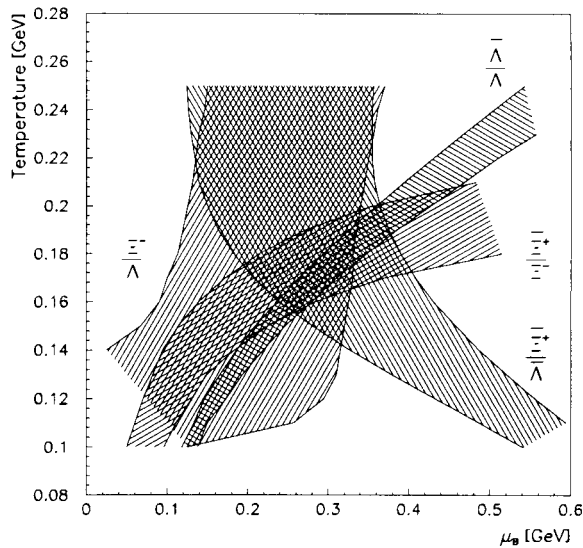

**Figure 15:** The $T - \mu_B$-plane for a strangeness neutral resonance gas with $\gamma_s = 1$. The values compatible with the measured particle ratios within one standard deviation (after resonance decays) are shown as hatched areas. The common overlap region defines the boundary of allowed values for $T$ and $\mu_B$. The Figure was taken from Ref. 173.

### 6.4 Thermal and Chemical Parameters from the AGS Experiments

The heavy ion experiments at the AGS in Brookhaven are performed at a lower beam momentum of about 10 – 15 A GeV. Comparing these experiments with the ones at CERN energies of 160 – 200 A GeV allows to study the beam energy dependence of particle production. We are interested in the differences in strange hadron production and in the resulting changes in the thermal fireball parameters. An analysis along these lines of the available data at the AGS from Si+Au (Pb) collisions was recently done in Ref. 26. Let us shortly discuss some small differences relative to the investigation of the CERN data presented above:

- The hadron gas used in Ref. 26 incorporates baryons up to masses of 2 GeV, but mesons only up to 1.5 GeV. In view of the lower temperature at the AGS (see below) this seems justified.

- In Ref. 26 a finite size correction[103] to the density of states in the partition function was applied. This correction is of the order $\sim \lambda_{th}/R$ where $\lambda_{th} = \sqrt{2\pi/mT}$ is the thermal wavelength of a particle with mass $m$ and $R$ is the linear dimension of the system. This correction is important for low temperatures and for light particles and mostly affects the ratios between a heavy and a light particle (e.g. $\pi/(p+n)$).

- For extensive observables the volume correction was taken into account via the excluded volume description[87]. This drops out from particle ratios.

- The calculation was done assuming absolute chemical equilibrium ($\gamma_s = 1$) from the beginning, i.e. no strangeness suppression was allowed for.

Table 7: Particle ratios calculated in a thermal model for two different temperatures, a baryon chemical potential of $\mu_B = 0.54$ GeV and a strange quark chemical potential $\mu_s$ fixed by the condition of strangeness neutrality, in comparison with experimental data (statistical errors in parentheses) from central Si + Au (Pb) collisions at 14.6 A GeV. The table was taken from Ref. 26.

| Particles | Thermal Model | | Data | | |
|---|---|---|---|---|---|
| | $T=120$ MeV | $T=140$ MeV | exp. ratio | rapidity | Ref. |
| $\pi/(p+n)$ | 1.29 | 1.34 | 1.05(0.05) | 0.6 - 2.8 | 139 |
| $d/(p+n)$ | $4.3 \cdot 10^{-2}$ | $5.8 \cdot 10^{-2}$ | $3.0(0.3) \cdot 10^{-2}$ | 0.4 - 1.6 | 139 |
| $\bar{p}/p$ | $1.47 \cdot 10^{-4}$ | $5.8 \cdot 10^{-4}$ | $4.5(0.5) \cdot 10^{-4}$ | 0.8 - 2.2 | 149 |
| $K^+/\pi^+$ | 0.23 | 0.27 | 0.19(0.02) | 0.6 - 2.2 | 139 |
| $K^-/\pi^-$ | $5.0 \cdot 10^{-2}$ | $6.2 \cdot 10^{-2}$ | $3.5(0.5) \cdot 10^{-2}$ | 0.6 - 2.3 | 139 |
| $K^0_s/\pi^+$ | 0.14 | 0.16 | $9.7(1.5) \cdot 10^{-2}$ | 2.0 - 3.5 | 144 |
| $K^+/K^-$ | 4.6 | 4.3 | 4.4(0.4) | 0.7 - 2.3 | 139 |
| $\Lambda/(p+n)$ | $9.5 \cdot 10^{-2}$ | 0.11 | $8.0(1.6) \cdot 10^{-2}$ | 1.4 - 2.9 | 144 |
| $\bar{\Lambda}/\Lambda$ | $8.8 \cdot 10^{-4}$ | $3.7 \cdot 10^{-3}$ | $2.0(0.8) \cdot 10^{-3}$ | 1.2 - 1.7 | 149 |
| $\phi/(K^++K^-)$ | $2.4 \cdot 10^{-2}$ | $3.6 \cdot 10^{-2}$ | $1.34(0.36) \cdot 10^{-2}$ | 1.2 - 2.0 | 149 |
| $\Xi^-/\Lambda$ | $6.4 \cdot 10^{-2}$ | $7.2 \cdot 10^{-2}$ | 0.12(0.02) | 1.4 - 2.9 | 146 |
| $\bar{d}/\bar{p}$ | $1.1 \cdot 10^{-5}$ | $4.7 \cdot 10^{-5}$ | $1.0(0.5) \cdot 10^{-5}$ | 2.0 | 108 |

Together with the requirement of strangeness neutrality, this last assumption reduces the number of input parameters to two: $T_f$ and $\mu_B$ ($\mu_q$). The temperature was fixed by an analysis of the measured $\Delta(1232)$ contribution to the proton yield. An investigation[104] showed that the $\Delta/p$-ratio is compatible with a temperature in the range of $T = 120$–$140$ MeV. This very useful information is, unfortunately, not available from the CERN experiments. If measured well, it allows an accurate extraction of the chemical freeze-out temperature independent of the baryon and strange chemical potentials.

The baryon chemical potential is fixed by a compromise between the measured $\pi/(p+n)$-ratio and considerations regarding the particle density at freeze-out[26]. The latter can be fixed from the size of the freeze-out volume as measured by HBT interferometry[105] and the measured baryon multiplicity. The resulting value was given as[26] $\mu_B = 540$ MeV.

The results of this analysis are shown in Table 7. In general the agreement between the thermal model ratios and the experimental ones is very good. The antibaryon/baryon ratios are very sensitive to the temperature for fixed $\mu_B$; in the given temperature interval they vary by a factor of 4 and cover a range which includes the experimental value.

In view of the different rapidity windows for the various measurements and the generic rapidity dependence of particle ratios such an agreement is not trivial. How-ever, most of the data cover a rather wide rapidity range around the central rapidity[d] of 1.4, and stopping is more efficient and longitudinal flow less important at AGS energies[106,107]. To a large extent the analysis of Ref. 26 thus resembles in character the global analysis of the $4\pi$ NA35 data at CERN in Section 6.2.

Taking a closer look at the strange particle ratios one observes an interesting systematic behavior. While within the chosen temperature interval the predictions for the two particle/antiparticle ratios $K^+/K^-$ and $\bar\Lambda/\Lambda$ overlap nicely with the experimental values inside the error bars, this is not the case for some of the other ratios containing strange quarks. Let us therefore look at those and divide the experimental mean values by the two model values. Neglecting the $\Xi/\Lambda$-ratio (see below) we obtain

$$\frac{\text{exp. value}}{\text{model value}} \Longrightarrow \begin{array}{ccc} \text{ratio} & T = 120 \text{ MeV} & T = 140 \text{ MeV} \\ K^+/\pi^+: & 0.82 & 0.70 \\ K^-/\pi^-: & 0.70 & 0.56 \\ K^0_s/\pi^+: & 0.70 & 0.61 \\ \Lambda/(p+n): & 0.84 & 0.72 \\ \Phi/(K^+ + K^-): & 0.56 & 0.37 \end{array}$$

The model ratios are systematically below the experimental values by a factor 0.6–0.8. Now all these ratios involve particles with one more strange quark or antiquark in the numerator than in the denominator. This indicates incomplete strangeness saturation and thus a violation of the model assumption of absolute chemical equilibrium. We can soften this assumption by assuming only relative chemical equilibrium and introducing as before the parameter $\gamma_s$ through Eq. (38). From the above table we extract a value of $\gamma_s = 0.7 \pm 0.1$; thus the level of strangeness saturation reached at the AGS is similar to the CERN value.

We left out the $\Xi/\Lambda$-ratio for a particular reason: neither does it follow the above systematics, nor does it agree with the predictions of any of the available event generators[146]. So far there is no model which can explain such a large ratio consistently with the other observations. This strange result cries out for independent experimental confirmation.

Braun-Munzinger et al.[26] have also tried to confirm the thermal parameters by comparing the measured transverse momentum spectra with the model predictions from a thermalized source with the same temperature undergoing longitudinal and transverse hydrodynamical flow[96,97]. Since the model was developed for symmetric systems and the new momentum spectra for Au+Au collisions[137,140] were not yet publicly available, this comparison was done in Ref. 26 for the lighter Si+Al system[e]. The $m_\perp$-spectra were described by Eq. (60) with $\alpha = 1$. After adding resonance decay contributions excellent agreement[26] is obtained for *all* particle spectra from the pions to the deuterons. The presence of transverse flow is very clearly seen in the slope of the experimental spectra for $m_\perp - m_0 > 0.3$ GeV (i.e. beyond the resonance decay region) which flattens considerably with increasing mass of the hadrons[95]; in the data presented in Ref. 26 the deuteron and pion slopes differ by about 50%. This is well reproduced by the assumed average flow velocity $\langle \beta_f \rangle = 0.39$ (0.33) for $T_f = 120$ (140) MeV, respectively.

---

[d]We define as central rapidity the value where the pion production peaks[139].

[e]A similar comparison of the $m_\perp$ spectra with E802 data from Si+Au collisions is shown in Ref. 107, with very similar parameters.

We conclude that a thermal model for Si+Au(Pb,Al) collisions at AGS energies works well and leads to the following thermal parameters:

$$T = 130 \pm 10 \text{ MeV} \quad \mu_B \approx 540 \text{ MeV}$$
$$\gamma_s = 0.7 \pm 0.1 \quad \mu_S = 121 \pm 14 \text{ MeV} \quad (98)$$

In terms of the fugacities used before this translates as

$$\lambda_q = 4.0 \pm 0.5 \quad ; \quad \lambda_s = 1.55 \pm 0.1 . \quad (99)$$

This result can be compared with the earlier, less elaborate thermal model interpretations of Refs. 23, 24. The authors of Ref. 23 consider only the two particle/antiparticle ratios $K^+/K^-$ and $\bar{\Lambda}/\Lambda$ from Table 7 as input. As discussed before, resonance decay contributions can be neglected for these ratios. Using the requirement of strangeness neutrality the fugacities and also the freeze-out temperature can be extracted[24]:

$$\lambda_q = 3.7 \pm 0.3 \quad ; \quad \lambda_s = 1.72 \pm 0.19 \quad ; \quad T_f = 127 \pm 8 \text{ MeV} . \quad (100)$$

This agrees very nicely with Eqs. (98), (99); the small differences are probably due to the fact that, in contrast to Ref. 23, the $\bar{\Lambda}/\Lambda$-ratio was not fitted exactly in Ref. 26.

## 7 Specific Entropy from Heavy Ion Experiments

The careful reader will have noticed the conspicuous absence of any discussion of the pion yields so far. Although the pion is the most abundantly produced hadron, especially so at CERN energies, it carries "no" conserved quantum numbers and is thus unsuitable for extracting information about the chemical potentials. In chemical equilibrium the pion multiplicity is a function of the temperature alone. However, of all the thermodynamic parameters used in Section 6, the chemical freeze-out temperature $T_f$ is the most uncertain one. This uncertainty affects the prediction of total pion multiplicities in a very direct way.

On the other hand the pions, being the most abundant of all produced hadrons, are the dominant carriers of entropy in the collision fireball. This was already discussed in Section 5.5 where we related the intensive quantity $S/B$ to the specific net charge $D_Q$ which is normalized to the total charged particle multiplicity and thus dominated by pion production.

In Section 6 we discussed extensively whether at freeze-out the collision zone can be described as a hadron resonance gas in thermal and some kind of relative chemical equilibrium. We will now discuss and test an additional aspect of this picture, namely its specific entropy content.

A measurement of $D_Q$ as a function of rapidity was performed for the S+Pb system at 200 A GeV by the EMU-05 collaboration[109]. The value at central rapidity is $D_Q = 0.085 \pm 0.01$. This can be compared with values calculated from the hadron gas model with the thermal parameters extracted from the WA85 data. (Both collision systems feature nearly the same number of target participants.) No detailed comparison has been made for the NA36 parameter values, but since the thermal parameters extracted from this experiment are similar, the conclusions would be the same.

The thermal model results for $D_Q$ and $S/B$ corresponding to the thermal parameters extracted from the chemical analysis are given in the last row Table 1. Only for scenario B, which is plagued by a strong violation of strangeness neutrality, the model produces a value of $D_Q$ which is compatible with the experimental value of $0.085 \pm 0.01$. In all other cases the calculated $D_Q$ is higher then the measured one, implying that the data contain a higher total charged multiplicity and larger specific entropy $S/B$ than provided by the hadron gas model[100]. The discrepancy is of the order of a factor 2.

The situation is similar for the NA35 experiments. The measured $D_Q$ is given in Table 3. The hadron gas values for the parameters from the scenario C(2.0) would be $D_Q = 0.15$ for the $4\pi$-data and $D_Q = 0.09$ at central rapidity. In other words, the measured negative particle multiplicity ($N^- = 98 \pm 5$ from Table 3) is higher than the predicted one ($N^- = 65$ from Table 5) – again the discrepancy is large, of the order of 50%. Similar conclusions have been reached earlier by Davidson et al.[15].

An independent way to arrive at this statement has recently been presented by Gaździcki[110]. He looked at the systematics of pion production in symmetric nuclear reactions at beam energies from 2 A GeV (BEVALAC) to 200 A GeV (CERN) and compared it with pion production in elementary nucleon-nucleon collisions in the same energy range[111]. According to his analysis, one sees for BEVALAC and AGS energies a suppression of pion production in nuclear collisions relative to nucleon-nucleon interactions[110]:

$$\Delta \frac{\langle \pi \rangle}{\langle N_P \rangle} = \frac{\langle \pi \rangle_{AA}}{\langle N_P \rangle_{AA}} - \frac{\langle \pi \rangle_{NN}}{\langle N_P \rangle_{NN}} \approx -0.35 \ . \tag{101}$$

Here $\langle \pi \rangle$ is the mean pion multiplicity for symmetric nuclear collisions (A+A) or nucleon-nucleon reactions (N+N), respectively, and $\langle N_P \rangle$ is the number of participating nucleons. Note that in the studied energy range (101) turns out to be valid independent of the center of mass energy $\sqrt{s_{NN}}$. Proceeding to the higher CERN energies, he obtains from an analysis of the NA35 data on S+S collisions[110] a value of

$$\Delta \frac{\langle \pi \rangle}{\langle N_P \rangle} = +0.9 \pm 0.6 \ . \tag{102}$$

Gaździcki also derives an absolute value for the entropy in the form $S'/\langle N_P \rangle = 6.8 \pm 0.6$, where $S'$ is the entropy in units of the entropy per pion[110]. Taking for the latter the thermal value of about 4 units per pion, we obtain $S/B = 27 \pm 2$. This is again about 50% higher than the hadron gas value resulting from the thermal model parameters for the S+S collisions given in Table 4, in agreement with our above conclusion.

This feature is generic for all heavy ion experiments at CERN energies. The hadron gas model, assuming relative chemical equilibrium with only the absolute strangeness level out of equilibrium, is not consistent with all experimental observations. As long as the model is forced to yield zero net strangeness, it predicts values of $D_Q$ well above the measured values, irrespective of the particular choice for the freeze-out temperature[25]. In other words, the specific entropy $S/B$ in the CERN data is large and cannot be reproduced by the hadron gas model.

At the lower AGS energies $D_Q$ has not been investigated experimentally. But we can consider the absolute pion production directly. Table 7 shows that the thermal model slightly *over*estimates the measured pion/nucleon ratio, but not at a level which would indicate serious disagreement. From the extracted thermal model parameters a value of $S/B = 13 \pm 1$ was calculated in Ref. 24; this is significantly lower than the CERN values. We conclude that at AGS energies the entropy production is low and can be described by the thermal model with a hadron resonance gas equation of state, while at CERN energies entropy production is dramatically higher and cannot be reproduced by the hadron gas model.

## 8 Discussion of the Results

We have shown the results of an interpretation of experimental particle abundances in a thermal hadron gas model, with special emphasis on strange particles. For the lower AGS energies the model yields an excellent description of all hadron spectra and particle abundances from Si+Al and Si+Au collisions. It will be interesting to see whether this can be confirmed for Au+Au collisions where some of the necessary data have already been taken; this largest presently available collision system at the AGS is expected to show even clearer signs of thermal and chemical equilibration, superimposed by collective expansion.

At the higher CERN energies, on the other hand, some features of the data can not be reproduced by the model. What exactly goes wrong is not yet quite clear; the interpretation of how the model fails depends to some extent on how it is applied to the data. The crucial parameter is the freeze-out temperature whose determination is a bottleneck in the analysis. Whereas the baryon and strange fugacities can be very precisely given by a few particle/antiparticle ratios (85), (86), the temperature is much harder to determine. One of the reasons is that one must distinguish between the *thermal* and the *chemical* freeze-out temperature. The *thermal* freeze-out temperature can be extracted from the slope of the $m_\perp$-spectra, although the separation of thermal and collective motion is not quite model independent. Dynamical arguments[97] suggest thermal freeze-out around $T_f \approx 150$ MeV. The *chemical* freeze-out temperature is reflected by the particle ratios; ratios between heavy (baryons) and a light hadrons (mesons) show the largest sensitivity to the freeze-out temperature.

This sensitivity can, however, be exploited only if prior to freeze-out chemical equilibrium was established to begin with. Whether this is true or not is an important dynamical question whose answer would provide crucial information about various time scales in nuclear collisions. We have argued that chemical equilibration happens in stages, and that the equilibration of the overall strangeness level and, in a hadronic environment, of the baryon+antibaryon↔meson processes should occur most slowly. If the latter processes have not equilibrated, the abovementioned meson/baryon ratios are useless for a determination of the freeze-out temperature.

We have therefore based our chemical analysis on particle ratios taken entirely from within the baryonic or from within the mesonic sector. It turns out that the baryonic ratios by themselves do not fix the freeze-out temperature very well; if there is a preference at all, it is that the (not very accurately known) triple/double strange ratios prefer a high freeze-out temperature (of the order of the $m_\perp$-slope of $\approx 230$ MeV).

Another (in principle) important constraint is the condition of overall strangeness neutrality for the decaying collision region. Unfortunately it can also be exploited only if the system is chemically equilibrated or if chemical equilibrium is broken in particular, well-controlled ways which we have labeled by *relative chemical equilibrium*. Assuming relative chemical equilibrium with respect to all processes except strange pair production, the strange and non-strange baryon/antibaryon and meson/meson ratios (except for those involving pions) are compatible with strangeness neutrality in a hadron resonance gas if the freeze-out temperature is around 190 – 200 MeV. Lower temperatures are inconsistent with strangeness neutrality in a hadron gas unless the meson/baryon equilibrium is broken, but even by allowing for such a possibility the disagreement with the measured pion yields cannot be resolved.

Summarizing these comments we conclude that the CERN data do not allow for a consistent and comprehensive interpretation within the hadron gas model, even after allowing for some simple patterns for breaking absolute chemical equilibrium. Still, the chemical parameters extracted from the analysis presented in the previous chapters show a number of strikingly consistent features which are worth being exposed in a systematic way. In spite of all the caveats above, these features provide an intuitive and internally consistent picture of some of the main characteristics of high energy nuclear collisions.

A priori nuclear collisions are characterized by the two controllable parameters $\sqrt{s}$, i.e. the center of mass energy per (participating) nucleon pair, and $B_{\mathrm{part}}$, the number of participating projectile and target baryons, which is related to the size of target and projectile nuclei and the impact parameter. Accordingly we will subdivide the observables into two categories, those which scale dominantly with $\sqrt{s}$ and those which show mostly a scaling with $B_{\mathrm{part}}$. We call these different scaling laws "energy scaling" and "size scaling", respectively.

Figure 16 summarizes graphically the thermal parameters extracted in Sections 6 and 7. The *initial* fireball temperature is related to the initial energy density and thus expected to exhibit mostly an energy scaling; a weaker size scaling arises from the different stopping power of light and heavy nuclei which transforms beam energy into internal excitation energy. But what is the expected scaling of the *freeze-out* temperature $T_{\mathrm{f}}$?

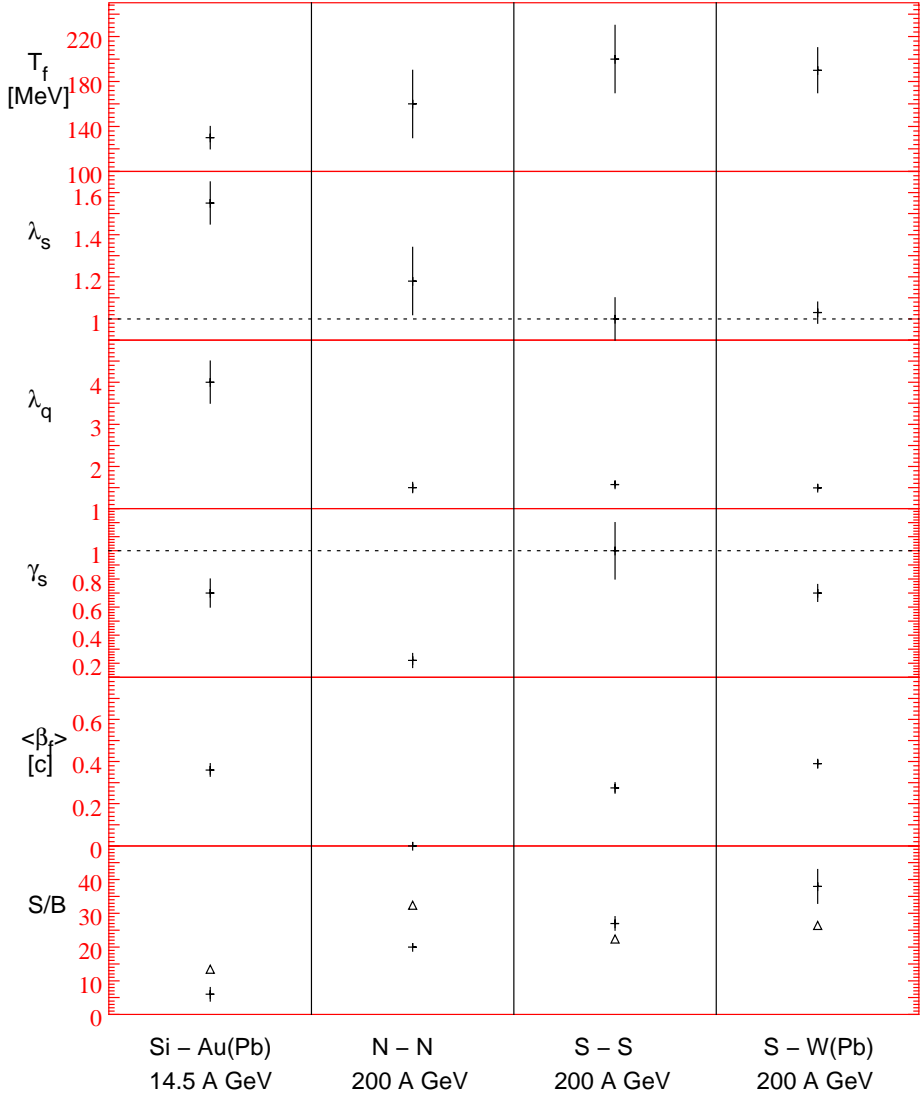

**Figure 16:** Summary of the thermal parameters extracted in Section 6. In the last row of diagrams, the triangles correspond to the value of $S/B$ in a hadron gas with the thermal parameters shown in the first 4 rows, while the crosses correspond to "measured" values of this quantity. The values of $S/B$ in the first three columns are from Ref. 110; note that the value plotted by a cross in the Si-Au column is really from Si-Al collisions[110]. The experimental $S/B$ value in the last column is from the $D_Q$ measurement[109] using Eq. (78).

If we assume that $T_f$ is determined by geometry and occurs when the mean free path $\lambda_F$ is of the order of the transverse radius $R_\perp$ of the fireball,

$$R_\perp(T_f) \approx \lambda_F(T_f) \equiv \frac{1}{n(T_f)\langle\sigma v\rangle(T_f)} \qquad (103)$$

where $n$ is the fireball density and $\langle \sigma v \rangle$ the thermal cross section (23) at freeze-out, then we expect $T_f$ to scale with the inverse of the system size. The reason is that both the particle density and the thermal cross sections rise with temperature (see e.g. Figure 4). Eq. (103) can depend on the beam energy only through the baryon stopping power via the relationship between the fireball density $n$ and $T_f$; its energy scaling is thus expected to be very weak.

This expectation is in stark contrast to the behavior seen in Figure 16 where $T_f$ shows a clear scaling with both increasing beam energy and increasing size of the system. Note that we plotted in Figure 16 the chemical freeze-out temperature as determined by the particle ratios and the strangeness neutrality condition (and not the measured slope parameter of the $m_\perp$-spectra). As seen, $T_f$ follows the scaling behavior expected for the initial temperature $T_0$ as discussed above. Somehow the final particle ratios seem to have a memory of the initial conditions of the fireball. Naively this would seem to contradict the thermal interpretation because thermalization by assumption implies memory loss via an increase in entropy.

However, in our discussion of thermal vs. chemical freeze-out we noted the rather strong evidence in the transverse momentum spectra for the presence of collective transverse flow at freeze-out. In the presence of flow the geometric freeze-out criterium (103) should be replaced by a dynamical one because collective expansion leads to dynamical decoupling before the geometric freeze-out criterium is satisfied[95,112]. Hydrodynamical simulations suggest that in nuclear collisions thermal freeze-out is entirely dominated by the collective dynamics[96,97]. It is more than likely that the same is true for chemical freeze-out because the only difference between the two freeze-out criteria results from the substitution of the total by the corresponding inelastic partial cross section. Since higher initial energy densities generate larger hydrodynamic flow and larger hydrodynamic flow leads to earlier dynamical freeze-out at a higher freeze-out temperatures[96,97], the qualitative behavior of $T_f$ seen in Figure 16 can be understood in this picture. In this sense the development of collective flow helps the system to "remember" the initial conditions of the fireball in spite of thermalization.

The two fugacities $\lambda_s$ and $\lambda_q$ fall off going from left to right in Figure 16. At the AGS the large value of $\lambda_q$ reflects the high baryon density caused by the nearly complete baryon stopping at these energies. A part, but not all of the strong decrease of $\lambda_q$ at CERN energies can be understood by the larger freeze-out temperatures there; on top of this trivial effect it reflects, however, the onset of partial transparency at CERN energies. Compared to the strong jump between the lower (first column) and the higher beam energies (second to fourth column), $\lambda_q$ varies very little with the system size. It should be noted, however, that the values in the second and third column are from $4\pi$ particle ratios, whereas the last column is from the WA85 midrapidity ratios. The analysis of Ref. 85 showed that $\lambda_q$ depends on rapidity and, at CERN energies, is lower at midrapidity than in the nuclear fragmentation regions. Comparing the $\lambda_q$ values at midrapidity from S+S (NA35), S+W (WA85) and S+Pb collisions (NA36) as they were given in Section 6, one can see a rising tendency with the size of the nuclear target. This reflects again the increasing baryon stopping power of larger nuclei which counteracts the onset of nuclear transparency at CERN energies.

Hence $\lambda_q$ is a measure for the efficiency of baryon number stopping of different size nuclei at different energies. At the relativistic heavy-ion collider RHIC ($\sqrt{s}$=200 GeV) one expects $\lambda_q \approx 1$ at central rapidity, independent of the size of the colliding nuclei.

In first approximation $\lambda_s$ must follow the behavior of $\lambda_q$ because of the strangeness neutrality relation. However, it is not all trivial that for all nuclear collisions at SPS energies the value of $\lambda_s$ is approximately 1, corresponding to a vanishing strange quark chemical potential. (In N+N collisions this seems to be different, but due to the large error bar a safe statement cannot be made.) Indeed, this appears to be the most stable result of the chemical analysis, and it is independent of the actual value of the freeze-out temperature and how the strangeness neutrality is actually realized between the baryons and mesons. We already commented before on the intriguing agreement of this value with the one prevalent in a strangeness neutral quark-gluon plasma. To what extent this may be more than a diabolical coincidence will be discussed in Section 9.2.

The strangeness suppression factor $\gamma_s$ shown in the fourth row of Figure 16 demonstrates clearly that the physics of nuclear collisions is different from N+N collisions. For nuclear collisions $\gamma_s$ is close to 1, while in the N+N case is much smaller, $\gamma_s \approx 0.2$. This size scaling of $\gamma_s$ was worked out in more detail in Ref. 22. The increase of $\gamma_s$ with $B_{\mathrm{part}}$ is a clear sign that strangeness equilibration is caused by secondary interactions. The large degree of strangeness saturation in nuclear collisions which is reflected by $\gamma_s \lesssim 1$ is a strong indication that unusually efficient and rapid mechanisms for strangeness production are at work.

In the second but last row of Figure 16 we have plotted the average transverse collective flow velocity at the point of *thermal* freeze-out, i.e. at the point of decoupling of the momentum spectra. For this we need to know the experimental slope of the $m_\perp$ spectra and the thermal freeze-out temperature. At the AGS the transverse flow and the thermal freeze-out temperature can be determined separately from the momentum spectra, because fortunately also momentum spectra of very heavy particles like the deuteron[26] are available; still, the determination is not yet very accurate, and lower thermal freeze-out temperatures down to $T \simeq 100$ MeV, compensated by correspondingly larger transverse flow velocities, still give acceptable fits to the $m_\perp$ spectra. The identity of the thermal and chemical freeze-out temperatures at the AGS is thus not rigorously proven, although it seems to work quite nicely. At CERN energies such a clear separation of thermal and collective motion from the spectra is not (yet) possible[f]. The transverse flow must be determined with the help of a theoretical model for the dynamics of the fireball and involves a freeze-out criterium. Such calculations are only available for the S+S case at SPS energies[96]. These calculations suggest thermal freeze-out at $T = 150 - 160$ MeV with $\langle \beta_f \rangle = 0.25 - 0.3$; this value has been indicated in the S+S column. For the N+N case collective hydrodynamical expansion is not expected, so we entered the value $\langle \beta_f \rangle = 0$ in the second column. The value in the last column was obtained by assuming (for lack of better knowledge) that thermal freeze-out occurs also in the heavier systems at $T = 150$ MeV (see scenario B in Section 6.1), by averaging over the $m_\perp$-slopes quoted by WA85[185,186] and NA36[173].

---

[f] A first deuteron spectrum was presented by the NA44 collaboration in Ref. 113, but these data have not yet been subjected to a full thermal+flow analysis.

Generically, the experimental $m_\perp$ spectra flatten for larger collision systems, indicating earlier thermal decoupling at a higher temperature and/or larger collective transverse flow. Hydrodynamical calculations[96] suggest a combination of both. Comparing Si+Au collisions at 14.5 A GeV and S+W collisions at 200 A GeV, both similar in size, we see a decrease in $\langle \beta_f \rangle$ combined with an increase in the thermal freeze-out temperature (reflected in the flatter $m_\perp$-slope). This is connected with the increasing nuclear transparency at higher energies; the resulting fast longitudinal expansion leads to earlier freeze-out when transverse motion has not yet developed as strongly.

Finally we consider the entropy per baryon $S/B$ in the last row of Figure 16. There are two marks in each plot. The crosses correspond to the "measured" entropy obtained from the pion multiplicity[110] or $D_Q$. The triangles correspond to the values in a thermal model with the parameters plotted in the first 4 rows. The experimental values for $S/B$ increase continuously from left to right in Figure 16. The strong rise of entropy production from AGS to SPS energies is expected. The observed size scaling can be understood in terms of increasing stopping power. We already discussed in Section 7, but want to point out again that the thermal model predictions lie *above* the measured values in Si+Au (AGS) collisions and N+N collisions, but *below* the measured values in the sulphur induced reactions at CERN.

In summary, the thermal quantities can roughly be divided into two classes: $T_f$, $\lambda_q$, $\lambda_s$, and $S/B$ scale dominantly with energy, while $\gamma_s$ and $\langle \beta_f \rangle$ scale mostly with size. We have seen that the qualitative behavior of these parameters is fully consistent with the expectations based on our present understanding of the heavy-ion collision dynamics. However, quantitatively the thermal hadron gas picture is plagued by internal consistencies. In the following Section we will pick up on these problems and use them to extract some further constraints on the early collision dynamics and the freeze-out process.

## 9 Consequences for Freeze-Out Models

In the preceding Sections 6–8 we presented an extensive study of (strange) particle abundances in thermal models. Since a thermalized system surrounded only by vacuum necessarily begins after a while to expand hydrodynamically, and any thermal model analysis should for consistency be done on the basis of a hydrodynamical evolution background. This introduces the flow velocity as an additional essential parameter into the analysis which must be determined consistently with the rest of the thermal parameters and with the dynamical evolution and freeze-out kinetics. In this light the strangeness signal for QGP formation acquires additional facets beyond the mere question of enhanced strange hadron production in ultrarelativistic heavy-ion collisions as originally discussed[33]. As we showed, also the temperature and chemical potentials at the chemical freeze-out point can be extracted from strange particle abundances once the effects from the decay of unstable resonances after freeze-out are properly taken into account. Combining the experimental information from the various strange particle abundances and from their momentum spectra we obtain valuable insight into the state of the system at freeze-out; using dynamical models we can then draw further conclusions also about earlier stages of the collision.

We saw that, contrary to the situation at the AGS, at CERN energies the assumption of a thermalized hadronic system in relative chemical equilibrium cannot describe *all* observables. There is a crucial inconsistency with the observed pion production which we interpreted in terms of a lack of entropy in the hadron gas model compared to the data. Furthermore, the chemical analysis points towards chemical freeze-out temperatures near 200 MeV which are problematic for any weakly interacting hadron gas picture and at variance with lattice QCD results on thermally equilibrated systems of strongly interacting particles. Up to now no fully consistent solution to these problems has been provided. However, there are a number of interesting theoretical suggestions in this connection which we would like to shortly discuss in this Section.

*9.1 Sequential Freeze-out*

An implicit assumption in the thermal and chemical analyses described in this review is that all particles freeze out simultaneously when the critical freeze-out temperature $T = T_f$ is reached. This is a technical requirement related to the problem that it hard to include the modifications of partial freeze-out of some selected particle species on the subsequent hydrodynamical evolution of the remaining system. But it is nevertheless possible that different particle species decouple from the system at different times or temperatures, depending on their average cross sections with the medium[114,115] (which at CERN energies consists mainly of pions). Such a sequential freeze-out can occur both for the chemical composition of the fireball and for the momentum distributions.

In Ref. 18 the idea of sequential freeze-out was applied to the measured strange and non-strange particle abundances. Freeze-out criteria of the type (103) lead to different freeze-out temperatures $T_f$ for particles with different cross sections $\langle \sigma v \rangle(T)$. Cleymans et al.[18] suggested that generically strange particles interact more weakly than non-strange hadrons and thus freeze out earlier. Assuming a typical factor 2 between these two types of cross sections and adiabatic spherical expansion, they argue that freeze-out of strange particles at $T = 185$ MeV is consistent with freeze-out of non-strange particles at $T = 130$ MeV. The conceptual difference between chemical and thermal freeze-out temperatures was not considered in this work, and thus no theoretical justification for the observed similar $m_\perp$-slopes of *all* hadrons could be given. This model is able to resolve the excess entropy problem in terms of a larger specific entropy $S/B$ of the hadron gas at the lower freeze-out temperature of the non-strange hadrons; at the higher freeze-out temperature of the strange hadrons it agrees with the values around $S/B \sim 20$ quoted above. It is not clear how this additional final state entropy arises in the model; implicitly it is generated between the strange particle and non-strange particle decoupling points. The question of viability of the hadron gas picture at the high strangeness freeze-out temperature is not addressed.

It is natural to ask whether there is any direct evidence for sequential freeze-out in nuclear collisions. It was suggested many years ago[114,115] that kaons should freeze out from a baryon rich environment earlier than most other hadrons, based on the very small $K^+N$ cross section. As far as thermal freeze-out is concerned, this idea can be tested by looking at the two-particle momentum distributions ("Hanbury-Brown/Twiss (HBT) interferometry"): naively, if the kaons decouple earlier than, say, the pions and the fireball is expanding, the $K^+K^+$ correlation function should reflect a smaller HBT radius than the $\pi^+\pi^+$ correlation function. Measurements of kaon and pion correlations at the AGS[152,153,116] and at the SPS[176,177] at first sight indeed seem to show such an effect. However, it is not clear whether this is indeed evidence for a smaller kaon freeze-out radius: Pion HBT radii are generically larger than kaon radii due to a larger contribution from delayed resonance decays which create a halo to the source. Furthermore, transverse collective expansion generates a characteristic $1/\sqrt{m_\perp}$ dependence of the HBT radii[117,118] which causes kaon radii to be smaller than pion radii if measured at the same $p_\perp$. The situation at the AGS has not yet been fully analysed; at the SPS no statistically significant differences have been observed between $K^+K^+$ and $K^-K^-$ correlations[176] (for $K^-$ the earlier freeze-out argument does not apply because it has a large cross section with nucleons), and the smaller kaon radius compared with the pion one is claimed to be consistent[177] with the expected effects from resonance decays and transverse flow.

Thus there is presently no clear direct experimental evidence for sequential thermal freeze-out. To test this idea further for chemical freeze-out requires extensive kinetic and hydrodynamic simulations.

*9.2 Sudden Hadronization Scenarios*

Another possibility to solve the inconsistencies of the thermal analysis is to postulate that that the particles do *not* decouple from an equilibrated system. However, even in that case one still has to explain the approximately thermal nature of the momentum spectra and the remaining order in the particle ratios which is described by the vanishing strange quark chemical potential, $\mu_s = 0$. A completely non-equilibrium approach to hadronization was recently initiated by the Budapest group[119,120]. Unfortunately, this model cannot predict the final hadronic momentum spectra and is not yet in a form where it can be implemented into dynamical calculations.

In Ref. 25 it was argued that the apparently large specific entropy and vanishing $\mu_s$ could be due to the existence of a QGP in the early stages of the collision which hadronizes suddenly into a hadronic system which immediately decouples. The existence of a QGP would naturally explain the excess specific entropy[100] and the vanishing $\mu_s$. Sudden hadronization followed by immediate decoupling was required in order to preserve the values of the chemical potentials across the phase transition and avoid their re-equilibration towards equilibrium hadron gas values before freeze-out. However, no consistent dynamical model which would produce the desired results was found in Ref. 25.

The idea of sudden QGP hadronization (also called "QGP break-up model") was recently picked up by several groups[121–123]. They revived an old idea by van Hove[124] who suggested that, if the hadronization phase transition is of first order, hadronization could occur dynamically through a deflagration or detonation[g] shock front. Such a violent process results in very rapid hadronization in which freely moving particles are ejected into the surrounding vacuum with large average velocities.

The hadronization of a QGP through a shock front must satisfy energy-momentum and baryon number conservation across the front. This defines the Taub adiabat[122]

$$\left(\frac{n_{\rm QGP}}{n_{\rm H}}\right)^2 = \frac{(\varepsilon_{\rm QGP} + P_{\rm QGP})(\varepsilon_{\rm QGP} + P_{\rm H})}{(\varepsilon_{\rm H} + P_{\rm H})(\varepsilon_{\rm H} + P_{\rm QGP})} \, , \qquad (104)$$

where $n$ is the baryon number density, $\varepsilon$ the energy density and $P$ the pressure in the initial QGP and final hadronic state, respectively. Even if for the final state an equilibrium hadron gas equation of state is assumed, Eq. (104) is not sufficient to determine the temperature and chemical potential change through the hadronization front. As a further constraint one often assumes entropy conservation which leads to the additional equation (Poisson adiabat)

$$\frac{S_{\rm QGP}}{B_{\rm QGP}} = \frac{S_{\rm H}}{B_{\rm H}} \, . \qquad (105)$$

For shock fronts Eq. (105) is generally not a good assumption, but it can be used to give upper or lower limits on the changes of thermodynamic parameters during the hadronization.

The idea behind the approaches in Refs. 121–123 is to see whether a detonation or deflagration scenario can provide a dynamical setting in which a superheated *equilibrated* hadron gas can arise in the final state which has $T \sim 200$ MeV, $\mu_{\rm s} = 0$, and is strangeness neutral, but which successfully fights its inherent tendency to return into the QGP state and rather decouples immediately due to its rapid dynamical expansion. However, these papers do not address the problem that such an *equilibrated* hadron gas would again yield too little specific entropy. In this sense they are complementary to the approaches of the previous subsection. No attempts have been made yet to include suitable deviations from thermodynamic equilibrium in the final state.

Eqs. (104) and (105) define a detonation or deflagration front in space-time which can be either timelike[123] or spacelike[121,122]. The time evolution is roughly as follows. The system starts from a QGP with a high initial temperature (assumed to be around 200 MeV or higher). Then it supercools to about $0.6 - 0.8 \, T_c$. Rapid longitudinal expansion and cooling of the QGP, combined with a very large nucleation time[125] of around 100 fm/c for hadron gas bubbles, prevent the system from hadronization at $T_c$. The strong supercooling is basically caused by the requirement (105): only if the QGP is strongly supercooled and the hadron gas is strongly superheated the larger specific entropy in the QGP phase can be absorbed by the outgoing hadrons.

---

[g]A shock front is called a detonation if in its rest frame the velocity of the incoming (QGP) matter is lower than that of the outgoing one (hadron gas), otherwise it is called a deflagration.

The supercooled QGP is mechanically unstable. Within the bag model, the QGP pressure turns negative[123] already at $T < 0.98\,T_c$. At $0.6-0.8\,T_c$ sudden hadronization via a deflagration/detonation shock sets in. The outcome is a superheated hadron gas with a temperature around 200 MeV. After the deflagration/detonation the particles can considered as frozen out[122,123].

Models of this type explain naturally some of the puzzling experimental results:

- $\mu_B$ and $\mu_s$ are nearly conserved during the detonating hadronization[121]. Thus the naturally vanishing strange quark chemical potential of a strangeness neutral QGP can be dynamically transferred to the final hadronic state.

- Lattice QCD calculations[42] favor a $T_c$ around 150 MeV at $\mu_B = 0$. However, strangeness neutrality in the thermal hadron gas model with $\mu_s = 0$ requires temperatures around $T_f = 190$ MeV. The $m_\perp$-slopes of all particles measured at CERN energies result in effective temperatures above 200 MeV. The superheating to about $1.3\,T_c$ in the calculation of Ref. 122 combined with the collective ejection of the hadrons from the shock front could explain all these features.

- A QGP is a natural source of a large specific entropy[100,110]. Instead of (105), a realistic detonation or deflagration scenario would lead to further entropy increase[124]. However, as already mentioned, this additional entropy can only be used profitably in comparison with the data if the final state is not modeled by an equilibrium hadron gas. In this respect all suggested models fail so far.

- Assuming boost-invariant longitudinal expansion, HBT interferometry results[160] suggest for sulphur induced collisions at 200 A GeV a total collision lifetime of about 4.5 – 6.5 fm/c. This agrees with the lifetime of the QGP phase until detonation from the supercooled state in Ref. 121. Assuming near-equilibrium hadronization by bubble nucleation would take much longer[125].

- The detonation into freely moving particles would naturally explain the high antiparticle yields seen in the experiments (see references in Table 9). During slow hadronization the produced antiparticles would be absorbed in the longlived hadron gas subphase during and after the phase transition[11].

These points seem make the detonation/deflagration scenario an attractive candidate for the dynamics of heavy-ion collisions at CERN energies. However, as far as we know, it only works if the hadronization is a strong first order phase transition. This is not supported by modern lattice QCD calculations, even when including finite net baryon number according to the best present knowledge[42]. Clearly much work is needed to supplement the present qualitative considerations by a fully consistent dynamical hadronization calculation.

## 10 Conclusions and Outlook

We have reviewed new developments concerning the strangeness signal in heavy-ion experiments. In spite of tremendous experimental progress a clear proof of QGP formation still does not exist. However, circumstantial evidence in favor of a deconfined quark-gluon state in the initial stages of the collision, in particular at CERN energies, keeps accumulating. A high degree of strangeness equilibration ($\gamma_s \sim 0.7 - 1$) indicates the opening of new, unusually rapid strangeness production channels. The mechanism could be QGP formation, but strong medium effects in a dense hadronic environment, leading to drastically reduced strange hadron masses and modified inelastic cross sections, have also been shown to reproduce some of the observations. More detailed and comprehensive studies should be performed for further differentiation.

The rather stable result of a vanishing strange quark chemical potential, $\mu_s = 0$, in all of the CERN experiments further points in the direction of a state in which the phase space for strange quarks and antiquarks is symmetric. In any type of confined state this symmetry is spoiled by a finite net baryon density because the excess of light quarks over light antiquarks rubs off on the strange quarks because they are clustered together with them into hadrons.

The observation of unexpectedly high pion production in heavy-ion collisions at CERN[110] points to a new efficient mechanism for entropy production as one passes from the lower BEVALAC and AGS energies to the SPS.

We showed that all of the AGS data and most of the SPS data can be efficiently described by a thermal hadron gas model. The model contains only a few parameters with obvious intuitive interpretation. As shown in Section 8, they provide an efficient parametrization of the multitude of data, resulting in a comprehensive understanding of the basic features of the collision dynamics in ultrarelativistic heavy-ion collisions. The model can be tested with high accuracy using the large body of available data on particle yields and momentum spectra. This test is successfully passed by the data from Si+Al/Au collisions at the AGS, providing convincing evidence for the formation of a thermally and chemically equilibrated state with a large degree of strangeness saturation and showing clear signs of collective behavior. But even where the model fails in its application to hadron production data at CERN, it still provides a useful parametrization of the failure, thus leading to useful insights and stimulating further model building, as discussed in Section 9.

The future will provide us with many new sets of data, especially from the larger collision systems Au+Au at the AGS and Pb+Pb at CERN. These will allow for further tests and confirmation of the picture developed so far. We expect even stronger evidence for equilibration and collective behavior at the AGS and perhaps a clarification of the remaining puzzles about the nature of the fireball at CERN.

On the theoretical side, a major effort in constructing realistic dynamical models for the hadronization of a QGP is required, in order to reach a comprehensive qualitative and quantitative understanding of the particle production process and to supplement the microscopic kinetic models presently under development[28] by a successful semiphenomenological macroscopic approach. Also the kinetics of the hadronic freeze-out process and its proper implementation into macroscopic models should be studied in more detail.

Finally, more efforts are needed to better understand the perturbative and non-perturbative contributions to strangeness production in a deconfined (equilibrium or non-equilibrium) quark-gluon state.

**Acknowledgments**

We would like to thank Marek Gaździcki, Jean Letessier, Ahmed Tounsi and especially Johann Rafelski for several years of collaboration on this subject. Many of new results reported in this overview were obtained by them or in collaboration with them. This work was supported by DFG, BMBF, and GSI.

## A  Appendix: Literature Guide to Strange Particle Measurements

Although this article is meant as a theoretical review on strangeness, the phenomenology of strangeness requires the input from and the comparison with experiment. For this reason we present here a list of experimental publications on strange particle production in <u>nuclear</u> collisions, organized by particle species and experiment number. The two tables summarize the experiments at the Brookhaven AGS and the CERN SPS, respectively. We concentrate on strange hadron spectra and multiplicities and left out articles on e.g. kaon interferometry.

The list mainly contains articles in refereed journals. Conference proceedings are mostly neglected with the exception of Quark Matter '91[126] and '93[127]. We recommend for updated strangeness data also the proceedings of Quark Matter '95[13] and Strangeness '95[12], of which a few contributions have already been included according the notes taken by one of the authors (UH) and which will be published soon.

The tables should contain all relevant publications from the last few years known to the authors; it is, however, quite likely that we missed one or the other important contribution for which we would like to apologize.

## References

1. R. C. Hwa (ed.), *Quark-Gluon Plasma*, World Scientific, Singapore, 1990.
2. P. V. Ruuskanen, *Nucl. Phys.* **A554** (1992) 169;
   P. V. Ruuskanen, in Ref. 1, p. 519.
3. J. Rafelski, in *Workshop on Future Relativistic Heavy Ion Experiments*, R. Stock and R. Bock (Eds.), GSI-Report 81–6 (1981), p. 282.
4. S. A. Chin and A. K. Kerman, *Phys. Rev. Lett.* **43** (1979) 1292.
5. E. Witten, *Phys. Rev.* **D30** (1984) 272.
6. C. Greiner, P. Koch, and H. Stöcker, *Phys. Rev. Lett.* **58** (1987) 1825;
   C. Greiner, D. H. Rischke, H. Stöcker, and P. Koch, *Phys. Rev.* **D38** (1988) 2797.

Table 8: References for strange particle measurements at the Brookhaven AGS in heavy ion collisions at ∼ 15 A GeV.

| particle | E802/E866 | E810 | E814 | E859 |
|---|---|---|---|---|
| $K_s^0$ | | 141 142 143 144 145 | | |
| $K^\pm$ | 128 129 130 131 132 133 134 135 136 137 138 139 | | 147 | 148 149 150 151 153 116 |
| $\phi$ | | | | 148 149 150 153 |
| $\Lambda$ | | 141 142 143 144 145 146 | | 149 |
| $\bar{\Lambda}$ | | | | 149 |
| $\Xi^-$ | | 146 | | |

Table 9: References of measured strange particles at the CERN SPS in heavy ion collisions at 200 A GeV.

| particle | NA34 | NA35 | NA36 | NA44 | WA85 | WA94 |
|---|---|---|---|---|---|---|
| $K_s^0$ | | 156 157 158 159 160 163 164 165 | 167 168 170 171 172 | | 184 186 188 | |
| $K^\pm$ | 154 155 | 160 161 162 163 164 165 | | 174 175 176 177 | 190 | |
| $\Lambda, \bar{\Lambda}$ | | 156 157 158 159 160 163 164 165 166 | 167 168 169 170 171 172 173 | | 178 179 181 183 184 186 188 | 191 |
| $\Xi^-, \bar{\Xi}^+$ | | | 172 173 | | 180 181 182 184 186 188 189 | 191 |
| $\Omega^-, \bar{\Omega}^+$ | | | | | 185 187 188 189 | |

7. R. L. Jaffe, *Phys. Rev. Lett.* **38** (1977) 195;
   C. B. Dover, *Nucl. Phys.* **A450** (1986) 95c;
   C. B. Dover, P. Koch, and M. May, *Phys. Rev.* **C40** (1989) 115.
8. J. Schaffner, C. Greiner, and H. Stöcker, *Phys. Rev.* **C46** (1992) 322.

9. P. Koch, *Prog. Part. Nucl. Phys.* **26** (1991) 253.
10. C. Greiner, A. Diener, J. Schaffner, and H. Stöcker, *Nucl. Phys.* **A566** (1994) 157c.
11. J. Rafelski, *Phys. Rep.* **88** (1982) 331.
12. Proceedings of the Workshop *Strangeness '95* (Tucson, Jan. 3 - 6, 1995), J. Rafelski et al. (Eds.), American Institute of Physics, New York, 1995, in press.
13. Proceedings of *Quark Matter '95*, A. Poskanzer et al. (Eds.), *Nucl. Phys.* **A** (1995), in press.
14. N. J. Davidson, H. G. Miller, R. M. Quick, and J. Cleymans, *Phys. Lett.* **B255** (1991) 105;
D. W. von Oertzen, N. J. Davidson, R. A. Ritchie, and H. G. Miller, *Phys. Lett.* **B274** (1992) 128.
15. N. J. Davidson, H. G. Miller, D. W. von Oertzen, and K. Redlich, *Z. Phys.* **C56** (1992) 319.
16. J. Letessier, A. Tounsi, and J. Rafelski, *Phys. Lett.* **B292** (1992) 417.
17. J. Cleymans and H. Satz, *Z. Phys.* **C57** (1993) 135.
18. J. Cleymans, K. Redlich, H. Satz, and E. Suhonen, *Z. Phys.* **C58** (1993) 347.
19. K. Redlich, J. Cleymans, H. Satz, and E. Suhonen, *Nucl. Phys.* **A566** (1994) 391c.
20. J. Sollfrank, M. Gaździcki, U. Heinz, and J. Rafelski, *Z. Phys.* **C61** (1994) 659.
21. J. Letessier, J. Rafelski, and A. Tounsi, *Phys. Lett.* **B321** (1994) 394.
22. J. Letessier, J. Rafelski, and A. Tounsi, *Phys. Lett.* **B323** (1994) 393.
23. J. Rafelski and M. Danos, *Phys. Rev.* **C50** (1994) 1684.
24. J. Letessier, J. Rafelski, and A. Tounsi, *Phys. Lett.* **B328** (1994) 499.
25. J. Letessier, A. Tounsi, U. Heinz, J. Sollfrank, and J. Rafelski, *Phys. Rev.* **D51** (1995) 3408.
26. P. Braun-Munziger, J. Stachel, J. P. Wessels, and N. Xu, *Phys. Lett.* **B344** (1995) 43.
27. P. Koch, in Ref. 1, p. 555.
28. K. Geiger, *Nucl. Phys.* **A566** (1994) 257c, and references therein; K. Geiger, CERN-TH.7313/94, *Phys. Rep.*, in press.
29. T. S. Biró et al., *Phys. Rev.* **C48** (1993) 1275.
30. P. Malhotra and R. Orava, *Z. Phys.* **C17** (1983) 85.
31. E. V. Shuryak, *Phys. Rev. Lett.* **68** (1992) 3270.
32. J. Rafelski and B. Müller, *Phys. Rev. Lett.* **48** (1982) 1066.
33. P. Koch, B. Müller and J. Rafelski, *Phys. Rep.* **142** (1986) 167.

34. E. Braaten and R. D. Pisarski, *Nucl. Phys.* **B337** (1990) 569.
35. T. Altherr and D. Seibert, *Phys. Lett.* **B313** (1993) 149;
    T. Altherr and D. Seibert, *Phys. Rev.* **C49** (1994) 1684.
36. N. Bilić, J. Cleymans, I. Dadić and D. Hislop, University of Cape Town preprint UCT-TP 213/94, submitted to *Phys. Rev.* **D**.
37. R. Bensel, Ph.D thesis, Universität Regensburg, in preparation.
38. T. S. Biró, P. Lévai and B. Müller, *Phys. Rev.* **D42** (1990) 3078.
39. A. Ukawa, *Nucl. Phys.* **A498** (1989) 227c.
40. T. Matsui, B. Svetitsky, and L. McLerran, *Phys. Rev.* **D34** (1986) 783;
    T. Matsui, B. Svetitsky, and L. McLerran, *Phys. Rev.* **D34** (1986) 2047.
41. R. D. Pisarski, *Phys. Rev.* **D47** (1993) 5589.
42. F. Karsch, in: *Particle Production in Highly Excited Matter*, H. H. Gutbrod and J. Rafelski (Eds.), Plenum, New York, 1993, p. 207;
    Proceedings of *Lattice '94*, F. Karsch (Eds.), *Nucl. Phys.* **B** (Proc. Suppl.), in press;
    F. Karsch, in Ref. 13.
43. R. Hagedorn, in *Hot Hadronic Matter: Theory and Experiment*, J. Letessier et al. (Eds.), Plenum, New York, 1995, and references therein.
44. R. D. Pisarski, *Phys. Lett.* **B110** (1982) 155.
45. G. E. Brown and M. Rho, *Phys. Rev. Lett.* **66** (1991) 2720.
46. R. J. Furnstahl, T. Hatsuda, and S. H. Lee, *Phys. Rev.* **D42** (1990) 1744.
47. V. Bernard, U. G. Meißner, and I. Zahed, *Phys. Rev. Lett.* **59** (1987) 966;
    V. Bernard and U. G. Meißner, *Phys. Lett.* **B227** (1989) 465.
48. T. Kunihiro, *Phys. Lett.* **B219** (1989) 363;
    T. Kunihiro, *Nucl. Phys.* **B351** (1991) 593.
49. M. Lutz, S. Klimt, and W. Weise, *Nucl. Phys.* **A542** (1992) 521.
50. V. Bernard, U. G. Meißner, and I. Zahed, *Phys. Rev.* **D38** (1988) 1551.
51. T. Hatsuda and Su. H. Lee, *Phys. Rev.* **C46** (1992) R34.
52. R.D. Pisarski, in Ref. 13.
53. G. E. Brown, C. M. Ko, and K. Kubodera, *Z. Phys.* **A341** (1992) 301.
54. C. M. Ko, Z. G. Wu, L. H. Xia, and G. E. Brown, *Phys. Rev. Lett.* **66** (1991) 2577;
    C. M. Ko, Z. G. Wu, L. H. Xia, and G. E. Brown, *Phys. Rev.* **C43** (1991) 1881.
55. C. M. Ko and B. H. Sa, *Phys. Lett.* **B258** (1991) 6.
56. C. M. Ko, M. Askawa, and P. Lévai, *Phys. Rev.* **C46** (1992) 1072.

57. A. M. Rossi et al., *Nucl. Phys.* **B84** (1975) 269;
    U. Becker et al., *Phys. Rev. Lett.* **37** (1976) 1731.
58. C. M. Ko and L. H. Xia, *Phys. Rev.* **C38** (1988) 179;
    C. M. Ko and L. H. Xia, *Phys. Lett.* **B222** (1989) 343.
59. W. Q. Chao, C. S. Gao, and Y. L. Zhu, *Nucl. Phys.* **A514** (1990) 734.
60. Y. Pang, T. J. Schlagel, and S. H. Kahana, *Nucl. Phys.* **A544** (1992) 435c.
61. R. Mattiello, H. Sorge, H. Stöcker, and W. Greiner, *Phys. Rev. Lett.* **63** (1989) 1459.
62. G. E. Brown and M. Rho, *Phys. Lett.* **B222** (1989) 324.
63. M. Lutz, A. Steiner, and W. Weise, *Phys. Lett.* **B278** (1992) 29.
64. S. H. Lee, G. E. Brown, D. P. Min, and M. Rho, *Nucl. Phys.* **A** (1995), in press.
65. M. Lutz, A. Steiner, and W. Weise, *Nucl. Phys.* **A574** (1994) 755.
66. T. Waas, diploma thesis, University of Regensburg, 1994, to be published.
67. T. Biró, W. Barz, B. Lukács, and J. Zimányi, *Phys. Rev.* **C27** (1983) 2695.
68. J. P. Guilland et al., NA38 Collaboration, *Nucl. Phys.* **A525** (1991) 449c.
69. C. M. Ko, P. Lévai, X. J. Qiu, and C. T. Li, *Phys. Rev.* **C45** (1992) 1400.
70. A. Shor, *Phys. Rev. Lett.* **54** (1985) 1122.
71. P. Koch, U. Heinz and J. Pišút, *Z. Phys.* **C47** (1990) 477.
72. P. Koch, U. Heinz and J. Pišút, *Phys. Lett.* **B243** (1990) 149.
73. U. Heinz and K. S. Lee, *Phys. Lett.* **B259** (1991) 162.
74. P. Bordalo, in Ref. 12.
75. J.P. Blaizot and R. Mendez Galain, *Phys. Lett.* **B271** (1991) 32.
76. K. Haglin, MSUCL-937, *Nucl. Phys.* **A**, in press.
77. J. Bolz, U. Ornik, and R. M. Weiner, *Phys. Rev.* **C46** (1992) 2047.
78. H. Sorge, *Nucl. Phys.* **A566** (1994) 633c.
79. K. Werner, *Phys. Rev. Lett.* **73** (1994) 1594; and report in Ref. 12; K. Werner and J. Aichelin, Nantes preprint SUBATECH-95-02.
80. N. K. Glendenning and S. A. Moszkowski, *Phys. Rev. Lett.* **67** (1991) 2414.
81. E. V. Shuryak, *Phys. Rev.* **D42** (1990) 1764.
82. V. Koch, SUNY-preprint SUNY-NTG 94-26;
    V. Koch, in Ref. 13.
83. J. Sollfrank, P. Koch, and U. Heinz, *Phys. Lett.* **B253** (1990) 256;
    J. Sollfrank, P. Koch, and U. Heinz, *Z. Phys.* **C52** (1991) 593.
84. J. Rafelski, *Phys. Lett.* **B262** (1991) 333.

85. C. Slotta, diploma thesis, Universität Regensburg, 1995;
    C. Slotta, J. Sollfrank, and U. Heinz, in Ref. 12.
86. W. Blümel, P. Koch, and U. Heinz, *Z. Phys.* **C63** (1994) 637;
    W. Blümel and U. Heinz, Regensburg preprint TPR–94–30, *Z. Phys.* **C** (1995), in press;
    W. Blümel, Ph.D. thesis, University of Regensburg, 1995.
87. R. Hagedorn and J. Rafelski, in *Statistical Mechanics of Quarks and Hadrons*, H. Satz (Ed.), North Holland, Amsterdam, 1981, p. 237.
88. Particle Data Group, *Review of Particle Properties*, *Phys. Rev.* **D50** (1994) 1173.
89. U. Heinz, K. S. Lee, and M. J. Rhoades-Brown, *Mod. Phys. Lett.* **A2** (1987) 153.
90. K. S. Lee, M. J. Rhoades-Brown, and U. Heinz, *Phys. Rev.* **C37** (1988) 1452.
91. C. Greiner and H. Stöcker, *Phys. Rev.* **D44** (1992) 3517.
92. K. S. Lee and U. Heinz, *Phys. Rev.* **D47** (1993) 2068.
93. J. Rafelski, *Phys. Lett.* **B190** (1987) 167.
94. E. Schnedermann, Ph.D. thesis, Universität Regensburg, 1992.
95. K. S. Lee, U. Heinz, and E. Schnedermann, *Z. Phys.* **C48** (1990) 525.
96. E. Schnedermann, J. Sollfrank, and U. Heinz, *Phys. Rev.* **C48** (1993) 2462;
    E. Schnedermann, J. Sollfrank, and U. Heinz, in: *Particle Production in Highly Excited Matter*, H. H. Gutbrod and J. Rafelski (Eds.), Plenum, New York, 1993, p. 175.
97. E. Schnedermann and U. Heinz, *Phys. Rev. Lett.* **69** (1992) 2908;
    E. Schnedermann and U. Heinz, *Phys. Rev.* **C47** (1993) 1738;
    E. Schnedermann and U. Heinz, *Phys. Rev.* **C50** (1994) 1675.
98. S. Wenig, Ph.D. thesis, Universität Frankfurt, GSI-Report GSI-90-23, 1990.
99. J.D. Bjørken, *Phys. Rev.* **D27** (1983) 140.
100. J. Letessier, A. Tounsi, U. Heinz, J. Sollfrank, and J. Rafelski, *Phys. Rev. Lett.* **70** (1993) 3530.
101. M. Gaździcki and O. Hansen, *Nucl. Phys.* **A528** (1991) 754.
102. J. Bächler et al., NA35 Collaboration, *Z. Phys.* **C51** (1991) 157;
    J. Bächler et al., NA35 Collaboration, *Phys. Rev. Lett.* **72** (1994) 1419.
103. H.R. Jaqama, A.Z. Mekjian, and L. Zamick, *Phys. Rev.* **C29** (1984) 2067.
104. T.K. Hemmick et al., E814 Collaboration, *Nucl. Phys.* **A566** (1994) 435c.
105. J. Barrette et al., E814 Collaboration, *Phys. Lett.* **B333** (1994) 33;
    N. Xu et al., E814 Collaboration, *Nucl. Phys.* **A566** (1994) 585c.

106. J. Barrette et al., E814 Collaboration, SUNY-RHI-94-9, submitted to *Phys. Rev.* **C**.

107. N. Xu et al., E814 Collaboration, contribution to the 10th Winter Workshop on Nuclear Dynamics, Snowbird, Utah, Jan. 15-21, 1994.

108. A. Aoki et al., E858 Collaboration, *Phys. Rev. Lett.* **69** (1992) 2345.

109. Y. Takahashi *et al.*, CERN–EMU05 Collaboration, private communication and to be published; a figure of their measurement of $D_Q$ as a function of rapidity is shown in Ref. 25.

110. M. Gaździcki, in: *Hot Hadronic Matter: Theory and Experiment*, J. Letessier et al. (Eds.), Plenum, New York, 1995;
M. Gaździcki, preprint IKF-HENPG/5-94, *Z. Phys.* **C**, in press;
M. Gaździcki, in Ref. 12.

111. M. Gaździcki and D. Röhrich, *Z. Phys.* **C65** (1995) 215.

112. J. P. Bondorf, S. I. A. Garpman, and J. Zimányi, *Nucl. Phys.* **A296** (1978) 320.

113. J. Simon-Gillo, NA44 Collaboration, in Ref. 13.

114. S. Nagamiya, *Phys. Rev. Lett.* **49** (1982) 1383.

115. U. Heinz, K. S. Lee, and M. J. Rhoades-Brown, *Phys. Rev. Lett.* **58** (1987) 2292.

116. V. Cianciolo et al., E859 Collaboration, in Ref. 13.

117. T. Csörgő and B. Lørstad, Univ. Lund preprint LUNFD6(NNFL-7082) (1994).

118. S. Chapman, P. Scotto, and U. Heinz, *Phys. Rev. Lett.* **74** (1995), in press;
S. Chapman, P. Scotto, and U. Heinz, *Heavy Ion Physics* **1** (1995) 1.

119. Z. Árvay et al., *Z. Phys.* **A348** (1994) 201.

120. T. S. Biró, P. Lévai, and J. Zimányi, *Phys. Lett.* **B347** (1995) 6.

121. N. Bilić, J. Cleymans, E. Suhonen, and D.W. von Oertzen, *Phys. Lett.* **B311** (1993) 266.

122. N. Bilić, J. Cleymans, K. Redlich and E. Suhonen, *Z. Phys.* **C63** (1994) 525.

123. T. Csörgő and L. P. Csernai, *Phys. Lett.* **B333** (1994) 494

124. L. van Hove, *Z. Phys.* **C21** (1983) 93.

125. L. P. Csernai and J. I. Kapusta, *Phys. Rev. Lett.* **69** (1992) 737; *Phys. Rev.* **D46** (1992) 1379.

126. Proceedings of *Quark Matter '91*, G. Baym et al. (Eds.), *Nucl. Phys.* **A544** (1992).

127. Proceedings of *Quark Matter '93*, E. Stenlund et al. (Eds.), *Nucl. Phys.* **A566** (1994).

**E802/E866**

128. T. Abbott et al., *Phys. Rev. Lett.* **64** (1990) 847.
129. T. Abbott et al., *Phys. Rev. Lett.* **66** (1991) 1567.
130. Y. Miake et al., *Nucl. Phys.* **A525** (1991) 231c.
131. J. B. Costales et al., *Nucl. Phys.* **A525** (1991) 455c.
132. M. A. Bloomer et al., *Nucl. Phys.* **A527** (1991) 595c.
133. T. Abbott et al., *Phys. Lett.* **B291** (1992) 341.
134. B.A. Cole et al., *Nucl. Phys.* **A544** (1992) 553c.
135. T. Abbott et al., *Phys. Rev. Lett.* **70** (1993) 1393.
136. M. Gonin et al., *Nucl. Phys.* **A553** (1993) 799c.
137. H. Hamagaki et al., *Nucl. Phys.* **A566** (1994) 27c.
138. M. Gonin et al., *Nucl. Phys.* **A566** (1994) 601c.
139. T. Abbott et al., *Phys. Rev.* **C50** (1994) 1024.
140. F. Videbaek et al., in Ref. 13.

### E810

141. S. E. Eiseman et al., *Phys. Lett.* **B248** (1990) 254.
142. W. A. Love et al., *Nucl. Phys.* **A525** (1991) 601c.
143. K. J. Foley et al., *Nucl. Phys.* **A544** (1992) 335c.
144. S. E. Eiseman et al., *Phys. Lett.* **B297** (1992) 44.
145. R. S. Longacre et al., *Nucl. Phys.* **A566** (1994) 167c.
146. S. E. Eiseman et al., *Phys. Lett.* **B325** (1994) 322.

### E814/E877

147. J. Stachel et al., *Nucl. Phys.* **A566** (1994) 183c.

### E859

148. W. A. Zajc et al., *Nucl. Phys.* **A544** (1992) 237c.
149. G. S. F. Stephans et al., *Nucl. Phys.* **A566** (1994) 269c.
150. Y. Wang et al., *Nucl. Phys.* **A566** (1994) 379c.
151. D. P. Morrison et al., *Nucl. Phys.* **A566** (1994) 457c.
152. O. E. Vossnack et al., *Nucl. Phys.* **A566** (1994) 535c.
153. B. Cole et al. in Ref. 13.

### NA34 Helios/2

154. H. van Hecke et al., *Nucl. Phys.* **A525** (1991) 227c.

155. T. Åkesson et al., *Phys. Lett.* **B273** (1992) 273.

### NA35

156. A. Bamberger et al., *Z. Phys.* **C43** (1989) 25.
157. J. Bartke et al., *Z. Phys.* **C48** (1990) 191.
158. H. Ströbele et al., *Nucl. Phys.* **A525** (1991) 59c.
159. R. Stock et al., *Nucl. Phys.* **A525** (1991) 221c
160. P. Seyboth et al., *Nucl. Phys.* **A544** (1992) 293c.
161. M. Kowalski et al., *Nucl. Phys.* **A544** (1992) 609c.
162. J. Bächler et al., *Z. Phys.* **C58** (1993) 367.
163. D. Röhrich et al., *Nucl. Phys.* **A566** (1994) 35c.
164. M. Gaździcki et al., *Nucl. Phys.* **A566** (1994) 503c.
165. T. Alber et al., *Z. Phys.* **C64** (1994) 195.
166. J. Günther et al., in Ref. 13.

### NA36

167. E. Andersen et al., *Phys. Lett.* **B294** (1992) 127.
168. D. E. Greiner et al., *Nucl. Phys.* **A544** (1992) 309c.
169. E. Andersen et al., *Phys. Rev.* **C46** (1992) 727.
170. E. Andersen et al., *Phys. Lett.* **B316** (1993) 603.
171. E. Andersen et al., *Nucl. Phys.* **A553** (1993) 817c.
172. J. M. Nelson et al., *Nucl. Phys.* **A566** (1994) 217c.
173. E. Andersen et al., *Phys. Lett.* **B327** (1994) 433.

### NA44

174. M. Sarabura et al., *Nucl. Phys.* **A544** (1992) 125c.
175. M. Murray et al., *Nucl. Phys.* **A566** (1994) 515c.
176. H. Beker et al., *Z. Phys.* **C64** (1994) 209.
177. H. Beker et al., CERN-PPE/94-119, submitted to *Phys. Rev. Lett.*

### WA85

178. S. Abatzis et al., *Nucl. Phys.* **A498** (1989) 369c.
179. S. Abatzis et al., *Phys. Lett.* **B244** (1990) 130.
180. S. Abatzis et al., *Phys. Lett.* **B259** (1991) 508.
181. S. Abatzis et al., *Phys. Lett.* **B270** (1991) 123.
182. D. Evans et al., *Nucl. Phys.* **A525** (1991) 441c.

183. S. Abatzis et al., *Nucl. Phys.* **A525** (1991) 445c.
184. J. B. Kinson et al., *Nucl. Phys.* **A544** (1992) 321c.
185. S. Abatzis et al., *Phys. Lett.* **B316** (1993) 615.
186. D. Evans et al., *Nucl. Phys.* **A566** (1994) 225c.
187. F. Antinori et al., *Nucl. Phys.* **A566** (1994) 491c.
188. S. Abatzis et al., Proceedings of the *27th Int. Conf. on High Energy Physics*, Glasgow, 1994, in press.
189. S. Abatzis et al., *Phys. Lett.* **B347** (1995) 158.
190. N. di Bari et al., in Ref. 12.

### WA94

191. A. C. Bayes et al., *Nucl. Phys.* **A566** (1994) 499c.